\documentclass[a4paper,11pt]{article}
\pdfoutput=1 
\usepackage{jcappub}
\usepackage{amsmath}
\usepackage{graphicx}
\usepackage{latexsym}
\usepackage{xspace}
\usepackage{color}
\usepackage[dvipsnames]{xcolor}
\usepackage{hyperref} 
\usepackage{bm}
\usepackage{relsize}
\usepackage{tabularx}
\usepackage{multirow}
\usepackage{amssymb}
\usepackage[table]{xcolor}
\usepackage{braket}
\usepackage[utf8]{inputenc}
\usepackage{slashed}
\usepackage{empheq}
\usepackage{cancel}
\usepackage{soul}
\usepackage[most]{tcolorbox}
\usepackage{capt-of}
\usepackage{xfrac} 
\usepackage[compat=1.1.0]{tikz-feynman}

\usepackage{booktabs}

\makeatletter
\gdef\@fpheader{}
\g@addto@macro\bfseries{\boldmath}
\makeatother

\usepackage{savesym}

\savesymbol{Ref}
\savesymbol{erf}

\newcommand{\dd}{\mathrm{d}}
\newcommand{\ee}{e}

\newcommand{\boldmathsymbol}[1]{{\ensuremath{\boldsymbol{#1}}}}

\newcommand{\bmk}{\boldmathsymbol{k}}

\newcommand{\bms}{\boldmathsymbol{s}}

\newcommand{\beq}{\begin{equation}}
\newcommand{\eeq}{\end{equation}}
\newcommand{\bea}{\begin{equation}\begin{aligned}}
\newcommand{\eea}{\end{aligned}\end{equation}}

\newlength{\wsingfig}
\newlength{\wdblefig}
\newlength{\wquadfig}
\newlength{\wtriplefig}
\newcommand{\Eq}[1]{Eq.~(\ref{#1})}
\newcommand{\Eqs}[1]{Eqs.~(\ref{#1})}
\newcommand{\Fig}[1]{Fig.~{\ref{#1}}}

\newcommand{\Sec}[1]{Sec.~\ref{#1}}

\newcommand{\App}[1]{Appendix~\ref{#1}}

\restoresymbol{phys}{Ref}
\restoresymbol{phys}{erf}

\usepackage[english]{babel}
\usepackage{numprint}
\usepackage{indentfirst}
\usepackage[T1]{fontenc}
\usepackage{lmodern}
\usepackage{eurosym}
\usepackage{lipsum}
\usepackage{textcomp}
\usepackage{wasysym}
\usepackage{textgreek}
\usepackage{listings}
\usepackage{url}
\usepackage[margin=1in,includefoot]{geometry}
\usepackage{soul}

\usepackage{xfrac} 
\usepackage{amsbsy}
\usepackage{amsmath}
\usepackage{adjustbox}
\usepackage{mathtools}
\usepackage{gensymb}
\usepackage{breqn}
\usepackage{cancel}
\usepackage{multirow}
\usepackage{amssymb}
\usepackage{physics}
\numberwithin{equation}{section}

\usepackage{xcolor}
\usepackage{amstext} 
\usepackage{array}   
\newcolumntype{L}{>{$}l<{$}} 

\definecolor{myblue}{RGB}{0, 100, 200}
\definecolor{myred}{RGB}{214, 39, 40}
\definecolor{mygreen}{RGB}{44, 160, 44}
\definecolor{mybrown}{RGB}{123, 64, 26}
\definecolor{mydarkblue}{RGB}{44, 77, 118}

\def\sgn{\mathrm{sgn}\,}

\usepackage[framemethod=default]{mdframed}
\newmdenv[backgroundcolor=gray!15,%
skipabove=5pt,%
skipbelow=5pt,%
leftmargin=2pt,%
rightmargin=2pt,%
innertopmargin=-5pt,%
innerbottommargin=5pt,%
innerleftmargin=5pt,%
innerrightmargin=5pt,%
splittopskip=0pt,%
splitbottomskip=0pt,%
linewidth=0pt,%
nobreak=true]%
{keyeqn}

\usepackage{latexsym}
\usepackage{graphicx}  
\usepackage{float} 
\usepackage{ifoddpage}
\graphicspath{ {images/} }

\usepackage{tikz}	
\usepackage{tikz-feynman}
\tikzfeynmanset{compat=1.0.0}

\subheader{}

\title{Open Effective Field Theory and the Physics of Cosmological Collider Signals
}

\author[a]{Thomas Colas,}
\affiliation[a]{Department of Applied Mathematics and Theoretical Physics, University of Cambridge, Wilberforce Road, Cambridge, CB3 0WA, UK}
\author[a,b]{Zhehan Qin,}
\affiliation[b]{Department of Physics, Tsinghua University, Beijing 100084, China}
\author[a]{and Xi Tong}
\emailAdd{tc683@cam.ac.uk, qzh21@mails.tsinghua.edu.cn, xt246@cam.ac.uk}

\begin{document}
	\sloppy

\abstract{
    We examine the origin of the cosmological collider signal using the framework of open effective field theories. Focusing on the single exchange of a massive scalar field, we demonstrate that the trispectrum splits cleanly into its local and non-local components once the heavy-field propagators are decomposed in the Keldysh basis. Integrating out the massive degree of freedom yields a single-field effective field theory for the light scalar that necessarily contains both unitary operators and non-unitary contributions associated with dissipation and stochastic noise. We show that the leading local signal in parity-preserving theories arises from the unitary part of this effective field theory, whereas the non-local signal is intrinsically associated with its stochastic sector. The effective field theory coefficients themselves are a priori non-analytic in the external kinematics; however, this non-analyticity can be softened when a scale hierarchy — such as the heavy-mass expansion — is imposed, up to spurious contributions that ultimately cancel in observables. Finally, we establish a connection between the cosmological collider signal and entropy production, linking the observable non-local signal to intrinsic properties of the quantum state, including its degree of mixedness.
} 
\
	
\maketitle

\section{Introduction}

The Cosmological Collider (CC) program \cite{Chen:2009zp, Baumann:2011nk, Noumi:2012vr, Arkani-Hamed:2015bza, Lee:2016vti} aims at probing the smallest scales of physical reality from the largest  scales of the universe. During inflation, the exponential expansion of the spacetime drives the production of heavy particles with mass up to a few times the Hubble parameter, reaching up to $H\lesssim 10^{13}$GeV, far exceeding that of any terrestrial colliders. These massive particles interact with the massless inflaton and source non-Gaussian correlations in the Cosmic Microwave Background (CMB) and Large-Scale Structure (LSS) with distinctive signatures known as \textit{cosmological collider signals}.
        
Mathematically speaking, these cosmological collider signals are non-analyticities in the kinematic variables characterising the non-Gaussian correlators. They appear in the soft limits as non-integer scaling of momentum ratios for light fields and oscillations for heavy fields. From the kinematic structure of these scaling/oscillating signals, one can uniquely pin down the mass, spin, speed of sound and chemical potential of extra degrees of freedom active during inflation predicted from high-energy particle physics and gravity (see recent developments in
\cite{Craig:2024qgy,McCulloch:2024hiz,Wu:2024wti,Aoki:2024uyi,Aoki:2024jha,Hubisz:2024xnj,Bodas:2024hih, Pajer:2024ckd,Chakraborty:2025myb,Pimentel:2025rds, An:2025mdb, Wang:2025qww, Bodas:2025wuk,Wang:2025qfh, Aoki:2025uff,Kumar:2025anx, Jazayeri:2025vlv}).
        
Analogous to the way new particles are searched for at terrestrial colliders, one starts with model building and tries to put forward a consistent model of massive particles interacting with the massless inflaton. These models often take the form of multi-field Effective Field Theories (EFTs) built around a pure state (the Bunch-Davies vacuum) that unitarily evolves in time. One then proceeds to compute the non-Gaussian correlators spontaneously generated out of such unitary time evolution. The predicted signals are then tested against observational data model-by-model via template matching \cite{Planck:2013wtn,Cabass:2024wob,Sohn:2024xzd,Philcox:2025wts,Philcox:2025bbo,Suman:2025vuf}. 

\begin{figure}[tbp]
    \centering
    \includegraphics[width=0.84\textwidth]{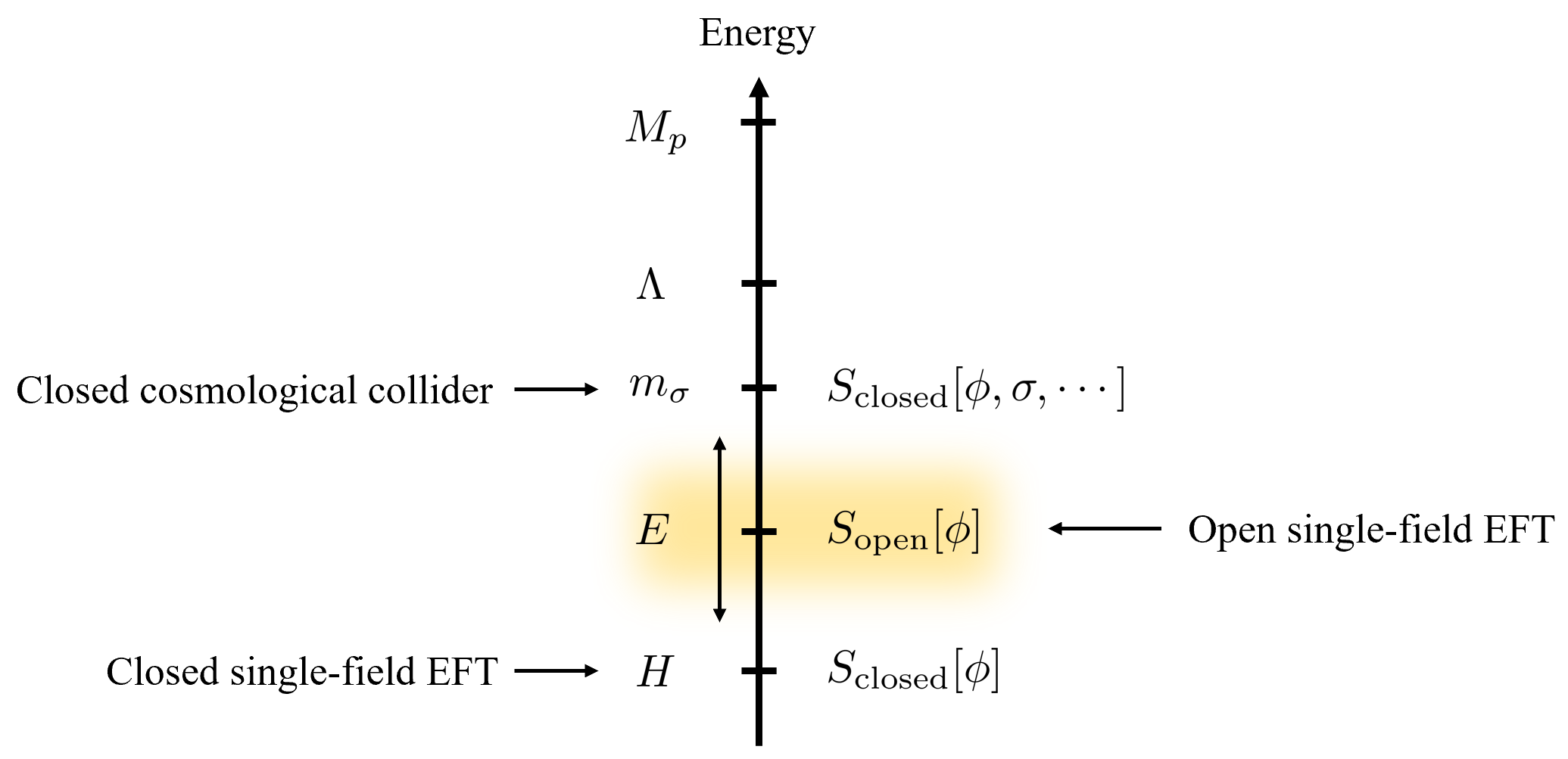}
    \caption{The \textcolor{YellowOrange}{open single-field EFT} interpolates between a closed multi-field EFT where cosmological collider physics is conventional studied and a closed single-field EFT where the heavy fields are completely integrated out. Here $M_p$ denotes Planck scale, $\Lambda$ denotes the cutoff of the closed multi-field EFT and $m_\sigma$ denotes collectively the mass of the heavy fields $\sigma$. At an energy scale $H<E<m_\sigma$, the \textcolor{YellowOrange}{open single-field EFT} is insufficient to explicitly resolve on-shell heavy particle states, but it does capture the stochastic disturbances due to a noisy environment of heavy fields.}
    \label{fig:cartoon}
\end{figure}

In Minkowski scattering experiments, heavy fields affect light degrees of freedom only through suppressed higher-derivative operators. This separation of scales and the unitarity of low-energy EFTs are justified by energy conservation, a cherished property lost in dynamical spacetimes such as inflation \cite{Burgess:2024heo}. Therefore, it becomes questionable whether multi-field \textit{closed} EFTs — evolving unitarily — are an under-fit or an over-fit to the true dynamics of fields during inflation. One potential way to approach this problem is to place, in between the single-field and multi-field closed EFTs, a single-field \textit{open} EFT — where IR unitarity is relaxed — to better understand the dialogue between an invisible massive-field environment and the massless-field system (see \Fig{fig:cartoon}). Such endeavours have also been made recently in \cite{Green:2024cmx, Cespedes:2025ple}.

The scope of this work, therefore, is to initiate a systematic framework for describing how heavy fields influence the dynamics and statistics of light degrees of freedom in cosmology. To further explore the effect of the cosmological collider signals on the light sector, we focus on the trispectrum (the four-point function in spatial Fourier space) of a massless scalar field $\varphi$, generated at tree level by the exchange of a massive scalar field $\sigma$. We compute this observable within the Schwinger--Keldysh formalism \cite{Schwinger:1960qe, Keldysh:1964ud}, both in Minkowski and in de Sitter spacetime. In this framework, the familiar Feynman propagator used in scattering amplitudes is replaced by two independent bulk-to-bulk propagators, $D^\sigma_{++}$ and $D^\sigma_{-+}$ (together with their complex conjugates $D^\sigma_{--}$ and $D^\sigma_{+-}$). As a result, the physics underlying a simple exchange diagram becomes considerably more intricate \cite{Chen:2017ryl}. We show that, fortunately, working in the so-called \textit{Keldysh basis} --- a rotation of the $+/-$ basis --- renders the physical interpretation more transparent. In this basis, the signal naturally decomposes into a conservative part, corresponding to the flat-space amplitude, a dissipative part, and a stochastic part. At leading order, these contributions map onto the \textit{local} and \textit{non-local} parts of the cosmological collider signal, offering a transparent link between the kinematic structure of the signal and its physical interpretation.

We also derive a single-field effective theory for $\varphi$ in which $\sigma$ has been integrated out. In order to reproduce the correct perturbative trispectrum at a given order in the coupling constant, this effective theory \textit{must be open} \cite{Salcedo:2024smn, Salcedo:2024nex, Salcedo:2025ezu, Colas:2025app}, implying that $\varphi$ experiences dissipation and noise induced by its coupling to $\sigma$. 
The resulting open EFT can be expressed in terms of local operators, at the price of introducing non-analytic structures in their coefficients — poles in flat space and branch cuts in de Sitter — which are softened when a scale hierarchy, such as the heavy-mass expansion, is imposed. We relate the emergence of dissipative and stochastic operators to particle production in the heavy sector, as captured by the evolution of entropy measures. This correspondence reproduces known results on the de- and re-coherence of massless fields coupled to heavy fields in de Sitter space \cite{Colas:2021llj, Colas:2022hlq, Colas:2022kfu, Colas:2024xjy, Burgess:2024eng, Colas:2024ysu, Burgess:2024heo}, highlighting the interplay between quantum information measures and cosmological collider signals.

This work builds upon recent advances in the characterisation of the CC signal \cite{Liu:2019fag, Tong:2021wai, Tong:2022cdz, Qin:2022lva, Stefanyszyn:2023qov, Qin:2023ejc, Qin:2023bjk, Bodas:2024hih, Aoki:2024jha, Aoki:2024uyi, Wang:2025qww, Bodas:2025wuk, Liu:2024xyi, Qin:2025xct, Cheung:2025dmc,Xianyu:2025lbk, Aoki:2025uff}, as well as on the recent developments in cosmological EFTs formulated on the Schwinger–Keldysh contour \cite{Mukohyama:2020lsu, Aoki:2021ffc, DuasoPueyo:2025lmq}. It offers a complementary perspective to \cite{Salcedo:2022aal, Green:2024cmx, Mahajan:2025iuz}. In particular, adopting the Keldysh basis proves especially powerful for exposing the underlying physics, as demonstrated in \cite{Chowdhury:2023arc, Ema:2024hkj, Ema:2025ftj}.

This work is organized as follows. In \Sec{sec:prelude}, we develop our analysis on a half-Minkowski background. This setup illustrates the role of the various propagators along the Schwinger--Keldysh contour and clarifies the structure of the single-field EFT obtained after integrating out the heavy field. In \Sec{sec:Keldysh}, we extend this framework to a massive field in a de Sitter background, leading to the characterisation of the cosmological collider signal summarised in Tab.\,\ref{tab_sum}. Finally, \Sec{sec:EFT} is devoted to constructing a time-local single-field EFT that reproduces the cosmological collider signal. The price to pay for such a construction is the appearance of non-analytic EFT coefficients, whose structure we discuss in detail. Concluding remarks are gathered in \Sec{sec:conclu}, followed by a series of Appendices where technical details are presented.

\subsection{Summary of the main results} 

\paragraph{General strategy.} We consider a massless and a massive scalar coupled through\footnote{Note that here the energy scale $\Lambda$ in the coupling could be the same as or higher than the cutoff scale of this two-field EFT.}
\begin{equation}\label{eq:modelintro}
    \mathcal L[\varphi,\sigma] =  -\frac12 (\partial_\mu \varphi)^2 - \frac12 (\partial_\mu \sigma)^2 - \frac12 M^2\sigma^2 + \frac{1}{2\Lambda} \dot\varphi^2\sigma,
\end{equation}
and aim to compute the tree-level $s$-channel four-point correlator
\begin{align}
    \mathcal I \equiv \langle \varphi_{\bm k_1}\varphi_{\bm k_2}\varphi_{\bm k_3}\varphi_{\bm k_4} \rangle_s',
\end{align}
where prime denotes that the momentum conservation factor $(2\pi)^3\delta(\bm k_1+\bm k_2+\bm k_3+\bm k_4)$ has been stripped off. The bulk-to-bulk propagators of the exchanged massive field $\sigma$ are decomposed into 
\begin{subequations}
    \begin{align}
            &D^\sigma_{\mp\pm}(k;\tau_1,\tau_2) = -i G_\sigma^K(k;\tau_1,\tau_2) \mp \frac i2 G_\sigma^\Delta(k;\tau_1,\tau_2),\\
            &D^\sigma_{\pm\pm}(k;\tau_1,\tau_2) = -i G_\sigma^K(k;\tau_1,\tau_2) \mp i G_\sigma^P(k;\tau_1,\tau_2).
        \end{align}
\end{subequations}
where $\tau$ is the time variable, either the cosmic time $t$ or the conformal time $\eta$, and $k = |\bmk|$. $G_\sigma^K$, $G_\sigma^\Delta$ and $G_\sigma^P$ are known as the Keldysh, Pauli-Jordan and principal-value propagators respectively. A first layer of analysis consists in decomposing $\mathcal I $ into the contributions of each of these propagators,
\begin{align}\label{eq:decomp}
    \mathcal I  = \mathcal I^P + \mathcal I^\Delta + \mathcal I^K.
\end{align}
As we will see, this already illustrates the underlying physical mechanisms behind each contribution. While $\mathcal I^P$ corresponds to the $2$-$2$ scattering process familiar from the in-out framework, $\mathcal I^\Delta$ and $\mathcal I^K$ have no analogue in a single-branch path integral. Specifically, $\mathcal I^\Delta$ encodes dissipation, arising when the $\varphi$ field loses energy to the $\sigma$ medium, whereas $\mathcal I^K$ captures noise, that is the on-shell fluctuations of $\sigma$ that source the dynamics of $\varphi$. These distinct contributions give rise to physically distinguishable signals, such as the local and non-local signals of the cosmological collider, as discussed below.

\begin{figure}[tbp]
    \centering
    \includegraphics[width=0.9\textwidth]{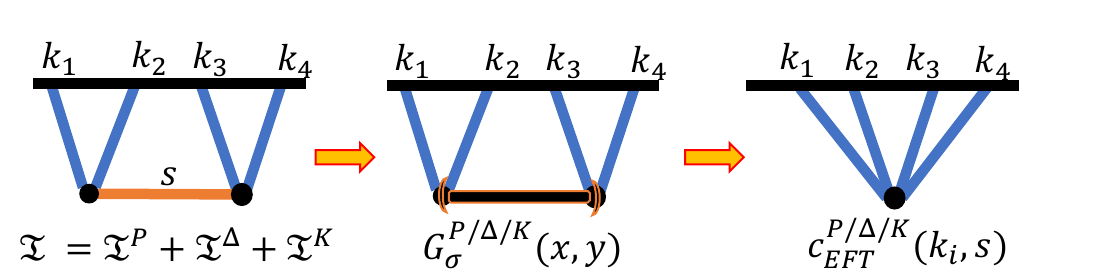}
    \caption{Illustration of the progression from a two-field microphysical model (\textit{Left}) to a single-field time-local open EFT (\textit{Right}). The trispectrum $\mathcal I$ is conveniently decomposed in the Keldysh basis into three contributions, each with a distinct physical origin. Integrating out the $\sigma$ field initially yields a non-local open EFT (\textit{Middle}), which incorporates dissipation, governed by the Pauli-Jordan propagator $G^\Delta_\sigma$, and noise, governed by the Keldysh propagator $G^K_\sigma$. At a given order in perturbation theory, one can employ the free equations of motion to recast this EFT in a time-local form. This procedure introduces non-analyticities in the EFT coefficients; however, in the presence of a suitable hierarchy of scales, these non-analyticities are systematically removed through a perturbative expansion in the hierarchy parameter. }
    \label{fig:EFT_steps}
\end{figure}

A second layer of analysis involves constructing a single-field EFT by integrating out the heavy field $\sigma$, as illustrated in \Fig{fig:EFT_steps}. This is naturally formulated in the Schwinger-Keldysh framework via the influence functional \cite{FEYNMAN1963118} (see \cite{Colas:2025app} for introductory lectures). Within this formalism, correlators of the $\varphi$ field are obtained as functional derivatives of the generating functional 
\begin{align}\label{eq:gen}
    \mathcal{Z}\left[J_r, J_a \right] = \int_{\mathrm{BD}}^\varphi \mathcal{D}\varphi_r \int_{\mathrm{BD}}^0 \mathcal{D}\varphi_a &  \ee^{iS_\varphi\left[ \varphi_r, \varphi_a\right]} \ee^{iS_{\mathrm{IF}}\left[ \varphi_r, \varphi_a\right]} \ee^{i \int \dd^4 x \sqrt{-g}\left(J_r\varphi_a + J_a\varphi_r \right)},\Bigg.
\end{align}
with respect to the sources $J_a$, where $\varphi_r = (\varphi_+ + \varphi_-)/2$ and $\varphi_a = \varphi_+ - \varphi_-$ give a convenient decomposition of the path integral variables $\varphi_\pm$ known as the Keldysh basis. $S^\varphi\left[ \varphi_r, \varphi_a\right]$ is the functional of the $\varphi$ only, while the effect of $\sigma$ on $\varphi$ is captured by $S_{\mathrm{IF}}\left[ \varphi_r, \varphi_a\right]$ which is formally defined by 
\begin{align}
    \ee^{iS_{\mathrm{IF}} \left[ \varphi_r, \varphi_a\right]} = \int_{\mathrm{BD}}^\sigma \mathcal{D}\sigma_r \int_{\mathrm{BD}}^0 \mathcal{D}\sigma_a  \ee^{iS_\sigma\left[ \sigma_r, \sigma_a\right]} \ee^{iS_{\mathrm{int}}\left[ \varphi_r, \varphi_a, \sigma_r, \sigma_a\right]}.\Bigg.
\end{align}
The explicit computation of $S_{\mathrm{IF}}\left[ \varphi_r, \varphi_a\right]$ at second order in the $\varphi$-$\sigma$ coupling constant in flat and de Sitter space is a technical result of this work. This procedure illustrates the emergence of effective quartic interactions in the $\varphi$ sector that can either be unitary, dissipative or stochastic, related to $G_\sigma^P$, $G_\sigma^\Delta$ and $G_\sigma^K$ respectively.

The resulting open theory is, in general, non-local. At second order in $1/\Lambda$, however, one can use the free propagators of $\varphi$ to recast the EFT entirely in terms of local operators. In this formulation, the non-locality is encoded in the intricate expression of the EFT coefficients, which can feature non-analytic structures such as folded poles in flat spacetime and branch cuts in de Sitter. In the presence of a scale hierarchy, these non-analyticities are effectively softened, up to spurious contributions that ultimately cancel in observables. Analysing the non-analytic structure of these coefficients constitutes the final series of results presented in this work.

\paragraph{Flat-space results.} In the Keldysh basis, the tree-level $s$-channel exchange
\begin{align}
    \mathcal I &= \frac{k_T+E_s}{8\Lambda^2E_sk_T(k_{12}+E_s)(k_{34}+E_s)},
\end{align}
where $E_s = \sqrt{M^2 + s^2}$ and we have defined the total energy $k_T\equiv k_{1234}$ with the shorthand $k_{ij\cdots} \equiv k_i+k_j+\cdots$, decomposes into 
\begin{align}
    \mathcal I^P
    =&-\frac{1}{16\Lambda^2k_T}\Big(\frac{1}{k_{12}^2-E_s^2}+\frac{1}{k_{34}^2-E_s^2}\Big) = \frac{1}{8\Lambda^2k_T} \frac{1}{M^2}
    + \frac{k_{12}^2+k_{34}^2-2s^2}{16\Lambda^2k_T}\frac{1}{M^4}
    + \cdots, \\
    \mathcal I^\Delta =&-\frac{k_T}{16\Lambda^2(k_{12}^2-E_s^2)(k_{34}^2-E_s^2)}  = -\frac{k_T}{16\Lambda^2} \frac{1}{M^4} +  \cdots, \\
    \mathcal I^K =&~ \frac{ k_{12}k_{34}}{8\Lambda^2E_s(k_{12}^2-E_s^2)(k_{34}^2-E_s^2)} = \frac{k_{12}k_{34}}{8\Lambda^2} \frac{1}{M^5} + \cdots.
\end{align}  
$\mathcal I^P$ has a \emph{total energy pole}, corresponding to the $2-2$ scattering amplitude (in flat spacetime) \cite{Arkani-Hamed:2017fdk, Goodhew:2020hob, Pajer:2020wxk, Salcedo:2022aal, Cespedes:2025dnq}, which only appears in a unitary theory. Conversely, the Pauli-Jordan and Keldysh propagators capture dissipation and noise, which are non-unitary effects. $\mathcal I^\Delta$ vanishes at $k_T=0$, and it would be interesting to explore whether this is a universal feature.\footnote{This feature holds for de Sitter with half-integer order parameter $\nu = \sqrt{9/4-M^2/H^2}$ for the intermediate field.} Since the traditional (unitary and local) EFT is the $\Box/M^2$ expansion of the heavy propagator, it could only have even powers of $1/M$ and each term contributes to a contact graph possessing the total energy pole. That is, the traditional EFT reproduces the contribution from $\mathcal I^P$. Corrections to this include even powers of $1/M$ from $\mathcal I^\Delta$ that vanishes at $k_T=0$, and odd powers of $1/M$ from $\mathcal I^K$.
        
The various contributions to the trispectrum can be recovered from a single-field open EFT of the form of \Eq{eq:gen} where the influence functional reads
\begin{align}
    S_{\mathrm{IF}}[\varphi_r,\varphi_a] =~ \int_{-\infty}^0 \dd t\int\mathcal D\bm k\,
            \bigg\{&c_1(k_1,k_2,E_s) \mathcal{O}_1^P(k_1,k_2,k_3,k_4) + c_2(k_1,k_2,E_s) \mathcal{O}_2^P(k_1,k_2,k_3,k_4) \nonumber \\
            -&c_1(k_1,k_2,E_s) \mathcal{O}_1^\Delta(k_1,k_2,k_3,k_4) - c_2(k_1,k_2,E_s) \mathcal{O}_2^\Delta(k_1,k_2,k_3,k_4)  \nonumber \\
            +&ic_3(k_1,k_2,E_s)\mathcal{O}_1^K(k_1,k_2,k_3,k_4) \bigg\}.
\end{align}
$\mathcal{O}_i^{P/\Delta/K}(k_1,k_2,k_3,k_4)$ denote local quartic operators constructed from $\varphi_r$, $\varphi_a$, and their time derivatives. Their explicit expressions are given in \Eqs{eq_flatIFH}, \eqref{eq_flatIFD}, and \eqref{eq_flatIFK}, though the details are not important for the present discussion. The measure $\int\mathcal D\bm k$ is defined in \eqref{eq_intKmeasure}. The EFT coefficients   
\begin{align}
    &c_1(k_i,k_j,E_s) \equiv \frac{-k_ik_j}{16\Lambda^2E_s}\left(\frac{1}{E_s+ k_{ij}} - \frac{1}{E_s + k_i -k_j} - \frac{1}{E_s-k_i+k_j} + \frac{1}{E_s-k_{ij}}\right),\\
    &c_2(k_i,k_j,E_s) \equiv  \frac{1}{16\Lambda^2E_s}\left(\frac{1}{E_s+ k_{ij}} + \frac{1}{E_s + k_i -k_j} + \frac{1}{E_s-k_i+k_j} + \frac{1}{E_s-k_{ij}}\right),\\
    &c_3(k_i,k_j,E_s) \equiv \frac{k_i}{8\Lambda^2E_s}\left(\frac{1}{E_s+ k_{ij}} + \frac{1}{E_s + k_i -k_j} - \frac{1}{E_s-k_i+k_j} - \frac{1}{E_s-k_{ij}}\right).
\end{align}
are reminiscent of the flat space contact bispectra computed in the Keldysh basis in \cite{Salcedo:2024smn}. Expanding in the heavy mass limit:
\begin{align}
    c_1(k_i,k_j,s) &= \frac{-k_i^2 k_j^2}{2\Lambda^2M^4} + \cdots,\\ 
    c_2(k_i,k_j,s)&= \frac{1}{4\Lambda^2M^2} + \frac{k_i^2+k_j^2- s^2}{4\Lambda^2M^4}  + \cdots,\\
    c_3(k_i,k_j,s) &= -\frac{k_i^2}{2\Lambda^2M^3} -\frac{k_i^2(2k_i^2+6 k_j^2-3s^2)}{4\Lambda^2M^5}  + \cdots,
\end{align}
where we find that each term, order by order in $1/M$, is analytic in all momenta, especially around $s\to 0$. Therefore, such an open EFT, though notably containing a non-unitary part, is actually local in both time and space under heavy mass expansion.


\paragraph{de Sitter results.}

In de Sitter space, the $s$-channel soft limit of the inflaton four-point function features characteristic oscillations 
        \begin{align}
            \lim_{s\to 0}\langle \varphi_{\bm k_1}\varphi_{\bm k_2}\varphi_{\bm k_3}\varphi_{\bm k_4} \rangle'\supset \mathcal{A}^{\rm L} \sin\left(\mu\ln\frac{k_{12}}{k_{34}}\right)+\mathcal{A}^{\rm NL} \cos\left(\mu\ln\frac{4k_{12} k_{34}}{s^2}+\delta\right)\,,
        \end{align}
where $\mu\equiv\sqrt{M^2/H^2-9/4}$ is the dimensionless mass of the $\sigma$ field and $\mathcal{A}^{\rm L}, \mathcal{A}^{\rm NL} \propto \Lambda^{-2} e^{-\pi\mu}$. The two types of functional dependence on the exchange momentum $s$ lead to the so-called \textit{local} signal (the first term), which is analytic in the $s\to 0$ limit, and the \textit{non-local} signal (the second term), which is non-analytic.
        
In the Keldysh basis, the decomposition \eqref{eq:decomp} allows us to show that these two families of cosmological collider signals have distinct origins and physical interpretations summarised in Tab.\,\ref{tab_sum}. Since the principal-value propagator $G^P_\sigma$ and the Pauli-Jordan propagators are analytic in the exchange momentum $s$ in the soft limit, they \textit{cannot} produce non-local signals. We deduce that the non-local signal must solely originate from the Keldysh propagator of the massive field $G^K_\sigma$. The local signal, on the other hand, can \textit{a priori} be produced by any propagators. However, the Pauli-Jordan propagator $G^\Delta_\sigma$ is further suppressed in Boltzmann factors. At last, we show in \Sec{eq:carving} that in parity-conserving theories, the local signal solely originates from the principal-value propagator $G^P_\sigma$. For parity-violating theories, the local signal comes from the principal-value propagator $G^P_\sigma$ in the parity-even sector, and from the Keldysh propagator $G^K_\sigma$ in the parity-odd sector. 

        \begin{table}[t] 
          \centering
           \vspace{2mm}
          \begin{tabular}{c|c|cc}
           \toprule[1.5pt]
             & Parity-even sector & Parity-odd sector \\ \hline
            Local signal& $G^P$ & $G^K$ $\Big.$ \\ \hline
            Non-local signal & $G^K$ & $G^K$ $\Big.$ \\
           \bottomrule[1.5pt] 
          \end{tabular}
          \caption{Origin of the leading cosmological collider signals. Note that the principal-value propagator $G^P$ as well as the leading parity-even local signals therein correspond to unitary effects in the open single-field EFT, whereas all other signals are associated with the Keldysh propagator $G^K$ and hence non-unitary diffusion.}
          \label{tab_sum}
        \end{table}

Explicitly, this can be seen by isolating the contributions to the four-point correlator $\mathcal I$ \eqref{eq_dS4ptinin} from each propagator of the exchanged field $\sigma$ in the Keldysh-r/a basis, $\mathcal I  = \mathcal I^P + \mathcal I^\Delta + \mathcal I^K$. Through a procedure detailed in \Sec{eq:carving}, we obtain 
        \begin{align}
           \mathcal{I}^P 
            =&~\frac{H^6}{16\Lambda^2k_1k_2k_3k_4 (k_{12}k_{34})^{5/2}}
            \bigg[ \frac{i\sinh\pi\mu}{2\pi}\, {\mathbf F}_+\left(\frac{s}{k_{12}}\right){\mathbf F}_-\left(\frac{s}{k_{34}}\right) + (\mu\to-\mu) \nonumber \\
            & \qquad\qquad\qquad\qquad\qquad\qquad~ + \mathbf B_>\left(\frac{s}{k_{12}},\frac{s}{k_{34}}\right)\bigg]\,, \\
           \mathcal{I}^K 
            =&~\frac{H^6}{32\pi\Lambda^2k_1k_2k_3k_4 (k_{12}k_{34})^{5/2}}\left[(1+i\sinh\pi\mu)
           {\mathbf F}_+\left(\frac{s}{k_{12}}\right){\mathbf F}_+\left(\frac{s}{k_{34}}\right)  \right. \nonumber\\
           & \qquad\qquad\qquad\qquad\qquad\qquad~ + \left. \cosh^2\pi\mu\,{\mathbf F}_+\left(\frac{s}{k_{12}}\right){\mathbf F}_-\left(\frac{s}{k_{34}}\right)   \right] + (\mu\to-\mu)\,,\\
           \mathcal{I}^\Delta 
            =&~\frac{-\sinh^2\pi\mu\, H^6}{32\pi\Lambda^2k_1k_2k_3k_4 (k_{12}k_{34})^{5/2}}
           {\mathbf F}_+\left(\frac{s}{k_{12}}\right){\mathbf F}_-\left(\frac{s}{k_{34}}\right) + (\mu\to-\mu)\,.
        \end{align}
where we have defined
        \begin{align}
        \mathbf F_\pm(r) = \left(\frac{r}2\right)^{\pm i\mu} \Gamma\left(\frac52\pm i\mu\right)\Gamma(\mp i\mu)\,{}_2\mathrm F_1\left[
        \begin{matrix}
        \frac54\pm\frac{i\mu}2,\frac74\pm\frac{i\mu}2\\
        1\pm i\mu
        \end{matrix}\middle| r^2
        \right],
        \end{align}
that captures the cosmological collider signal through a logarithmic oscillation encoded in the factor $r^{\pm i\mu}$, and 
        \begin{align}
        \mathbf B_>(r_1,r_2) = &~ 24\left(\frac{r_1}{r_2}\right)^{5/2}\sum_{n_1,n_2=0}^\infty  \frac{(-1)^{n_{12}}}{n_1!n_2!}\left(\frac{r_1}{2}\right)^{2n_{12}}\frac{(-i\mu)_{-n_1}(+i\mu)_{-n_2}}{i\mu(2n_2+\frac52-i\mu)} \nonumber\\
        &\times {}_2\mathrm F_1\left[
        \begin{matrix}
        2n_2+\frac52-i\mu,2n_{12}+5 \\ 2n_2+\frac72-i\mu
        \end{matrix}\middle| -\frac{r_1}{r_2}
            \right] + (\mu\to-\mu)\,,
        \end{align}
which is fully analytic in the region $0\leq r_1<r_2 < 1$ (except for the trivial scaling factor $(r_1/r_2)^{5/2}$) and thus plays the role of background that decays as a power law in $\mu$. Through the combinations of $\mathbf F_\pm$, we can check that only $\mathcal{I}^K$ can produce the non-local signal, while all three propagators contribute to the local signal. However, the leading local signal in the large mass limit comes from $\mathcal{I}^P$. 

\begin{figure}[tbp]
    \centering
    \includegraphics[width=0.9\textwidth]{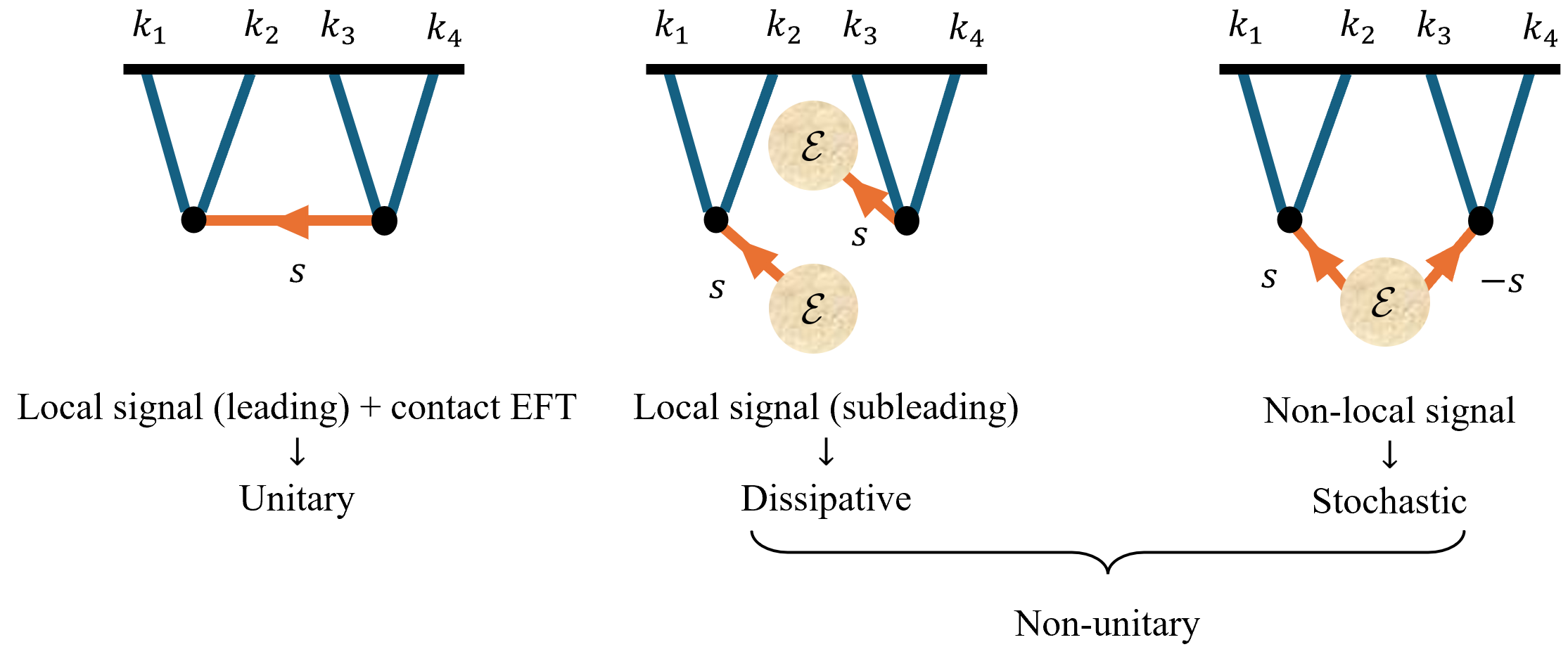}\\
    \caption{An open perspective of cosmological collider signals. Assuming parity invariance, the leading local signals along with the infinite tower of contact EFT operators map to unitary effects in the single-field open EFT. The subleading local signals, on the other hand, manifest themselves as dissipation whereas the non-local signals appear as stochastic noise. Here $\mathcal{E}$ denotes the heavy-field environment and the arrows denote the momentum flow.} 
    \label{fig:OpenPerspectiveOnCCS}
\end{figure}

The cosmological collider signals due to the massive field (the environment) and their impact on the massless field (the system) can be neatly summarised by \Fig{fig:OpenPerspectiveOnCCS}. Assuming parity, the leading local signal derives from the conservative action-reaction process that happen within the system, albeit conducted through a massive medium. Consequently, it affects the system in a unitary fashion. The subleading local signal, on the other hand, derives from non-conservative processes involving entropy injection and information loss, both conducted by on-shell massive particles flowing in and out of the system. Similarly, the non-local signals come from processes where a pair of massive particles in the environment decay into massless particles, injecting entropy into the system. Both signals therefore map to non-unitary effects in the low-energy open EFT of massless fields.

In Fourier space, the influence functional obtained from \Eq{eq:modelintro} when integrating out $\sigma$ is 
\begin{align}
    S_{\mathrm{IF}}[\varphi_r,\varphi_a] =~ \int_{-\infty}^0 \dd \eta\int\mathcal D\bm k\,
    \bigg\{&c_1(k_1,k_2,s;\eta) \widetilde{\mathcal{O}}_1^P + c_2(k_1,k_2,s;\eta) \widetilde{\mathcal{O}}_2^P + c_3(k_1,k_2,s;\eta) \widetilde{\mathcal{O}}_3^P \\
    -&c_1(k_1,k_2,s;\eta) \widetilde{\mathcal{O}}_1^\Delta + c_2(k_1,k_2,s;\eta) \widetilde{\mathcal{O}}_2^\Delta + c_3(k_1,k_2,s;\eta) \widetilde{\mathcal{O}}_3^\Delta   \nonumber \\
    +&i\tilde{c}_1(k_1,k_2,s;\eta) \widetilde{\mathcal{O}}_1^K + i\tilde{c}_2(k_1,k_2,s;\eta) \widetilde{\mathcal{O}}_2^K + \tilde{c}_3(k_1,k_2,s;\eta) \widetilde{\mathcal{O}}_3^K\bigg\}. \nonumber 
\end{align}
where $\widetilde{\mathcal{O}}_i^{P/\Delta/K}(k_1,k_2,k_3,k_4)$ are local quartic operators built of $\varphi_r$ and $\varphi_a$ and their time derivatives whose precise form can be found in \Eqs{eq_dSIFH}, \eqref{eq_dSIFD} and \eqref{eq_dSIFK}. In this case, the EFT coefficients can be computed explicitly, as in the flat space case, and simplified to reproduce the leading cosmological collider signal, in which case they take the form 
\begin{align}
        \lim_{s\to 0}c_1(k_1,k_2,s;\eta) =& \left[ -\frac{i k_1k_2\eta^2}{32\Lambda^2\mu} e^{i k_{12}\eta}E_{i\mu-3/2}(i k_{12}\eta) +(\mu\to-\mu) \right] \Bigg. \nonumber\\
        &  +(k_1\to-k_1) +(k_2\to-k_2) + (k_{1,2} \to -k_{1,2})\,, \\
        \lim_{s\to 0}c_2(k_1,k_2,s;\eta) =& \left[ -\frac{i (1-i k_1\eta)(1-i k_2\eta)}{32\Lambda^2H^2\mu k_1k_2\eta^2} e^{i k_{12}\eta}E_{i\mu-3/2}(i k_{12}\eta) +(\mu\to-\mu) \right] \Bigg. \nonumber\\
        &  +(k_1\to-k_1) +(k_2\to-k_2) + (k_{1,2} \to -k_{1,2})\,,\\
        \lim_{s\to 0}c_3(k_1,k_2,s;\eta) =& \left[ -\frac{i k_1(1-i k_2\eta)}{32\Lambda^2H\mu k_2} e^{i k_{12}\eta}E_{i\mu-3/2}(i k_{12}\eta) +(\mu\to-\mu) \right] \Bigg. \nonumber\\
        &  +(k_1\to-k_1) +(k_2\to-k_2) + (k_{1,2} \to -k_{1,2})\,,
        \end{align}    
        and
    \begin{align}
    \lim_{s\to 0}\tilde c_1(k_1,k_2,s;\eta) =& \bigg\{\frac{k_1k_2\eta^2}{16\pi\Lambda^2}  \left[ \frac{\pi\coth \pi\mu}{\mu} - \left(-\frac{s\eta}{2}\right)^{-2i\mu}\Gamma^2(i\mu)  \right] e^{i k_{12}\eta}E_{i\mu-3/2}(i k_{12}\eta) \nonumber\\
    & \qquad+(\mu\to-\mu) \bigg\}  +(k_1\to-k_1) +(k_2\to-k_2) + (k_{1,2} \to -k_{1,2})\,, \\
    \lim_{s\to 0}\tilde c_2(k_1,k_2,s;\eta) =& \bigg\{\frac{(1-i k_1\eta)(1-i k_2\eta)}{16\pi\Lambda^2H^2k_1k_2\eta^2}  \left[ \frac{\pi\coth \pi\mu}{\mu} -\left(-\frac{s\eta}{2}\right)^{-2i\mu}\Gamma^2(i\mu) \right] e^{i k_{12}\eta}E_{i\mu-3/2}(i k_{12}\eta) \nonumber\\
    & \qquad +(\mu\to-\mu) \bigg\}  +(k_1\to-k_1) +(k_2\to-k_2) + (k_{1,2} \to -k_{1,2})\,, \\
    \lim_{s\to 0}\tilde c_3(k_1,k_2,s;\eta) =& \bigg\{\frac{k_1(1-i k_2\eta)}{16\pi\Lambda^2Hk_2} \left[ \frac{\pi\coth \pi\mu}{\mu} -\left(-\frac{s\eta}{2}\right)^{-2i\mu}\Gamma^2(i\mu) \right] e^{i k_{12}\eta}E_{i\mu-3/2}(i k_{12}\eta) \nonumber\\
    & \qquad +(\mu\to-\mu) \bigg\} +(k_1\to-k_1) +(k_2\to-k_2) + (k_{1,2} \to -k_{1,2})\,.
    \end{align}
We demonstrate in \Sec{subsec:recover} that the trispectrum computed from the open EFT perfectly matches the top-down calculation from the two-field model.

Finally, we relate the linear entropy of a mode --- a measure of the information shared between the $\varphi$ and $\sigma$ fields --- to the decomposition of the trispectrum signal, $\mathcal I = \mathcal I^P + \mathcal I^\Delta + \mathcal I^K$. We find that only the $\mathcal I^\Delta + \mathcal I^K$ components contribute to the variation of the linear entropy, consistent with their identification as the non-unitary sector of the EFT. Moreover, in contrast with the dominant local and non-local CC signals, changes in purity constitute a subleading effect proportional to $\exp(-2\pi \mu)$, making their observational detection challenging.

\section{Prelude: flat-space intuition}\label{sec:prelude}

Before diving into dS spacetime, let us first study the same problem but in a much simpler background, namely the Minkowski spacetime, whose metric reads
\begin{equation}
    \mathrm ds^2 = -\mathrm d t^2+ \mathrm d \bm x^2.
\end{equation}
We consider the following two-field toy model with the Lagrangian
\begin{equation}
\label{eq_flatL}
    \mathcal L[\varphi,\sigma] =  -\frac12 (\partial_\mu \varphi)^2 - \frac12 (\partial_\mu \sigma)^2 - \frac12 M^2\sigma^2 + \frac{1}{2\Lambda} \dot\varphi^2\sigma,
\end{equation}
with $\varphi$ being a massless scalar and $\sigma$ a massive scalar with mass $M$. Here $\Lambda$ is the cutoff scale for the dim-5 operator $\dot\varphi^2\sigma$, and we denote derivatives with respect to physical time $t$ by overdots. After quantisation, scalar fields can be expanded in terms of creation/annihilation operators. For instance,
\begin{align}
\sigma(t,\bm x) = \int \frac{\mathrm d^3\bm k}{(2\pi)^3}\,e^{i\bm k\cdot\bm x}\Big[u_\sigma(k,t)a_{_\sigma,\bm k} + u_\sigma^*(k,t)a_{_\sigma,-\bm k}^\dagger\Big],
\end{align}
where $u_\sigma(k,t)$ is the mode function of $\sigma$ determined by solving the Klein-Gordon equation with appropriate initial conditions. Assuming the Bunch-Davis (BD) vacuum, the mode function is given by
\begin{align}
 \label{eq_flatusigma}
 u_\sigma(k,t) = \frac{e^{-i E_kt}}{\sqrt{2E_k}},\qquad E_k \equiv \sqrt{M^2+k^2}.
\end{align}
Similarly, the energy of massless $\varphi$ is $E_k=k$, so the mode function is
\begin{align}
 \label{eq_flatuphi}
 u_\varphi(k,t) = \frac{e^{-i k t}}{\sqrt{2k}}.
\end{align}
 
\begin{figure}[tbp]
    \centering
    \includegraphics[width=0.5\textwidth]{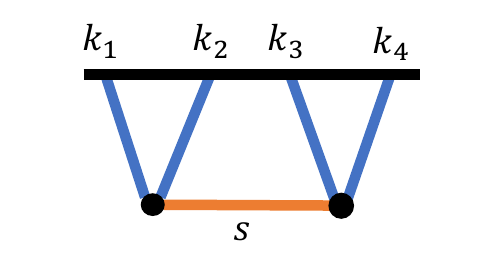}
    \caption{In-in diagrammatics of tree-level $s$-channel four-point correlator with a massive exchange. The massless scalar $\varphi$ is represented in \textit{blue} and the massive scalar $\sigma$ in \textit{orange}. 
    }
    \label{fig:Tpm}
\end{figure}

Below we will compute the $s$-channel four-point correlator:
\begin{align}
\label{eq_flatI}
\mathcal I \equiv \langle \varphi_{\bm k_1}\varphi_{\bm k_2}\varphi_{\bm k_3}\varphi_{\bm k_4} \rangle_s',
\end{align}
where we use a prime to denote the momentum conservation factor $(2\pi)^3\delta(\bm k_1+\bm k_2+\bm k_3+\bm k_4)$ has been stripped off.
Focusing on this process, we study the diagrams presented in \Fig{fig:Tpm}. Our goal is to characterise and physically interpret the influence of the heavy field $\sigma$ (the \textit{environment}) while focusing on the massless field $\varphi$ (the \textit{system}).

\subsection{Schwinger-Keldysh formalism}\label{subsec:flatSK}

We proceed with the calculation via Schwinger-Keldysh formalism. Following \cite{Schwinger:1960qe, Keldysh:1964ud} (also see \cite{2016RPPh...79i6001S} for a review), we double all the field content, and the equal-time correlator (defined at $t=0$) can be expressed as the path integral\footnote{
Here the spacetime integral should be understood as $\int \dd^4x \equiv \int_{-\infty}^0\dd t\int \dd^3\bm x$.}
\begin{align}
\label{eq_flatPathInt}
\langle \mathcal O[\varphi,\sigma]\rangle = \int_{\text{BD}}^\varphi \mathcal D\varphi_+\int_{\text{BD}}^\varphi \mathcal D\varphi_- \int_{\text{BD}}^\sigma \mathcal D\sigma_+\int_{\text{BD}}^\sigma \mathcal D\sigma_-\,
 \mathcal O[\varphi,\sigma]e^{i \int \dd^4x \, ( \mathcal L[\varphi_+,\sigma_+] - \mathcal L[\varphi_-,\sigma_-] )}.
\end{align}
By perturbation expansion of the interaction term in the Lagrangian \eqref{eq_flatL}, we can compute the momentum-space four-point correlator $\mathcal I$ \eqref{eq_flatI} as an ``in-in'' integral
\begin{align}
\label{eq_flat4ptinin}
\mathcal I =&~\frac{1}{\Lambda^2} \sum_{\mathsf a,\mathsf b=\pm} (-\mathsf a\mathsf b) \int_{-\infty}^0 \mathrm dt_1\mathrm dt_2\, \partial_{t_1}K_{\mathsf a}^\varphi(k_1,t_1)\times \partial_{t_1}K_{\mathsf a}^\varphi(k_2,t_1)\nonumber\\
&\times D_{\mathsf a\mathsf b}^\sigma(s;t_1,t_2)\times \partial_{t_2}K_{\mathsf b}^\varphi(k_3,t_2)\times \partial_{t_2}K_{\mathsf b}^\varphi(k_4,t_2).
\end{align}
Here $D_{\mathsf a\mathsf b}^\sigma(k;t_1,t_2) = \langle \sigma_{\mathsf a}(t_1,\bm k)\sigma_{\mathsf b}(t_2,-\bm k)\rangle'$ are the bulk-to-bulk propagators of $\sigma$\footnote{We follow the notations in \cite{Chen:2017ryl}. Notice that the four bulk-to-bulk propagators are not linearly independent due to the identity $D_{++}+D_{--} = D_{+-} + D_{-+}$.}
\begin{subequations}
\label{eq_flatD}
\begin{align}
\label{eq_flatDmp}
&~D_{-+}^\sigma(k;t_1,t_2) = u_\sigma(k,t_1)u_\sigma^*(k,t_2) = \frac{e^{-i E_k (t_1-t_2)}}{2E_k},\\
\label{eq_flatDpm}
&~D_{+-}^\sigma(k;t_1,t_2) = \big[D_{-+}^\sigma(k;t_1,t_2)\big]^* = \frac{e^{+i E_k (t_1-t_2)}}{2E_k},\\
\label{eq_flatDpp}
&~D_{\pm\pm}^\sigma(k;t_1,t_2) = D_{\mp\pm}^\sigma(k;t_1,t_2)\theta(t_1-t_2) + D_{\pm\mp}^\sigma(k;t_1,t_2)\theta(t_2-t_1), \bigg.
\end{align}
\end{subequations}
and
$K_\pm^\varphi$ are the bulk-to-boundary propagators of $\varphi$
\begin{align}
\label{eq_flatK}
K_+^\varphi(k,t) = u^*_\varphi(k,t)u_\varphi(k,0) = \frac{e^{i kt}}{2k},\qquad K_-^\varphi(k,t)= \big[ K_+^\varphi(k,t) \big]^* = \frac{e^{-i kt}}{2k}.
\end{align}

We then plug the propagators \eqref{eq_flatD} and \eqref{eq_flatK} into \Eq{eq_flat4ptinin}, and it is straightforward to compute this ``in-in'' integral. The result is
\begin{align}
\label{eq_flatIresult}
\mathcal I = \frac{k_T+E_s}{8\Lambda^2E_sk_T(k_{12}+E_s)(k_{34}+E_s)},
\end{align}
where we have defined the total energy $k_T\equiv k_{1234}$ with the shorthand $k_{ij\cdots} \equiv k_i+k_j+\cdots$.

Let us consider the heavy mass limit of \Eq{eq_flatIresult}. As noted in \cite{Salcedo:2022aal, Green:2024cmx}, it is impossible to recover all terms in \Eq{eq_flatIresult} in the traditional $\Box/M^2$ expansion of the heavy propagator. To see this, we expand \Eq{eq_flatIresult} in $M\gg k_i$ for all momenta (recall that $E_s = \sqrt{M^2+s^2}$)
\begin{align}
 \mathcal I = \frac{1}{8\Lambda^2  k_T } \frac{1}{M^2} - \frac{(k_{12} k_{34} + s^2)}{8\Lambda^2 k_T }  \frac{1}{M^4} + \frac{k_{12} k_{34} }{8\Lambda^2 } \frac{1}{M^5}+ \mathcal{O}\left( \frac{1}{M^6}\right),
\end{align}
where we find some odd powers of $1/M$ that cannot be produced from the $\Box/M^2$ expansion. The origin of these terms has been attributed in \cite{Salcedo:2022aal, Green:2024cmx} to non-Hermitian boundary terms in the single-field EFT for $\varphi$. We below discuss their physical origin and their systematic embedding in an open EFT construction.

\subsubsection{Implications from Keldysh basis}
This puzzle is first considered through the perspective of the Keldysh basis, which provides a physical interpretation of the origin of the odd powers of $1/M$. Finite-time QFT is conveniently organized in the Keldysh basis of fields, which is a linear combination of $+/-$ fields on the two Schwinger contours \cite{kamenev_2011}. The retarded and advanced fields are respectively defined as
\begin{align}
\label{eq_sigmaRA}
\sigma_r = \frac{\sigma_+ + \sigma_-}{2},\qquad \sigma_a = \sigma_+ - \sigma_-,
\end{align}
and the propagators can be easily derived from the bulk-to-bulk propagators \eqref{eq_flatD}
\begin{subequations}
\label{eq_GRAK}
\begin{align}
\label{eq_GR}
-i G_\sigma^R(k;t_1,t_2) \equiv &~\langle \sigma_r(t_1,\bm k) \sigma_a(t_2,-\bm k) \rangle'  = D^\sigma_{++}(k;t_1,t_2)-D^\sigma_{+-}(k;t_1,t_2),\\
\label{eq_GA}
-i G_\sigma^A(k;t_1,t_2) \equiv &~\langle \sigma_a(t_1,\bm k) \sigma_r(t_2,-\bm k) \rangle' = D^\sigma_{++}(k;t_1,t_2)-D^\sigma_{-+}(k;t_1,t_2),\\
\label{eq_GK}
-i G_\sigma^K(k;t_1,t_2) \equiv &~\langle \sigma_r(t_1,\bm k) \sigma_r(t_2,-\bm k) \rangle'
= \frac12 \big[D_{-+}^\sigma(k;t_1,t_2)+D_{+-}^\sigma(k;t_1,t_2) \big],\\
\label{eq_G0}
&~\langle \sigma_a(t_1,\bm k) \sigma_a(t_2,-\bm k) \rangle' = 0.
\end{align}
\end{subequations}
Notice that both the \emph{retarded propagator} $G^R_\sigma$ and the \emph{advanced propagator} $G^A_\sigma$ are the Green's functions satisfying:
\begin{equation}
    \big[\partial_{t_1}^2+(k^2+M^2)\big]G^{R/A}_\sigma(k;t_1,t_2) = \delta(t_1-t_2),
\end{equation}
but are only supported on $t_1>t_2$ and $t_1<t_2$, respectively. On the other hand, the \emph{Keldysh propagator} $G^K_\sigma$ is not a Green's function\footnote{In general, the Keldysh propagator obeys an inhomogeneous differential equation sourced by the environmental noise. In this precise case, it reduces to the sourceless Klein-Gordon equation.}.

For later convenience, we further define the \emph{principal-value propagator} $G^P$ and the \emph{Pauli-Jordan propagator}\footnote{$G^\Delta_\sigma$ is also known under the name of \textit{Schwinger function}. We here use \textit{propagator} in a loose sense - referring to the fact these functions encode how information propagates in the environment $\sigma$.} $G^\Delta_\sigma$ as a linear combination of retarded/advanced propagators \cite{Birrell:1982ix}:
\begin{subequations}
\label{eq_GHD}
\begin{align}
\label{eq_GH}
&G_\sigma^P(k;t_1,t_2) = \frac12\big[G_\sigma^R(k;t_1,t_2) + G_\sigma^A(k;t_1,t_2)\big]
=\frac i2\big[D^\sigma_{++}(k;t_1,t_2)-D^\sigma_{--}(k;t_1,t_2)\big],
\\
\label{eq_GD}
&G_\sigma^\Delta(k;t_1,t_2) = G_\sigma^R(k;t_1,t_2)-G_\sigma^A(k;t_1,t_2) = i \big[ D_{-+}^\sigma(k;t_1,t_2) - D_{+-}^\sigma(k;t_1,t_2)\big]\,.
\end{align}
\end{subequations}
where $\mathrm{sgn}(x)\equiv\theta(x)-\theta(-x)$ denotes the sign function. In terms of the mode function $u_\sigma(k,t)$ and its complex conjugate, we can write down the three linearly independent propagators:
\begin{subequations}
\label{eq_flatGHDK}
\begin{align}
\label{eq_flatGH}
&G_\sigma^P(k;t_1,t_2) = \frac i2  \big[ u_\sigma(k,t_1) u^*_\sigma(k,t_2) - u^*_\sigma(k,t_1) u_\sigma(k,t_2) \big] \text{sgn}(t_1-t_2),\\
\label{eq_flatGD}
&G_\sigma^\Delta(k;t_1,t_2) = i \big[ u_\sigma(k,t_1) u^*_\sigma(k,t_2) - u^*_\sigma(k,t_1) u_\sigma(k,t_2) \big],\\
\label{eq_flatGK}
&G_\sigma^K(k;t_1,t_2) =  \frac i2 \big[ u_\sigma(k,t_1) u^*_\sigma(k,t_2)+u^*_\sigma(k,t_1) u_\sigma(k,t_2) \big].
\end{align}
\end{subequations}
Finally, we insert the explicit expression for the mode function \eqref{eq_flatusigma} and obtain:
\begin{subequations}
\label{eq_flatGHDKex}
\begin{align}
\label{eq_flatGHex}
&G_\sigma^P(k;t_1,t_2) = \frac{\sin E_k(t_1-t_2)}{2E_k} \text{sgn}(t_1-t_2),\\
\label{eq_flatGDex}
&G_\sigma^\Delta(k;t_1,t_2) = \frac{\sin E_k(t_1-t_2)}{E_k},\\
\label{eq_flatGKex}
&G_\sigma^K(k;t_1,t_2) =  \frac{i \cos E_k(t_1-t_2)}{2E_k}.
\end{align}
\end{subequations}
We summarise some basic properties of the propagators $G^P_\sigma$, $G^\Delta_\sigma$, and $G^K_\sigma$ in Tab.\,\ref{tab_prop}, including that they are either real or (pure) imaginary; either symmetric or asymmetric under $t_1\leftrightarrow t_2$; and either factorised or nested in time order. By factorised, we mean that it can be written by a finite sum of factorised terms, and by nested, we mean that time integrals involving this propagator exhibit nested time integrals due to $\text{sgn}(t_1-t_2)$. These characteristics can be easily verified from the explicit expressions \eqref{eq_flatGHDKex}, but they are actually fundamentally defined properties from \eqref{eq_flatGHDK}, and thus remain valid in arbitrary spacetime.
\begin{table}[t] 
  \centering
   \caption{Basic properties of propagators in the Keldysh basis}
   \vspace{2mm}
  \begin{tabular}{ccccc}
   \toprule[1.5pt]
    Propagators &Real/Imaginary & Symmetric/Asymmetric & Factorised/Nested \\ \hline
    $G^P(k;t_1,t_2)$ & R & S & N\\
    $G^\Delta(k;t_1,t_2)$ & R & A & F\\
    $ G^K(k;t_1,t_2)$ & I & S & F\\
   \bottomrule[1.5pt] 
  \end{tabular}
  \label{tab_prop}
\end{table}

To gain some insights from the Keldysh basis, we turn to the new basis of propagators \eqref{eq_flatGHDK} for $\sigma$ in the ``in-in'' integral \eqref{eq_flat4ptinin}, and extract the contribution from each one. More explicitly, the four ``in-in'' propagators \eqref{eq_flatD} can be expressed in this basis:
\begin{subequations}
\label{eq_sub}
\begin{align}
\label{eq_sub1}
&D^\sigma_{\mp\pm}(k;t_1,t_2) = -i G_\sigma^K(k;t_1,t_2) \mp \frac i2 G_\sigma^\Delta(k;t_1,t_2),\\
\label{eq_sub2}
&D^\sigma_{\pm\pm}(k;t_1,t_2) = -i G_\sigma^K(k;t_1,t_2) \mp i G_\sigma^P(k;t_1,t_2).
\end{align}
\end{subequations}
Therefore, we can substitute the bulk-to-bulk propagator $D^\sigma_{\mathsf a\mathsf b}$ with \Eq{eq_sub} in the integral \eqref{eq_flat4ptinin} to obtain the corresponding contributions.
In particular, the principal-value propagator $G^P_\sigma$ contributes to $\pm\pm$ branches
\begin{align}
\label{eq_flatIH}
\mathcal I^P =
&~\frac{1}{\Lambda^2} \sum_{\mathsf a=\pm} (-1)\int_{-\infty}^0 \mathrm dt_1\mathrm dt_2\, \partial_{t_1}K_{\mathsf a}^\varphi(k_1,t_1)\times \partial_{t_1}K_{\mathsf a}^\varphi(k_2,t_1)\nonumber\\
&\times \left[- \mathsf a i G_\sigma^P(s;t_1,t_2) \right]\times \partial_{t_2}K_{\mathsf a}^\varphi(k_3,t_2)\times \partial_{t_2}K_{\mathsf a}^\varphi(k_4,t_2)\nonumber\\
=&-\frac{1}{16\Lambda^2k_T}\Big(\frac{1}{k_{12}^2-E_s^2}+\frac{1}{k_{34}^2-E_s^2}\Big),
\end{align}
and the Pauli-Jordan propagator $G^\Delta_\sigma$ contributes to $\pm\mp$ branches
\begin{align}
\label{eq_flatID}
\mathcal I^\Delta =
&~\frac{1}{\Lambda^2} \sum_{\mathsf a=\pm} \int_{-\infty}^0 \mathrm dt_1\mathrm dt_2\, \partial_{t_1}K_{\mathsf a}^\varphi(k_1,t_1)\times \partial_{t_1}K_{\mathsf a}^\varphi(k_2,t_1)\nonumber\\
&\times \left[   \frac{\mathsf ai}2 G_\sigma^\Delta(s;t_1,t_2) \right]\times \partial_{t_2}K_{-\mathsf a}^\varphi(k_3,t_2)\times \partial_{t_2}K_{-\mathsf a}^\varphi(k_4,t_2)\nonumber\\
=&-\frac{k_T}{16\Lambda^2(k_{12}^2-E_s^2)(k_{34}^2-E_s^2)}. 
\end{align}
Finally, the Keldysh propagator $G^K_\sigma$ contributes to all SK branches
\begin{align}
\label{eq_flatIK}
\mathcal I^K =
&~\frac{1}{\Lambda^2}\sum_{\mathsf a,\mathsf b=\pm} (-\mathsf a\mathsf b)\int_{-\infty}^0 \mathrm dt_1\mathrm dt_2\, \partial_{t_1}K_{\mathsf a}^\varphi(k_1,t_1)\times \partial_{t_1}K_{\mathsf a}^\varphi(k_2,t_1)\nonumber\\
&\times \left[ -i G_\sigma^K(s;t_1,t_2) \right]\times \partial_{t_2}K_{\mathsf b}^\varphi(k_3,t_2)\times \partial_{t_2}K_{\mathsf b}^\varphi(k_4,t_2)\nonumber\\
=&~ \frac{ k_{12}k_{34}}{8\Lambda^2E_s(k_{12}^2-E_s^2)(k_{34}^2-E_s^2)}.
\end{align}
Indeed, one can easily recover the full correlator \eqref{eq_flatIresult} by summing up \Eqs{eq_flatIH}-\eqref{eq_flatIK}.

\subsubsection{Discussion}
There are two main distinctions among the above three contributions \eqref{eq_flatIH}-\eqref{eq_flatIK}.
First, it is only $\mathcal I^P$ that possesses the \emph{total energy pole} (or the $k_T$ pole), namely it diverges when $k_T\to 0$.
This is because the (residue of) total energy pole corresponds to a unitary scattering amplitude (in flat spacetime) \cite{Arkani-Hamed:2017fdk, Goodhew:2020hob, Pajer:2020wxk, Salcedo:2022aal, Cespedes:2025dnq}, and thus it could only appear in a unitary theory. As we will see in \Sec{subsec:flatEFT}, when integrating out $\sigma$, the principal-value propagator encodes all the information about the unitary single-field EFT for $\varphi$. Conversely, the Pauli-Jordan and Keldysh propagators capture dissipation and noise, which are non-unitary effects. It explains why $\mathcal I^P$ is unique in possessing this $k_T$ pole.
This can be also understood from a technical perspective: both $G^\Delta_\sigma$ and $G^K_\sigma$ are factorised in time, see Tab.\,\ref{tab_prop}, so they cannot maintain singularity in $k_T$ which is not factorisable. We also observe that $\mathcal I^\Delta$ vanishes at $k_T=0$, and it would be interesting to explore the underlying reason and whether this is a universal feature. We have verified that this feature persists in de Sitter for half-integer values of the order parameter $\sqrt{9/4 - M^2}$ — although further work is required to obtain a fully general result. 
Aside from the total energy pole, we find all these three parts have \emph{partial energy poles} ($k_{12}+E_s=0$ or $k_{34}+E_s=0$) and \emph{folded poles} ($k_{12}-E_s=0$ or $k_{34}-E_s=0$), while the latter will be cancelled when the three parts add together to the full correlator \eqref{eq_flatIresult} due to the BD initial condition.

Second, let us expand each contribution in the heavy mass limit $M\to \infty$:
\begin{align}
&\mathcal I^P = \frac{1}{8\Lambda^2k_T} \frac{1}{M^2}
+ \frac{k_{12}^2+k_{34}^2-2s^2}{16\Lambda^2k_T}\frac{1}{M^4}
+ \mathcal O\Big(\frac{1}{M^6}\Big),\\
&\mathcal I^\Delta = -\frac{k_T}{16\Lambda^2} \frac{1}{M^4} + \mathcal O\Big(\frac{1}{M^6}\Big),\\
&\mathcal I^K = \frac{k_{12}k_{34}}{8\Lambda^2} \frac{1}{M^5} + \mathcal O\Big(\frac{1}{M^7}\Big).
\end{align}
We can find that both $\mathcal I^P$ and $\mathcal I^\Delta$ contain even powers in $1/M$, while $\mathcal I^K$ only contains odd powers. 
This can be traced back to the fact that $G^P_\sigma$ and $G^\Delta_\sigma$ vanish in the coincident time limit while $G^K_\sigma$ does not, see \Eq{eq_flatGHDKex}.
Explicitly, $G^P_\sigma$ and $G^\Delta_\sigma$ are built from the $\sin$ function, while $G^K_\sigma$ follows from the $\cos$ function.
Since the traditional (unitary and local) EFT is the $\Box/M^2$ expansion of the heavy propagator, it could only have even powers of $1/M$ and each term contributes to a contact graph possessing the total energy pole. That is, the traditional EFT reproduces the contribution from the principal-value propagator \eqref{eq_flatIH}. Corrections to this include even powers of $1/M$ from the Pauli-Jordan propagator \eqref{eq_flatID} that vanishes at $k_T=0$, and odd powers of $1/M$ from the Keldysh propagator \eqref{eq_flatIK}.

\subsection{Flat-space open effective field theory}\label{subsec:flatEFT}

Having seen the implication of $\sigma$ on the $\varphi$ observables from the Keldysh basis, we now construct an open EFT (see \cite{Burgess:2022rdo, Colas:2024lse} for short reviews) for the massless field $\varphi$.
That is, we define the retarded and advanced fields for $\varphi$ similarly to \Eq{eq_sigmaRA}:
\begin{equation}
\varphi_r = \frac{\varphi_+ + \varphi_-}{2},\qquad \varphi_a = \varphi_+ - \varphi_-.
\end{equation}
Under this basis, the path integral \eqref{eq_flatPathInt} becomes
\begin{align}
\langle \mathcal O[\varphi,\sigma] \rangle &= \int_{\text{BD}}^\varphi \mathcal D\varphi_r\int_{\text{BD}}^0 \mathcal D\varphi_a\int_{\text{BD}}^\sigma \mathcal D\sigma_r\int_{\text{BD}}^0 \mathcal D\sigma_a\, \mathcal O[\varphi,\sigma]e^{i S_0^\varphi[\varphi_r,\varphi_a]+i S_0^\sigma[\sigma_r,\sigma_a]}e^{i S_{\text{int}}[\varphi_r,\varphi_a,\sigma_r,\sigma_a]}, \bigg.
\end{align}
where the free action reads\footnote{Here the propagators are written in position space and are the Fourier transform of the form in \Eq{eq_flatGHDK}
\begin{equation}
G(x,y) = \int \frac{\dd^3\bm k}{(2\pi)^3} G(k;t_1,t_2) e^{i \bm k\cdot (\bm x-\bm y)},
\end{equation}
where we have omitted the superscript ($P$, $\Delta$, $K$) and subscript ($\varphi$, $\sigma$) for the propagator. The retarded and advanced propagators in position space are defined similarly.
}
\begin{align}
S_0^\varphi[\varphi_r,\varphi_a] =& -\frac12 \int \dd^4x\,
\begin{pmatrix}
\varphi_r & \varphi_a
\end{pmatrix}
\begin{pmatrix}
G^K_\varphi & G^R_\varphi \\ G^A_\varphi & 0
\end{pmatrix}^{-1}
\begin{pmatrix}
\varphi_r \\ \varphi_a
\end{pmatrix},\\
S_0^\sigma[\sigma_r,\sigma_a] =& -\frac12\int \dd^4x \,
\begin{pmatrix}
\sigma_r & \sigma_a
\end{pmatrix}
\begin{pmatrix}
G^K_\sigma & G^R_\sigma \\ G^A_\sigma & 0
\end{pmatrix}^{-1}
\begin{pmatrix}
\sigma_r \\ \sigma_a
\end{pmatrix},
\end{align}
and the interaction part is
\begin{align}
S_{\text{int}}[\varphi_r,\varphi_a,\sigma_r,\sigma_a] =&~\frac{1}{2\Lambda} \int \dd^4x \, ( \dot\varphi_+^2\sigma_+ - \dot\varphi_-^2\sigma_-)\nonumber\\
=&~\frac{1}{\Lambda} \int \dd^4x \, \Big( \dot\varphi_r\dot\varphi_a\sigma_r + \frac12 \dot\varphi_r^2\sigma_a + \frac18 \dot\varphi_a^2\sigma_a \Big).
\end{align}

We now aim to integrate out the field $\sigma$ and to obtain the single field open EFT for $\varphi$ only.
In practice, we follow a procedure similar to the one described in \cite{Proukakis:2024pua}. We first define the \emph{influence functional} $S_{\text{IF}}$ \cite{FEYNMAN1963118}
\begin{align}
    	\ee^{iS_{\mathrm{IF}} \left[ \pi_r, \pi_a\right]} =   \int_{\mathrm{BD}}^\sigma \mathcal{D}\sigma_r \int_{\mathrm{BD}}^0 \mathcal{D}\sigma_a\,  \ee^{iS^\sigma_{0}\left[ \sigma_r, \sigma_a\right]} \ee^{iS_{\mathrm{int}}\left[ \varphi_r, \varphi_a, \sigma_r, \sigma_a\right]},
        \end{align}
and then expand $S_{\mathrm{int}}$ to the second order in $1/\Lambda$ to obtain
\begin{align}
    \ee^{iS_{\mathrm{IF}}}  = 1 + i \langle S_{\mathrm{int}} \rangle_\sigma - \frac{1}{2} \langle S^2_{\mathrm{int}}
    \rangle_\sigma + \cdots .
\end{align}
The theory being linear in $\sigma$ and having removed the tadpole contribution, we assume for the moment that $\langle S_{\mathrm{int}} \rangle_\sigma = 0$. We then identify the influence functional in the leading order
\begin{equation}
S_{\mathrm{IF}} \simeq \frac i2 \langle S^2_{\mathrm{int}} \rangle_\sigma.
\end{equation}
Therefore, the second-order effective action is simply obtained by replacing internal $\sigma$ legs by its propagators, leading to
\begin{align}
\label{eq_flatIF}
            &S_{\mathrm{IF}}[\varphi_r,\varphi_a] = ~\frac{1}{2\Lambda^2} \int \dd^4x \int \dd^4y \bigg\{ \dot\varphi_r(x) \dot\varphi_a(x)  \times G^{R}_\sigma(x,y) \times
              \left[ \frac12\dot\varphi_r^2(y) + \frac{1}{8}\dot\varphi_a^2(y)\right] \\
            +&\left[ \frac12\dot\varphi_r^2(x)+\frac18\dot\varphi_a^2(x)\right] \times G^{A}_\sigma(x,y) \times \dot\varphi_r(y)\dot\varphi_a(y) + \dot\varphi_r(x) \dot\varphi_a(x)  \times G^{K}_\sigma(x,y) \times \dot\varphi_r(y)\dot\varphi_a(y)
            \bigg\}. \nonumber
        \end{align}
Notice that when doing the contraction, the two vertices give $1/\Lambda^2$, and the correlation functions of (free) $\sigma_{r/a}$ have an extra $-i$ compared to the corresponding propagators \eqref{eq_GRAK}.
We emphasize that the last term in \Eq{eq_flatIF} is manifestly non-unitary: it contains an even number of advanced fields, and thus if going back to the original +/- basis, it can never be written in a factorised form \cite{Salcedo:2024nex}.

The first two terms in \Eq{eq_flatIF} are more subtle. As shown in \Sec{subsec:flatSK}, it is more intuitive to express $G^{R/A}$ in terms of $G^{P/\Delta}$ using \Eq{eq_GHD}, where the influence functional becomes:
 \begin{align}
\label{eq_flatIFHD}
            &S_{\mathrm{IF}}[\varphi_r,\varphi_a] = ~\frac{1}{2\Lambda^2} \int \dd^4x \int \dd^4y \bigg\{ \dot\varphi_r(x) \dot\varphi_a(x)  \times 2G^{P}_\sigma(x,y) \times
              \left[ \frac12\dot\varphi_r^2(y) + \frac{1}{8}\dot\varphi_a^2(y)\right] \\
            +&\dot\varphi_r(x) \dot\varphi_a(x)  \times G^{\Delta}_\sigma(x,y) \times
              \left[ \frac12\dot\varphi_r^2(y) + \frac{1}{8}\dot\varphi_a^2(y)\right] + \dot\varphi_r(x) \dot\varphi_a(x)  \times G^{K}_\sigma(x,y) \times \dot\varphi_r(y)\dot\varphi_a(y)
            \bigg\},\nonumber
        \end{align}
where we have used $G^P_\sigma(y,x) = G^P_\sigma(x,y)$ and $G^\Delta_\sigma(y,x) = -G^\Delta_\sigma(x,y)$ to simplify the expression. Now the three terms in the non-local EFT action \eqref{eq_flatIFHD} have clear physical meanings:
\begin{enumerate}
\item The term of principal-value propagator $G^P_\sigma$ is a unitary EFT. Written back in the $+/-$ basis, this term can be expressed as $S_{\text{unit}}[\varphi_+]-S_{\text{unit}}[\varphi_-]$ with
\begin{equation}
S_{\text{unit}}[\varphi_\pm] = \frac{1}{8\Lambda^2}\int \dd^4x \int \dd^4y\, \dot\varphi^2_\pm(x)\dot\varphi^2_\pm(y) G_\sigma^P(x,y).
\end{equation}
If we further expand $G^P_\sigma$ in the heavy mass limit $M\to \infty$ and integrate over $y$, we will recover the traditional (unitary and local) EFT.
\item As a contrast, the term of Pauli-Jordan propagator $G^\Delta_\sigma$ is a non-unitary contribution which we interpret as the \emph{dissipation}. It cannot be decomposed into factorisable contributions in the $+/-$ basis and encodes energy exchange between $\varphi$ and $\sigma$ \cite{Calzetta:2008iqa}. The asymmetric property of $G^\Delta_\sigma$ discussed in Tab.\,\ref{tab_prop} may relate to the fact that dissipation generally breaks time translation symmetry, creating an effective arrow of time.\footnote{Conversely, the symmetry property of $G^P_\sigma$ makes it a natural candidate to control the unitary/Hamiltonian evolution which is time-reversal symmetric.}
\item Finally, the term of Keldysh propagator $G^K_\sigma$ is the \emph{noise} term that is also non-unitary. Physically, it originates from the fluctuations of the $\sigma$ medium sourcing the $\varphi$ dynamics. This can be seen from the fact that $G^K_\sigma$ controls the amplitude of the perturbations in the $\sigma$ field (that is the power spectrum), which makes it a natural candidate to encode the environmental noise onto the system. Note that $G^K_\sigma$ controls an operator quadratic in the advanced field $\varphi_a$ which fulfils the so-called \textit{non-equilibrium constraints} \cite{Liu:2018kfw} $S_{\mathrm{IF}}[\varphi_r,\varphi_a] = S^*_{\mathrm{IF}}[\varphi_r,-\varphi_a] $ and $\Im S_{\mathrm{IF}}[\varphi_r,\varphi_a] \geq 0$ due to the properties of $G^K_\sigma$. It illustrates the fact that the single-field effective description originates from a unitary two-field partial UV completion.
\end{enumerate} 

\begin{tcolorbox}[%
        enhanced, 
        breakable,
        skin first=enhanced,
        skin middle=enhanced,
        skin last=enhanced,
        before upper={\parindent15pt},
        ]{}

        \vspace{0.05in}

\paragraph{Summary.}

\Eq{eq_flatIFHD} represents the second-order effects of $\sigma$ on the dynamics of $\varphi$. It can be decomposed into three distinctive effects:\\
\begin{enumerate}
    \item  The line controlled by $G_\sigma^P(x,y)$ is \textit{unitary} and corresponds to the generation of an effective (non-local) vertex in the Lagrangian of $\varphi$, sometimes called the \textit{Lamb shift}.
    \item The line controlled by $G^\Delta_\sigma(x,y)$ is \textit{non-unitary} and corresponds to the \textit{dissipative} evolution of $\varphi$ through the $\sigma$ medium.
    \item The line controlled by $G_\sigma^K(x,y)$ is \textit{non-unitary} and corresponds to the \textit{noise} generated by fluctuations of the $\sigma$ medium backreacting on the evolution of $\varphi$.
\end{enumerate}

\end{tcolorbox}

\subsubsection{Towards a time-local open EFT} 

Notice that all the three terms in the influence functional $S_{\text{IF}}$ \eqref{eq_flatIFHD} are non-local.
However, it is possible to make them localised under the heavy mass expansion. As a first hint, the contribution from $G^H_\sigma$ is equivalent to the traditional EFT, where each term in the interaction is local under the heavy mass expansion.

The key is to evolve the heavy field $\sigma$ at different times back and forth with the free equation of motion, which is valid because we are only focusing on the leading order in $1/\Lambda$.
That is to say, we can evolve $\varphi_{r,a}(t,\bm k)$ (we turn to $(t,\bm k)$ coordinates for later convenience) and their derivatives by solutions to the Klein-Gordon equation:
\begin{align}
&\varphi_{r,a}(\tilde t,\bm k) = \cos[k(t-\tilde t)]\varphi_{r,a}(t,\bm k) -  \frac{\sin[k(t-\tilde t)]}{k}\dot\varphi_{r,a}(t,\bm k),\\
\label{eq_flatevolve}
&\dot\varphi_{r,a}(\tilde t,\bm k) = k \sin[k(t-\tilde t)]\varphi_{r,a}(t,\bm k)+\cos[k(t-\tilde t)]\dot\varphi_{r,a}(t,\bm k),
\end{align}
with which we can ``pinch'' the two interaction vertices to the same spacetime point, and thus the influence functional is manifest as an integral of a local operator. We leave all the intermediate steps in \App{app:localEFT} and present the final result below.

As in \Eq{eq_flatIFHD}, the influence functional can be decomposed into three terms:
\begin{align}
S_{\mathrm{IF}}[\varphi_r,\varphi_a] =&~ S_{\mathrm{IF}}^P[\varphi_r,\varphi_a]+S_{\mathrm{IF}}^\Delta[\varphi_r,\varphi_a]+S_{\mathrm{IF}}^K[\varphi_r,\varphi_a],
\end{align}
where, using the shorthand $\varphi_{i} \equiv \varphi(t,\bm k_i)$, the contribution from $G^P_\sigma$ is:
\begin{align}
\label{eq_flatIFH}
S_{\mathrm{IF}}^P[\varphi_r,\varphi_a] = &~ \int_{-\infty}^0 \dd t\int\mathcal D\bm k\,\\
\bigg\{&c_1(k_1,k_2,E_s)\left[ \varphi_{r,1}\varphi_{a,2}
\Big( \dot\varphi_{r,3}\dot\varphi_{r,4}+\frac14\dot\varphi_{a,3}\dot\varphi_{a,4}\Big)
+\Big(\varphi_{r,1}\varphi_{r,2}+\frac14 \varphi_{a,1}\varphi_{a,2}\Big)\dot\varphi_{r,3}\dot\varphi_{a,4} \right]\nonumber\\
+&c_2(k_1,k_2,E_s)\left[ \dot\varphi_{r,1}\dot\varphi_{a,2}
\Big( \dot\varphi_{r,3}\dot\varphi_{r,4}+\frac14\dot\varphi_{a,3}\dot\varphi_{a,4}\Big)
+\Big(\dot\varphi_{r,1}\dot\varphi_{r,2}+\frac14 \dot\varphi_{a,1}\dot\varphi_{a,2}\Big)\dot\varphi_{r,3}\dot\varphi_{a,4} \right]\bigg\}, \nonumber
\end{align}
the contribution from $G^\Delta_\sigma$ is:
\begin{align}
\label{eq_flatIFD}
S_{\mathrm{IF}}^\Delta[\varphi_r,\varphi_a] = & - \int_{-\infty}^0 \dd t\int\mathcal D\bm k\,\\
\bigg\{&c_1(k_1,k_2,E_s)\left[ \varphi_{r,1}\varphi_{a,2}
\Big( \dot\varphi_{r,3}\dot\varphi_{r,4}+\frac14\dot\varphi_{a,3}\dot\varphi_{a,4}\Big)
-\Big(\varphi_{r,1}\varphi_{r,2}+\frac14 \varphi_{a,1}\varphi_{a,2}\Big)\dot\varphi_{r,3}\dot\varphi_{a,4} \right]\nonumber\\
+&c_2(k_1,k_2,E_s)\left[ \dot\varphi_{r,1}\dot\varphi_{a,2}
\Big( \dot\varphi_{r,3}\dot\varphi_{r,4}+\frac14\dot\varphi_{a,3}\dot\varphi_{a,4}\Big)
-\Big(\dot\varphi_{r,1}\dot\varphi_{r,2}+\frac14 \dot\varphi_{a,1}\dot\varphi_{a,2}\Big)\dot\varphi_{r,3}\dot\varphi_{a,4} \right]\bigg\}, \nonumber
\end{align}
and the contribution from $G^K_\sigma$ is:
\begin{align}
\label{eq_flatIFK}
S_{\mathrm{IF}}^K[\varphi_r,\varphi_a] = & ~ i \int_{-\infty}^0 \dd t\int\mathcal D\bm k\,c_3(k_1,k_2,E_s) \left(\varphi_{r,1}\dot\varphi_{a,2}+\varphi_{a,1}\dot\varphi_{r,2}\right)\dot\varphi_{r,3}\dot\varphi_{a,4}.
\end{align}
Note the $i$ prefactor in the last expression, chosen such that $c_3(k_1,k_2,E_s)$ is manifestly real, following the non-equilibrium constraint $S_{\mathrm{IF}}[\varphi_r,\varphi_a] = S^*_{\mathrm{IF}}[\varphi_r,-\varphi_a]$ \cite{Liu:2018kfw}. 
Here we have only considered the $s$ channel and the measure $\int\mathcal D\bm k$ is defined as in \eqref{eq_intKmeasure}, while the three EFT coefficients are given by:
\begin{align}
\label{eq_flatc1}
&c_1(k_i,k_j,E_s) \equiv \frac{-k_ik_j}{16\Lambda^2E_s}\left(\frac{1}{E_s+ k_{ij}} - \frac{1}{E_s + k_i -k_j} - \frac{1}{E_s-k_i+k_j} + \frac{1}{E_s-k_{ij}}\right),\\
\label{eq_flatc2}
&c_2(k_i,k_j,E_s) \equiv  \frac{1}{16\Lambda^2E_s}\left(\frac{1}{E_s+ k_{ij}} + \frac{1}{E_s + k_i -k_j} + \frac{1}{E_s-k_i+k_j} + \frac{1}{E_s-k_{ij}}\right),\\
\label{eq_flatc3}
&c_3(k_i,k_j,E_s) \equiv \frac{k_i}{8\Lambda^2E_s}\left(\frac{1}{E_s+ k_{ij}} + \frac{1}{E_s + k_i -k_j} - \frac{1}{E_s-k_i+k_j} - \frac{1}{E_s-k_{ij}}\right).
\end{align}
As a consistency check, we use the influence functional \eqref{eq_flatIFH}-\eqref{eq_flatIFK} to compute the contributions from the three propagators to the correlator $\mathcal I$ in \App{app:trispectrum}, and the results perfectly recover \Eqs{eq_flatIH}-\eqref{eq_flatIK}. 
Note that the structure of the EFT coefficients is reminiscent of the flat space contact bispectra computed in the Keldysh basis in \cite{Salcedo:2024smn}. We come back to this point in \Sec{sec:conclu}. 
Finally, we can further expand \Eqs{eq_flatc1}-\eqref{eq_flatc3} in the heavy mass limit:
\begin{align}
\label{eq_c1expand}
    c_1(k_i,k_j,s) &= \frac{-k_i^2 k_j^2}{2\Lambda^2M^4} + \mathcal O\Big(\frac{1}{M^6}\Big),\\ 
    \label{eq_c2expand}
    c_2(k_i,k_j,s)&= \frac{1}{4\Lambda^2M^2} + \frac{k_i^2+k_j^2- s^2}{4\Lambda^2M^4} + \mathcal O\Big(\frac{1}{M^6}\Big),\\
    \label{eq_c3expand}
    c_3(k_i,k_j,s) &= -\frac{k_i^2}{2\Lambda^2M^3} -\frac{k_i^2(2k_i^2+6 k_j^2-3s^2)}{4\Lambda^2M^5} + \mathcal O\Big(\frac{1}{M^7}\Big),
\end{align}
where we find that each term, order by order in $1/M$, is analytic in all momenta, especially around $s\to 0$. Therefore, such an open EFT, though notably containing a non-unitary part, is actually local in both time and space under heavy mass expansion.


\section{Cosmological collider physics in the Keldysh basis}\label{sec:Keldysh}
    
With the flat-space theory properly understood, we now move on to its inflationary counterpart. We approximate the spacetime background by the de Sitter metric in the Poincaré patch:
\begin{align}
    \dd s^2=a^2(\eta)\left(-\dd \eta^2 + \dd {\bm x}^2\right)\,,\quad a(\eta)=-\frac{1}{H\eta}\,,
\end{align}
where $-\infty<\eta<0$ denotes the conformal time and $H$ is the Hubble parameter. The two-field model \eqref{eq_flatL} can now be embedded in de Sitter by dressing the operators with appropriate powers of the scale factor,
\begin{align}
    \label{eq:S_dSL}
	S[\varphi,\sigma] & = \frac{1}{2}\int \dd \eta\; \dd^3 \bm{x} \left\{ \; a^2\left[\varphi^{\prime 2 } - \left( \partial_i \varphi \right)^2\right]+\; a^2\left[\sigma^{\prime 2 } - \left( \partial_i \sigma \right)^2\right] -a^4 M^2\sigma^2 +\frac{1}{\Lambda}a^2\varphi^{\prime 2}\sigma\right\},
\end{align}
where prime denotes taking derivative with respect to the conformal time. We identify $\varphi$ with the inflaton perturbations and $\sigma$ a heavy field of mass $M/H > 3/2$ that is active during inflation. A remarkable observational signature of such heavy fields is the characteristic oscillations in the soft limit of the non-Gaussian inflaton correlators. For instance, the $s$-channel soft limit of the inflaton four-point function in the model \eqref{eq:S_dSL} schematically reads 
\begin{align}
    \lim_{s\to 0}\langle \varphi_{\bm k_1}\varphi_{\bm k_2}\varphi_{\bm k_3}\varphi_{\bm k_4} \rangle'\supset \mathcal{A}^{\rm L} \sin\left(\mu\ln\frac{k_{12}}{k_{34}}\right)+\mathcal{A}^{\rm NL} \cos\left(\mu\ln\frac{4k_{12} k_{34}}{s^2}+\delta\right),\label{CCphysicsInANutshell}
\end{align}
where $\mu\equiv\sqrt{M^2/H^2-9/4}$ is the dimensionless mass of the $\sigma$ field and $\mathcal{A}^{\rm L}, \mathcal{A}^{\rm NL} \propto \Lambda^{-2} e^{-\pi\mu}$. A measurement of the oscillation frequency thus amounts to a direct detection of the $\sigma$-particle mass, thereby laying the foundation of the \textit{cosmological collider physics} paradigm \cite{Chen:2009zp, Baumann:2011nk, Noumi:2012vr, Arkani-Hamed:2015bza, Lee:2016vti}.\footnote{On the other hand, if the exchanged field is still massive but light, namely when $0<M/H<3/2$, the mass parameter $\mu$ becomes pure imaginary. As a result, instead of logarithmic oscillations, the four-point function will possess a non-analytic scaling behaviour in the soft limit.} Note that at four-point level, there are two types of cosmological collider signals that correspond to the first and second term of \eqref{CCphysicsInANutshell}. They are distinguished by the functional dependence on the exchange momentum $s$. The \textit{local} signal (the first term) is analytic in the $s\to 0$ limit while the \textit{non-local} signal (the second term) is non-analytic. In the $+/-$ basis of Schwinger-Keldysh formalism, one can show that these two families of cosmological collider signals have distinct origins and physical interpretations. The local signal comes from vertex particle production in the non-linear theory while the non-local signal comes from particle pair production in the linear theory \cite{Tong:2021wai}. Understanding the bulk physics of these cosmological collider signals in the $+/-$ basis allows one to formulate a cutting rule to efficiently extract the relevant contribution in the evaluation of the inflaton correlator \cite{Tong:2021wai,Qin:2023bjk,Qin:2023nhv}.

In this section, we shall revisit the cosmological collider signals from a new perspective using the Keldysh-$r/a$ basis. More specifically, we will leverage on the properties of the Keldysh-basis propagators to pin down the cosmological collider signals from general Schwinger-Keldysh diagrams that compute the inflaton correlator. The information about the origin of cosmological collider signals will be essential for classifying their impact on the inflaton sector and formulating top-down inflaton EFTs in the next section. It provides a complementary analysis to the cutting rule expressed in the Keldysh-$r/a$ basis explored in \cite{Ema:2024hkj}.

\subsection{Cutting rules in the $+/-$ basis revisited}\label{CRpmRevisionSubsection}
        
Let us express the Schwinger-Keldysh generating functional associated with the model \eqref{eq:S_dSL}. Analogous to the flat spacetime example, the generating functional can be written equivalently in two different basis i.e. the $+/-$ basis and the $r/a$ basis. The former makes the singularity structure of correlators more manifest, and is therefore suitable for extracting the cosmological collider signals out of the perturbative diagrammatics. The latter, on the other hand, better elucidates the physical impacts of the heavy field (environment) on the massless inflaton (system). For this reason, we shall first compute the cosmological collider signals in the model \eqref{eq:S_dSL} using the $+/-$ basis, and gradually transmutate to the $r/a$ basis to gain new insights.

In the $+/-$ basis, the generating functional reads 
\begin{align}
     \mathcal{Z}\left[J^\varphi_+, J^\varphi_-; J^\sigma_+, J^\sigma_- \right] &=  \int_{\mathrm{BD}}^{\varphi} \mathcal{D}\varphi_+ \int_{\mathrm{BD}}^{\varphi} \mathcal{D}\varphi_-  \int_{\mathrm{BD}}^{\sigma} \mathcal{D}\sigma_+ \int_{\mathrm{BD}}^{\sigma} \mathcal{D}\sigma_-  \\
    &\qquad \ee^{i \int \dd^4 x \sqrt{-g}\left(J^\varphi_+ \varphi_+ - J^\varphi_- \varphi_- + J^\sigma_+\sigma_+ - J^\sigma_-\sigma_- \right)} \ee^{iS\left[ \varphi_+, \sigma_+\right]-iS\left[ \varphi_-, \sigma_-\right]}\,. \nonumber \Bigg.
\end{align}
Here we assume that in the UV theory with two fields, the quantum fluctuations start out at $\eta\to-\infty$ in a pure state which we take to be the Bunch-Davies (BD) vacuum. This BD state then evolves unitarily through time before reaching an end state at $\eta\to 0$ specified by arbitrary boundary values $\varphi_\pm = \varphi,~\sigma_\pm =\sigma$ that are integrated over. This object allows us to extract correlators of the theory. In particular, the free theory propagators are given by
\begin{align}
    D_{\sf ab}^X(k;\eta_1,\eta_2)\equiv -{\sf a b}\frac{\delta^2}{\delta J^X_{\sf a}(\eta_1)\delta J^X_{\sf b}(\eta_2)} \ln \mathcal{Z}\Bigg|_{J=\Lambda^{-1}=0}\,,\quad X=\varphi,\sigma\text{  and  }{\sf a,b}=\pm\,.
\end{align}
In terms of their mode functions with a BD initial condition,
\begin{align}
	\label{eq:modefctvp}	u_{\varphi}(k,\eta) & = \frac{H}{\sqrt{2 k^3}} \left( 1 +i k \eta \right)  e^{- i k \eta}\,, \\
	\label{eq:modefctvs}	u_{\sigma}(k,\eta) & =-\frac{i\sqrt{\pi}}{2} e^{i\pi\left(\frac{i\mu}{2}+\frac{1}{4}\right)} H (-\eta)^{3/2} H_{i\mu}^{(1)}(-k\eta)\,,\quad \mu\equiv \sqrt{\frac{M^2}{H^2}-\frac{9}{4}}\,,
\end{align}
the bulk-bulk propagators for the massive field in the $+/-$ basis are given by
\begin{align}
    \label{eq_dSDmp}
    &~D_{-+}^\sigma(k;\eta_1,\eta_2) = u_\sigma(k,\eta_1)u_\sigma^*(k,\eta_2) \bigg. 
    \\
    \label{eq_dSDpm}
    &~D_{+-}^\sigma(k;\eta_1,\eta_2) = u^*_\sigma(k,\eta_1)u_\sigma(k,\eta_2) \bigg. 
    \\
    \label{eq_dSDpp}
    &~D_{\pm\pm}^\sigma(k;\eta_1,\eta_2) = D_{\mp \pm}^\sigma(k;\eta_1,\eta_2)\theta(\eta_1-\eta_2) + D_{\pm \mp}^\sigma(k;\eta_1,\eta_2)\theta(\eta_2-\eta_1)\,. \bigg. 
\end{align}
along with the bulk-boundary propagators of the inflaton,
\begin{align}
    \label{eq_dSK}
    K_+^\varphi(k,\eta) &= u^*_\varphi(k,\eta)u_\varphi(k,0) = \frac{H^2}{2k^3}(1-i k \eta) e^{i k \eta}\,,\\
    K_-^\varphi(k,\eta)&= u_\varphi(k,\eta)u^*_\varphi(k,0) = \frac{H^2}{2k^3}(1+i k \eta) e^{-i k \eta}\,.
\end{align}
The propagators in the $+/-$ basis admit simple behaviours in the early-time limit $\eta\to-\infty$ as they have an oscillation frequency in the conformal time $\eta$ that is purely positive or negative,
\begin{align}
    D_{{\sf a}\pm}(k;\eta',\eta)\,,~K_{\pm}(k,\eta)\sim e^{\pm i k\eta}\,\quad \text{when} \quad \eta\to -\infty\,.
\end{align}
Therefore, the $+/-$ basis is better suited for studying UV-dominated physics such as the singularity structure in kinematics. In the late-time limit $\eta\to 0$, the $+/-$ basis propagators branch out multiple frequencies due to particle production. Thus the late-time physics is more intertwined in this basis. 

Moving to the interacting theory, each interaction vertex comes with two ``colourings'' i.e. $+$ and $-$. For instance, in our model \eqref{eq:S_dSL}, the relevant part of the action is
\begin{align}
    S\left[ \varphi_+, \sigma_+\right]-S\left[ \varphi_-, \sigma_-\right]\supset \int \dd \eta \dd^3 x \frac{a^2}{2\Lambda}\left(\varphi^{\prime 2}_+ \sigma_+ -\varphi^{\prime 2}_- \sigma_-\right)\,.
\end{align}
To compute the $n$-point boundary correlator of inflatons, one draws all diagrams with $n$ external inflaton lines and considers all possible colourings of the diagrams with $+$ or $-$ vertices (for a diagram with $V$ vertices, there are $2^V$ coloured diagrams) before summing them altogether. In our case, the four-point correlator of inflatons 
\begin{align}
    \mathcal{I}\equiv&\,\langle \varphi_{\bm k_1}\varphi_{\bm k_2}\varphi_{\bm k_3}\varphi_{\bm k_4} \rangle_s'
\end{align}
is thus calculated by 
\begin{align}
    \nonumber   \mathcal{I} =&~\frac{1}{\Lambda^2} \sum_{\mathsf a,\mathsf b=\pm} (-\mathsf a\mathsf b) \int_{-\infty}^0 \mathrm d\eta_1 \,a^2(\eta_1)\, \mathrm d\eta_2 \,a^2(\eta_2)\, \partial_{\eta_1}K_{\mathsf a}^\varphi(k_1,\eta_1)\times \partial_{\eta_1}K_{\mathsf a}^\varphi(k_2,\eta_1) \nonumber \\
    &\qquad \qquad \qquad \times D_{\mathsf a\mathsf b}^\sigma(s;\eta_1,\eta_2)\times \partial_{\eta_2}K_{\mathsf b}^\varphi(k_3,\eta_2)\times \partial_{\eta_2}K_{\mathsf b}^\varphi(k_4,\eta_2),\\
    \equiv&\, \mathcal{I}_{++}+\mathcal{I}_{--}+\mathcal{I}_{-+}+\mathcal{I}_{+-}\,.\label{eq_dS4ptinin} \bigg.
\end{align}
In a parity-symmetric theory, reversing the colouring is equivalent to a complex conjugation, which reduces $\mathcal{I}_{--}=\mathcal{I}_{++}^*$ and $\mathcal{I}_{+-}=\mathcal{I}_{-+}^*$. Therefore, we will focus on $\mathcal{I}_{++}$ and $\mathcal{I}_{-+}$ without loss of generality. According to the cutting rule, the factorised diagram $\mathcal{I}_{-+}$ is highly suppressed by a full Boltzmann factor \cite{Tong:2021wai}, 
\begin{align}\label{eq:suppr}
    \mathcal{I}_{-+}\sim e^{-2\pi\mu}\,,
\end{align}
and therefore can be discarded in the large-mass limit $\mu\gg 1$. The time-ordered diagram $\mathcal{I}_{++}$, on the other hand, contains a background piece that decays as a power law in $\mu$ and a signal piece that is suppressed by half the Boltzmann factor,
\begin{align}
    \mathcal{I}_{++}\sim \mu^{-2}\times \text{background} +e^{-\pi\mu} \times\text{signal}\,.
\end{align}
Consequently, the dominant cosmological collider signals always come from the time-ordered diagram $\mathcal{I}_{++}$, which is mediated by the time-ordered propagator $D_{++}^\sigma$ of the massive field. Due to the time-ordering $\theta$-functions in $D_{++}^\sigma$, the integrals over time are nested and are technically challenging to calculate. However, if we only focus on the leading cosmological collider signals therein, there is an efficient cutting rule allowing them to be analytically computed \cite{Tong:2021wai}. To single out the signal contribution, one performs the following cutting rule: 
\begin{itemize}
    \item If $k_{12}>k_{34}$, replace $D_{++}^\sigma$ with $D_{-+}^\sigma$;
    \item If $k_{12}<k_{34}$, replace $D_{++}^\sigma$ with $D_{+-}^\sigma$.
\end{itemize}
These replacements are approximations to the original integral $\mathcal{I}_{++}$ that preserve all the signals while discarding the irrelevant background. It serves as a cutting rule since the nested time integrals are now cut into two factorised integrals on the left and right vertices, making analytical computation much easier. This cutting rule can also be summarised as
\begin{align}
    D_{++}^\sigma \, \to \, \theta(k_{12}-k_{34}) D_{-+} + \theta(k_{34}-k_{12}) D_{+-}\,.
\end{align}
To see how the cutting rule works, we subtract them to obtain the remainder that is missed by the cutting rule. Using \Eq{eq_dSDpp}, one finds
\begin{align}
    \nonumber D_{++}^\sigma &-\left[ \theta(k_{12}-k_{34}) D_{-+} + \theta(k_{34}-k_{12}) D_{+-}\right] \bigg.\\
    =&\left[\theta(\eta_1-\eta_2)-\theta(k_{12}-k_{34})\right] D_{-+} + \left[\theta(\eta_2-\eta_1)-\theta(k_{34}-k_{12})\right] D_{+-}\,. \bigg.
\end{align}
Let us estimate the magnitude of this left over piece.
The cosmological collider signals come from the resonant production/decay at each vertices. Such resonance events happen locally in time when the physical energies of the inflatons matches the rest mass of the heavy field \cite{Chen:2015lza, Qin:2025xct},\footnote{One can understand this as a temporary restoration of energy conservation, since $k_1/a(\eta_1)+k_2/a(\eta_1)=M$. More generally, the resonances are determined from the saddle points in the complex time plane \cite{Qin:2025xct}.}
\begin{align}
    -k_{12}\eta_1\sim \mu\,,\quad -k_{34}\eta_2\sim \mu\,,
\end{align}
which implies the following inequality,
\begin{align}
    \frac{\eta_1-\eta_2}{k_{12}-k_{34}}=\frac{-\eta_1}{k_{34}}>0\,.\label{NecessaryInequalityFromResonance}
\end{align}
In other words, the remainder vanishes at the necessary histories for particle production/decay, because \eqref{NecessaryInequalityFromResonance} implies
\begin{align}
    \theta(\eta_1-\eta_2)-\theta(k_{12}-k_{34})=\theta(\eta_2-\eta_1)-\theta(k_{34}-k_{12})=0\,.
\end{align}
With the cut result, it is straightforward to complete the time integrals to obtain
\begin{align}
    \nonumber(\mathcal{I}_{++})_{\rm signal}=&\,-\frac{H^6}{\Lambda^2}\frac{\pi^2(16\mu^4+40\mu^2+9)^2\sech^2(\pi\mu)}{2^{19} }\frac{1}{k_1 k_2 k_3 k_4 s^5} \\
    &\times\left[F\left(-\frac{k_{12}}{s}\right)F\left(\frac{k_{34}}{s}\right)\theta(k_{12}-k_{34})+F\left(\frac{k_{12}}{s}\right)F\left(-\frac{k_{34}}{s}\right)\theta(k_{34}-k_{12})\right]\,,
\end{align}
where
\begin{align}
    F(x)\equiv {}_2{\rm F}_1\Bigg[\begin{array}{c} \frac{5}{2}-i\mu, \frac{5}{2}+i\mu\\[2pt] 3 \end{array}\Bigg|\,\frac{1}{2}-\frac{x}{2}\Bigg]
\end{align}
is the Hypergeometric function. Finally, expanding the result at the soft limit $s\to 0$ with $\mu\gg 1$, we obtain the desired simplified form of cosmological signals (c.f. \eqref{CCphysicsInANutshell}),
\begin{align}
    (\mathcal{I}_{++})_{\rm signal}\propto \frac{\mu^3 e^{-\pi\mu} }{\Lambda^2}\left[{\rm sn}(k_{12}-k_{34})\sin\left(\mu\ln\frac{k_{12}}{k_{34}}\right)-\cos\left(\mu\ln\frac{4k_{12} k_{34}}{s^2}+\mathcal{O}\left(\mu^{-1}\right)\right)\right]\,.
\end{align}
As expected, both the non-local signal and the local signal are suppressed by half the Boltzmann factor as well as the couplings, $\mathcal{A}^{\rm L},\mathcal{A}^{\rm NL}\sim \Lambda^{-2} e^{-\pi\mu}$. The fact that $|\mathcal{A}^{\rm L}|\,=|\mathcal{A}^{\rm NL}|$ in this model is a special consequence of having a dS-invariant dispersion relation for the heavy field $\sigma$. In more general cases with chemical potential, for instance, the amplitude of local and non-local signals would be different i.e. $|\mathcal{A}^{\rm L}|\,\neq|\mathcal{A}^{\rm NL}|$ \cite{Tong:2021wai, Qin:2022fbv}. We note that the power-law enhancement $\mu^3$ is model-dependent and generally increases with the number of derivatives in the interactions. The phase of the non-local signal is also model-dependent, while that of the local signal always remains zero.

\subsection{Lessons from propagators in the Keldysh-$r/a$ basis}

While the cutting rule gives a intuitive physical picture on the heavy field sector, it does not pin down the exact source of individual signals and explain their physical impact on the massless inflaton sector. To achieve these goals, we switch gear in this subsection and introduce the Keldysh-$r/a$ field basis,\footnote{So far we do so only for the heavy field. We will introduce the $r/a$ basis for the inflaton field as well in the next section.}
\begin{align}
    \sigma_r &= \frac{\sigma_+ + \sigma_-}{2}, \quad \sigma_a = \sigma_+ - \sigma_-\,.
\end{align}
As streamlined in \Sec{sec:prelude}, the retarded, advanced and Keldysh propagators are given by (c.f. \eqref{eq_GRAK})
\begin{align}
	G^R_\sigma(s;\eta_1,\eta_2)&= i\left[u_\sigma(s,\eta_1)u_\sigma^*(s,\eta_2)-u_\sigma^*(s,\eta_1)u_\sigma(s,\eta_2)\right]\theta(\eta_1-\eta_2)\,,\bigg.\\
    G^A_\sigma(s;\eta_1,\eta_2)&= i\left[u_\sigma^*(s,\eta_1)u_\sigma(s,\eta_2)-u_\sigma(s,\eta_1)u_\sigma^*(s,\eta_2)\right]\theta(\eta_2-\eta_1)\,,\bigg.\\
    G^K_\sigma(s;\eta_1,\eta_2)&= \frac{i}{2}\left[u_\sigma(s,\eta_1)u_\sigma^*(s,\eta_2)+u_\sigma^*(s,\eta_1)u_\sigma(s,\eta_2)\right]\,.\bigg.
\end{align}
In the $r/a$ basis, the propagators clearly separate into different categories. The retarded/advance propagators are proportional to the free-theory field commutator in the operator formalism
\begin{align}
	G^R_\sigma,\,G^A_\sigma\propto \langle[\widehat{\sigma}(\bm s, \eta_1),\widehat{\sigma}(-\bm s, \eta_2)]\rangle'=u_\sigma(s,\eta_1)u_\sigma^*(s,\eta_2)-u_\sigma^*(s,\eta_1)u_\sigma(s,\eta_2)\,,
\end{align}
and are only supported inside the forward/backward light cone. The Keldysh propagator, on the other hand, is proportional to the anticommutator,
\begin{align}
	G^K_\sigma\propto \langle\{\widehat{\sigma}(\bm s, \eta_1),\widehat{\sigma}(-\bm s, \eta_2)\}\rangle'=u_\sigma(s,\eta_1)u_\sigma^*(s,\eta_2)+u_\sigma^*(s,\eta_1)u_\sigma(s,\eta_2)\,,
\end{align}
and is supported everywhere in spacetime. The retarded and advanced propagators are also intrinsically non-factorisable due to the time-ordering $\theta$-function. However, this does not suggest that any combination of them are non-factorisable. To single out the irreducible non-separability of the propagators, one can extract the symmetric and antisymmetric part of the retarded and advanced propagators as the principal-value propagator and the Pauli-Jordan propagator,
\begin{align}
	G^P_\sigma(s;\eta_1,\eta_2)&\equiv \frac{1}{2}\left[G^R_\sigma(s;\eta_1,\eta_2)+G^A_\sigma(s;\eta_1,\eta_2)\right]\,,\\
	G^\Delta_\sigma(s;\eta_1,\eta_2)&\equiv G^R_\sigma(s;\eta_1,\eta_2)-G^A_\sigma(s;\eta_1,\eta_2)\,.
\end{align}
Thus the space of heavy field propagators can be equivalently spanned by the following three linearly independent propagators,
\begin{align}
	\label{eq_dSGH}
	&G_\sigma^P(s;\eta_1,\eta_2) = \frac i2  \big[ u_\sigma(s,\eta_1) u^*_\sigma(s,\eta_2) - u^*_\sigma(s,\eta_1) u_\sigma(s,\eta_2) \big] \text{sgn}(\eta_1-\eta_2)\,,\\
	\label{eq_dSGD}
	&G_\sigma^\Delta(s;\eta_1,\eta_2) = i \big[ u_\sigma(s,\eta_1) u^*_\sigma(s,\eta_2) - u^*_\sigma(s,\eta_1) u_\sigma(s,\eta_2) \big]\,,\\
	\label{eq_dSGK}
	&G_\sigma^K(s;\eta_1,\eta_2) =  \frac i2 \big[ u_\sigma(s,\eta_1) u^*_\sigma(s,\eta_2)+u^*_\sigma(s,\eta_1) u_\sigma(s,\eta_2) \big]\,.
\end{align}
In this basis, the principal-value propagator is the only function of time that is non-factorisable. We will see later that this irreducible non-factorisability is associated with the infinite tower of unitary contact interactions in the low-energy single-field EFT.

To uncover more useful properties of these Keldysh $r/a$-basis propagators, we plug in the heavy field mode function \eqref{eq:modefctvs} and perform a late-time expansion,
\begin{align}
	\nonumber G^P_\sigma(s;\eta_1,\eta_2)&=\frac{i\pi  H^2}{4\sinh \pi\mu} (\eta_1 \eta_2)^{3/2} \left[J_{i \mu }(-s  \eta_1 ) J_{-i \mu }(-s \eta_2)-J_{-i \mu }(-s \eta_1 ) J_{i \mu }(-s \eta_2)\right]{\rm sn}(\eta_1-\eta_2)\\
	&=\frac{H^2}{2\mu} (\eta_1 \eta_2)^{3/2}\left[\sin \left(\mu\ln \frac{\eta_2}{\eta_1}\right)+\mathcal{O}(s^2\eta_1^2,\,s^2\eta_2^2)\right]{\rm sn}(\eta_1-\eta_2)\,.\label{eq:dSHadamardPropIRExpansion}
\end{align}
Examining the structure of \eqref{eq:dSHadamardPropIRExpansion}, we notice two important properties that will turn out to be crucial for the analysis on cosmological collider signals in the next subsection:

\begin{itemize}
	\item The principal-value propagator $G^P_\sigma$ is analytic in the momentum variable $s$ and admit a convergent Taylor expansion at the origin $s=0$. Mathematically speaking, the analyticity follows from the power series representation of the Bessel-$J$ function,
	\begin{align}
		J_\alpha(z)=\left(\frac{z}{2}\right)^\alpha \sum_{n=0}^\infty \frac{(-1)^n}{n!\Gamma(\alpha+n+1)}\left(\frac{z}{2}\right)^{2n}\,,\quad \alpha=i\mu\,,
	\end{align}
	where the branching point at $z=0$ is cancelled by the multiplication of $J_{-\alpha}$ in $G^P_\sigma$. From a physical perspective, the analyticity of $G^P_\sigma$ at $s=0$ can be traced back to the locality of the $\sigma$ Lagrangian. In fact, all causal propagators ($G^{R/A}$ or equivalently $G^{P/\Delta}$) that are proportional to the field commutator enjoy this property. In Appendix \ref{zeroMomentumAnalyticityAppendix}, we show that the analyticity of the causal propagators in the vicinity of zero momentum is generic in any spacetime given the locality requirement.
	\item The principal-value propagator $G^P_\sigma$ is purely real in Lorentzian time, and purely imaginary in Euclidean time after a Wick rotation $\eta=i\chi$. The former fact is true in any spacetime background by construction, whereas the latter is specific to (3+1)-dimensional de Sitter. In fact, one can easily check that in $d$ spatial dimensions, the Wick-rotated principal-value propagator has a complex phase $G^P_\sigma\propto e^{-i\pi d/2}$. 
\end{itemize}
Since the Pauli-Jordan propagator $G^\Delta_\sigma$ differs from the principal-value propagator $G^P_\sigma$ by a trivial sign function, it shares the same properties above. Moving on to the Keldysh propagator, we obtain a slightly more complicated structure,
\begin{align}
	\nonumber \nonumber G^K_\sigma(s;\eta_1,\eta_2)&=\frac{i \pi  H^2(\eta_1 \eta_2)^{3/2}}{4\sinh^2( \pi\mu)}  \Bigg\{\cosh (\pi \mu) \bigg[J_{i \mu }(-s
	   \eta_1 ) J_{-i \mu }(-s \eta_2)+J_{-i \mu }(-s
	   \eta_1 ) J_{i \mu }(-s \eta_2)\bigg]\\
	&\qquad\qquad\qquad\qquad -\bigg[ J_{i \mu }(-s \eta_1 ) J_{i \mu }(-s \eta_2)+
	   J_{-i \mu }(-s \eta_1 ) J_{-i \mu }(-s \eta_2)\bigg]\Bigg\}\,,\label{eq:dSKeldyshProp}\\
	\nonumber&=\frac{i H^2(\eta_1 \eta_2)^{3/2}}{2 \mu}  \Bigg[\coth (\pi \mu) \cos\left(\mu\ln\frac{\eta_1}{\eta_2}\right)\\
	   &\qquad\qquad\qquad -\csch(\pi\mu)\cos\left(\mu\ln\frac{s^2\eta_1\eta_2}{4}+\delta\right)+\mathcal{O}(s^2\eta_1^2,s^2\eta_2^2)\Bigg]\,,\label{eq:lateTimeKeldyshProp}
\end{align}
with the phase parameter $\delta\equiv 2\arg \Gamma(1-i\mu)$. From \eqref{eq:lateTimeKeldyshProp}, we again make two useful observations:
\begin{itemize}
	\item The Keldysh propagator has a piece (the first lines of \eqref{eq:dSKeldyshProp} and \eqref{eq:lateTimeKeldyshProp}) that is analytic at $s=0$ and a piece (the second lines of \eqref{eq:dSKeldyshProp} and \eqref{eq:lateTimeKeldyshProp}) that is non-analytic at $s=0$.
	\item In Lorentzian time, the Keldysh propagator is purely imaginary by construction. After a Wick rotation $\eta=i\chi$, however, its two pieces behave differently in Euclidean time. The analytic piece is purely real whereas the non-analytic piece is complex in general.
\end{itemize}
We shall see in the next subsection that these analyticity and reality properties are important to extract cosmological collider signals.

\subsection{Carving out the cosmological collider signals}\label{eq:carving}

In this subsection, we shall use the analyticity and reality properties found in the previous subsection to pin down the source of cosmological collider signals and inspect their physical impact on the massless sector. Coming back to the time-ordered diagram,
\begin{align}
    \nonumber   \mathcal{I}_{++} =& -\frac{1}{\Lambda^2} \int_{-\infty}^0 \mathrm d\eta_1 \mathrm d\eta_2\, a^2(\eta_1) \,a^2(\eta_2)\, \partial_{\eta_1}K_+^\varphi(k_1,\eta_1)\times \partial_{\eta_1}K_+^\varphi(k_2,\eta_1) \nonumber \\
    &\qquad \qquad \qquad \times D_{++}^\sigma(s;\eta_1,\eta_2)\times \partial_{\eta_2}K_+^\varphi(k_3,\eta_2)\times \partial_{\eta_2}K_+^\varphi(k_4,\eta_2)\,,
\end{align}
we now decompose the time-ordered massive propagator into a principal-value propagator and a Keldysh propagator,
\begin{align}
	D_{++}^\sigma(s;\eta_1,\eta_2)=-i\left[G^P_\sigma(s;\eta_1,\eta_2)+G^K_\sigma(s;\eta_1,\eta_2)\right]\,.
\end{align}
Using the properties of these propagators, the two types of cosmological collider signals can then be located separately:
\begin{itemize}
	\item The non-local signal is identified by its non-analyticity structure. i.e. a branching point at the origin of the exchange momentum $s=0$. Such kinematic singularity can be traced back to the characteristic time dependence at the propagator level,
	\begin{align}
		\cos\left(\mu\ln\frac{s^2\eta_1\eta_2}{4}+\cdots\right)\quad\Rightarrow \quad\cos\left(\mu\ln\frac{4k_{12} k_{34}}{s^2}+\cdots\right)\,.
	\end{align}
	Since the principal-value propagator $G^P_\sigma$ is analytic in the exchange momentum $s$ in the soft limit, we immediately deduce that the principal-value propagator \textit{cannot} produce non-local signals. Consequently, the non-local signal must solely originate from the Keldysh propagator of the massive field.
	
	\item The local signal, on the other hand, can be identified using the reality property in Euclidean time. Under a Wick rotation $\eta=i\chi$, the integral becomes
	\begin{align}
	    \nonumber   \mathcal{I}_{++} =& \frac{-i}{\Lambda^2} \int_0^\infty \mathrm \dd\chi_1 \dd\chi_2  \,a^2(i\chi_1) a^2(i\chi_2)\, \partial_{\chi_1}K_+^\varphi(k_1,i\chi_1)\times \partial_{\chi_1}K_+^\varphi(k_2,i\chi_1) \nonumber \\
	    &\qquad \times \left(G^P_\sigma(s;i\chi_1,i\chi_2)+G^K_\sigma(s;i\chi_1,i\chi_2)\right)\times \partial_{\chi_2}K_+^\varphi(k_3,i\chi_2)\times \partial_{\chi_2}K_+^\varphi(k_4,i\chi_2)\,.\label{IppAfterWick}
	\end{align}
	The final correlator consisting of the combination $\mathcal{I}_{++}+\mathcal{I}_{--}$ amounts to taking the real part of \eqref{IppAfterWick}. Note that the Wick-rotated scale factors and massless propagators are both purely real,
	\begin{align}
		a^2(i\chi_1)a^2(i\chi_2)&=\frac{1}{H^4\chi_1^2\chi_2^2}\,,\\
		K_+^\varphi(k,i\chi)&=\frac{H^2}{2k^3}(1+k\chi)e^{-k\chi}\,.
	\end{align}
	suggesting that only the imaginary (in Euclidean time) component in the principal-value and Keldysh propagators can enter the observable correlator. Focusing on the local signal, which derives from the following time dependences in the propagators,
	\begin{align}
		\sin\left(\mu\ln\frac{\eta_1}{\eta_2}+\cdots\right)\quad\Rightarrow \quad\sin\left(\mu\ln\frac{k_{12}}{k_{34}}+\cdots\right)\,,
	\end{align}
	we notice that both the principal-value and Keldysh propagators contain such time dependences and are equally capable of producing local signals. However, the relevant component of the Keldysh propagator (as discussed below \eqref{eq:lateTimeKeldyshProp}) is purely real in Euclidean time, and thereby drops out of the final correlator. Thus the Keldysh propagator \textit{cannot} produce local signals in the final correlator. As a result, we deduce that in parity-conserving theories, such as our toy model \eqref{eq:S_dSL}, the local signal solely originates from the principal-value propagator.
\end{itemize}

These arguments can be verified by explicitly isolating the contributions to the four-point correlator $\mathcal I$ \eqref{eq_dS4ptinin} from each propagator of the exchanged field $\sigma$ in the Keldysh-r/a basis, by substituting $D_{\mathsf a\mathsf b}^\sigma$ with $G^{P/\Delta/K}_\sigma$ during the standard in-in integral. This top-down calculation will also serve as a consistency check of the bottom-up EFT calculation, as will be detailed in \Sec{sec:EFT}.
Similar to the flat-space counterpart, the principal-value propagator $G^P_\sigma$ contributes to $\pm\pm$ branches (c.f. Eq.~\eqref{eq_flatIH}):
\begin{align}
    \nonumber   \mathcal{I}^P =&~\frac{1}{\Lambda^2} \sum_{\mathsf a=\pm} (-1) \int_{-\infty}^0 \mathrm d\eta_1 \,a^2(\eta_1)\, \mathrm d\eta_2 \,a^2(\eta_2)\, \partial_{\eta_1}K_{\mathsf a}^\varphi(k_1,\eta_1)\times \partial_{\eta_1}K_{\mathsf a}^\varphi(k_2,\eta_1)\\
    &\qquad \qquad \qquad \qquad \times \big[-\mathsf a i G_\sigma^P(s;\eta_1,\eta_2)\big] \times \partial_{\eta_2}K_{\mathsf b}^\varphi(k_3,\eta_2)\times \partial_{\eta_2}K_{\mathsf b}^\varphi(k_4,\eta_2) 
\end{align}
that is\footnote{Such integrals can be computed using the Partial Mellin-Barnes representation method (see, e.g., \cite{Qin:2022lva,Qin:2022fbv,Qin:2024gtr}). We omit the intermediate steps, which are analogous to those detailed in \cite{Qin:2022fbv}, and provide only the final results here.}
\begin{align}\label{eq_IPindS}
    \mathcal{I}^P =&~\frac{H^6}{16\Lambda^2k_1k_2k_3k_4 (k_{12}k_{34})^{5/2}}
    \left[ \frac{i\sinh\pi\mu}{2\pi}\, {\mathbf F}_+\left(\frac{s}{k_{12}}\right){\mathbf F}_-\left(\frac{s}{k_{34}}\right) + (\mu\to-\mu) + \mathbf B_>\left(\frac{s}{k_{12}},\frac{s}{k_{34}}\right)\right]\,,
\end{align}
where we have defined:
\begin{align}
\mathbf F_\pm(r) = \left(\frac{r}2\right)^{\pm i\mu} \Gamma\left(\frac52\pm i\mu\right)\Gamma(\mp i\mu)\,{}_2\mathrm F_1\left[
\begin{matrix}
\frac54\pm\frac{i\mu}2,\frac74\pm\frac{i\mu}2\\
1\pm i\mu
\end{matrix}\middle| r^2
\right]
\end{align}
that characterises the cosmological collider signal, namely the logarithmic oscillation encoded in the factor $r^{\pm i\mu}$, and 
\begin{align}\label{eq_Bn_1}
\mathbf B_>(r_1,r_2) = &~ 24 \left(\frac{r_1}{r_2}\right)^{5/2}\sum_{n_1,n_2=0}^\infty  \frac{(-1)^{n_{12}}}{n_1!n_2!}\left(\frac{r_1}{2}\right)^{2n_{12}}\frac{(-i\mu)_{-n_1}(+i\mu)_{-n_2}}{i\mu(2n_2+\frac52-i\mu)} \nonumber\\
&\times {}_2\mathrm F_1\left[
\begin{matrix}
2n_2+\frac52-i\mu,2n_{12}+5 \\ 2n_2+\frac72-i\mu
\end{matrix}\middle| -\frac{r_1}{r_2}
    \right] + (\mu\to-\mu)\,,
\end{align}
which is fully analytic in the region $0\leq r_1<r_2 < 1$ (except for the trivial scaling factor $(r_1/r_2)^{5/2}$) and thus plays the role of background.\footnote{One might be more familiar with the double series expression that was originally found in \cite{Arkani-Hamed:2018kmz}, see also \cite{Qin:2022fbv}
\begin{align}\label{eq_Bn_2}
\mathbf B_>(r_1,r_2) = ~ \mathcal O_1\mathcal O_2 \sum_{m,n=0}^\infty \frac{(-1)^n(n+1)_{2m}}{2^{2m+1}(\frac{n}2+\frac{1}4+\frac{i\mu}2)_{m+1}(\frac{n}2+\frac{1}4-\frac{i\mu}2)_{m+1}}r_1^{2m+1}\left(\frac{r_1}{r_2}\right)^{n}\,,
\end{align}
where the weight-shifting operators are
\begin{align}
\mathcal O_i\equiv r_i^{3/2}\partial_{r_i}^2+r_i^{1/2}\partial_{r_i}\,.
\end{align}
We note that the expressions \eqref{eq_Bn_1} and \eqref{eq_Bn_2} are equal so long as $0<r_1<r_2<1$, but the former one has a larger convergent region and is valid in the full physical region $0<r_{1,2}<1$.
} We notice that while $\mathcal I^P$ should be symmetric in $k_{12}\leftrightarrow k_{34}$ by construction, this is not manifest in the expression \eqref{eq_IPindS}. Once $r_{1}$ and $r_2$ are interchanged, the term $\mathbf B_>(r_2,r_1)$ ceases to be analytic in the region $0\leq r_1<r_2<1$. In particular, non-analytic terms arise around $r_1/r_2\to 0$, where the argument of the hypergeometric function is $-r_2/r_1\to \infty$, hitting a branch point. Consequently, one can identify the cosmological collider signal and the background with the first two terms and the third term in Eq.~\eqref{eq_IPindS} respectively, \textit{only} in the region $r_1<r_2$. In the opposite case where $r_2<r_1$, one may instead exchange $k_{12}\leftrightarrow k_{34}$ to obtain the correct form of signal, that is
\begin{align}
\mathcal I^P_{\text{signal}}&=\left[\frac{i \sinh\pi\mu\, H^6}{32\pi \Lambda^2k_1k_2k_3k_4 (k_{12}k_{34})^{5/2}}
     {\mathbf F}_+\left(\frac{s}{k_{12}}\right){\mathbf F}_-\left(\frac{s}{k_{34}}\right) + (\mu\to-\mu) \right]\theta(k_{12}-k_{34}) \nonumber \\
     &\qquad  + (k_{12}\leftrightarrow k_{34}) \bigg.\,.
     \label{eq_topdownIPsignal}
\end{align}

Following the similar procedure, we can identify the contributions from $G^\Delta_\sigma$ and $G^K_\sigma$ (c.f. Eqs.~\eqref{eq_flatID} and \eqref{eq_flatIK}), which are
\begin{align}
    \nonumber   \mathcal{I}^\Delta =&~\frac{1}{\Lambda^2} \sum_{\mathsf a=\pm} \int_{-\infty}^0 \mathrm d\eta_1 \,a^2(\eta_1)\, \mathrm d\eta_2 \,a^2(\eta_2)\, \partial_{\eta_1}K_{\mathsf a}^\varphi(k_1,\eta_1)\times \partial_{\eta_1}K_{\mathsf a}^\varphi(k_2,\eta_1)\\
    &\qquad \qquad \qquad \times \left[\frac{\mathsf a i}2 G_\sigma^\Delta(s;\eta_1,\eta_2)\right] \times \partial_{\eta_2}K_{\mathsf b}^\varphi(k_3,\eta_2)\times \partial_{\eta_2}K_{\mathsf b}^\varphi(k_4,\eta_2), 
\end{align}
leading to 
\begin{align}
   \mathcal{I}^\Delta =&~\frac{-\sinh^2\pi\mu\, H^6}{32\pi\Lambda^2k_1k_2k_3k_4 (k_{12}k_{34})^{5/2}}
   {\mathbf F}_+\left(\frac{s}{k_{12}}\right){\mathbf F}_-\left(\frac{s}{k_{34}}\right) + (\mu\to-\mu)\,,
   \label{eq_topdownID}
   \end{align}
   and
   \begin{align}
     \nonumber   \mathcal{I}^K =&~\frac{1}{\Lambda^2} \sum_{\mathsf a=\pm}(-\mathsf a\mathsf b) \int_{-\infty}^0 \mathrm d\eta_1 \,a^2(\eta_1)\, \mathrm d\eta_2 \,a^2(\eta_2)\, \partial_{\eta_1}K_{\mathsf a}^\varphi(k_1,\eta_1)\times \partial_{\eta_1}K_{\mathsf a}^\varphi(k_2,\eta_1)\\
    &\qquad \qquad \qquad \times \left[-i  G_\sigma^K(s;\eta_1,\eta_2)\right] \times \partial_{\eta_2}K_{\mathsf b}^\varphi(k_3,\eta_2)\times \partial_{\eta_2}K_{\mathsf b}^\varphi(k_4,\eta_2), 
\end{align}
reading
\begin{align}
    \mathcal{I}^K =&~\frac{H^6}{32\pi\Lambda^2k_1k_2k_3k_4 (k_{12}k_{34})^{5/2}}\left[(1+i\sinh\pi\mu)
   {\mathbf F}_+\left(\frac{s}{k_{12}}\right){\mathbf F}_+\left(\frac{s}{k_{34}}\right)  \right. \nonumber\\
   & \qquad\qquad\qquad\qquad\qquad\qquad~ + \left. \cosh^2\pi\mu\,{\mathbf F}_+\left(\frac{s}{k_{12}}\right){\mathbf F}_-\left(\frac{s}{k_{34}}\right)   \right] + (\mu\to-\mu)\,.
     \label{eq_topdownIK}
\end{align}
From the combinations of $\mathbf F_\pm$, we can verify that only $G^K_\sigma$ generates the non-local signal, whereas all three propagators contribute to the local signal. In the large-mass limit, however, the dominant local contribution arises from $G^P_\sigma$: the apparently large local terms (of order $1$, rather than $e^{-\pi\mu}$) in $\mathcal I^\Delta$ and $\mathcal I^K$ cancel exactly due to the identity $-\sinh^2\pi\mu + \cosh^2\pi\mu = 1$. This confirms our characterisation of the origin of the different types of cosmological collider signals.

\paragraph{Generalisation.} With the origin of signals understood in our toy model, we can straightforwardly generalise to a much broader family of parity-violating theories. Consider a heavy field $\sigma_{i_1\cdots i_S}$ with integer spin $S=0,1,2,\cdots$ and mass $M$ that interacts with a massless field $\varphi$ via a triple vertex,
\begin{align}
	S[\varphi,\sigma_{i_1\cdots i_S}]=S_0^\varphi[\varphi]+S_0^\sigma[\sigma_{i_1\cdots i_S}]+\frac{1}{2}\int \dd \eta\,\dd^3 \bm{x} \,a^4(\eta) \left[\mathcal{\hat V}\left(\frac{\partial_\eta}{a(\eta)}, \frac{\partial_i}{a(\eta)}\right) \varphi\varphi\sigma\right]_{\text{contract}}\,,\label{generalMasslessMassiveSpinningInteractingTheory}
\end{align}
where $\mathcal{\hat V}$ denotes a general differential operator expandable as a real polynomial in derivatives. The detailed form of this interaction vertex is unimportant except that to maintain rotational invariance, all spatial indices must be contracted using either the Kronecker symbol $\delta_{ij}$ or the Levi-Civita symbol $\epsilon_{ijk}$. Vertices that involve an even (odd) number of Levi-Civita symbols are, by definition, even (odd) under parity. $S_0^\varphi$ and $S_0^\sigma$ denote the free Lagrangians of the massless scalar and massive (spinning) field. For simplicity, we shall assume that the free theory of the spinning field $S_0^\sigma$ conserves parity. The four-point correlator in the $s$-channel is then given by an exchange diagram as in the toy example,
\begin{align}
	\langle \varphi_{\bm k_1}\varphi_{\bm k_2}\varphi_{\bm k_3}\varphi_{\bm k_4} \rangle_s'=\mathcal{I}_{++}+\mathcal{I}_{--}+\mathcal{I}_{-+}+\mathcal{I}_{+-}\,,
\end{align}
where
\begin{align}
	\nonumber\mathcal{I}_{++}=& - \int_{-\infty}^0 \mathrm d\eta_1 \mathrm d\eta_2\, a^4(\eta_1) \,a^4(\eta_2)\,\Bigg[\mathcal{\hat V}\left(\frac{\partial_{\eta_1}}{a(\eta_1)}, \frac{-i\{\bm{k}\}_L}{a(\eta_1)}\right) \mathcal{\hat V}\left(\frac{\partial_{\eta_2}}{a(\eta_2)}, \frac{-i\{\bm{k}\}_R}{a(\eta_2)}\right)\\
	& \qquad \quad K_+^\varphi(k_1,\eta_1)\times K_+^\varphi(k_2,\eta_1) \times D_{++}^\sigma(s;\eta_1,\eta_2)\times K_+^\varphi(k_3,\eta_2)\times K_+^\varphi(k_4,\eta_2)\Bigg]_{\text{contract}}\,.
\end{align}
Here $\{\bm{k}\}_L$ and $\{\bm{k}\}_R$ collectively denote the momentum vectors flowing out of the left and right vertices. Going to Euclidean time by a Wick rotation $\eta=i\chi$ and decomposing the massive field propagator, we obtain an expression that is otherwise purely imaginary apart from the decomposed massive propagator,
\begin{align}
	\nonumber\mathcal{I}_{++}=&  -i\int_{0}^\infty \mathrm d\chi_1 \mathrm d\chi_2\, a^4(\chi_1) \,a^4(\chi_2)\,\Bigg[\mathcal{\hat V}\left(\frac{\partial_{\chi_1}}{a(\chi_1)}, \frac{\{\bm{k}\}_L}{a(\chi_1)}\right) \mathcal{\hat V}\left(\frac{\partial_{\chi_2}}{a(\chi_2)}, \frac{\{\bm{k}\}_R}{a(\chi_2)}\right)\\
	\nonumber& \qquad \quad \times K_+^\varphi(k_1,i\chi_1)\times K_+^\varphi(k_2,i\chi_1) \times \left(G^H_\sigma(s;i\chi_1,i\chi_2)+G^K_\sigma(s;i\chi_1,i\chi_2)\right)\\
	&\qquad \quad \times K_+^\varphi(k_3,i\chi_2)\times K_+^\varphi(k_4,i\chi_2)\Bigg]_{\text{contract}}\,.
\end{align}
For a parity-conserving free theory of the massive spinning field $\sigma_{i_1\cdots i_S}$, the mode function remains the same as a de Sitter-invariant massive scalar \eqref{eq:modefctvs} with the same analyticity and reality properties for the principal-value and Keldysh propagators. We are therefore led to the same conclusion that the non-local signal originates from the Keldysh component $G^K_\sigma$. On the other hand, the local signal requires a more careful treatment when parity is involved since the reality argument is sensitive to parity. In general, conjugating Schwinger-Keldysh colouring $+\leftrightarrow -$ is equivalent to a complex conjugation combined with a momentum reversal \cite{Liu:2019fag},
\begin{align}
	\mathcal{I}_{\sf -a \,-b}=\left(\mathcal{I}_{\sf a b}\right)^*\Big|_{\{\bm{k}\}\to -\{\bm{k}\}}\,.
\end{align}
Consequently, the parity-even and parity-odd parts of the exchange diagram transform as
\begin{align}
	\mathcal{I}_{--}^{\rm PE}=+\left(\mathcal{I}_{++}^{\rm PE}\right)^*\,,\\
\mathcal{I}_{--}^{\rm PO}=-\left(\mathcal{I}_{++}^{\rm PO}\right)^*\,,
\end{align}
under the Schwinger-Keldysh colouring conjugation. This implies that in the parity-even (odd) sector, it is the real (imaginary) part of $\mathcal{I}_{++}^{\rm PO}$ that enters the final correlator,
\begin{align}
	\langle \varphi_{\bm k_1}\varphi_{\bm k_2}\varphi_{\bm k_3}\varphi_{\bm k_4}\rangle^{\rm PE}&\simeq \mathcal{I}_{++}^{\rm PE}+\mathcal{I}_{--}^{\rm PE}= 2\Re \mathcal{I}_{++}^{\rm PE}\,,\\
	\langle \varphi_{\bm k_1}\varphi_{\bm k_2}\varphi_{\bm k_3}\varphi_{\bm k_4}\rangle^{\rm PO}&\simeq \mathcal{I}_{++}^{\rm PO}+\mathcal{I}_{--}^{\rm PO}= 2i\Im \mathcal{I}_{++}^{\rm PO}\,.
\end{align}
Interestingly, this implies that the local signal in final correlators originates differently under parity: it comes from the principal-value propagator $G^P_\sigma$ in the parity-even sector (as in the toy example), and from the Keldysh propagator $G^K_\sigma$ in the parity-odd sector. We note in passing that in the parity-odd sector, both signals originate from the Keldysh propagator of the massive field and thus inherit its factorisability. This is consistent with the parity-odd factorisation theorem, which asserts that parity-odd correlators must factorise as a whole in general circumstances \cite{Stefanyszyn:2023qov,Stefanyszyn:2024msm}. 
This conclusion could readily be generalised to cases where the free theory violates parity via the inclusion of the helical chemical potential, which could be most easily seen if we treat the chemical potential perturbatively and resum.
We close this section by summarising the origin of leading cosmological collider signals for general theories of the type \eqref{generalMasslessMassiveSpinningInteractingTheory} in Tab.\,\ref{tab_sum}.


\section{An open effective field theory perspective}\label{sec:EFT}
        
We now aim at integrating the field $\sigma$ and considering the single-field open EFT for $\varphi$ only, following the same approach we used in the flat space case in \Sec{subsec:flatEFT}. In particular, we focus on the effective quartic interactions generated and discuss whether the terms obtained can be encountered in a unitary theory or not. Formally, we perform the path integral over the $\sigma$ field, leading to 
\begin{align}
    \mathcal{Z}\left[J_r, J_a \right] = \int_{\mathrm{BD}}^\varphi \mathcal{D}\varphi_r \int_{\mathrm{BD}}^0 \mathcal{D}\varphi_a & \ee^{i \int \dd^4 x \sqrt{-g}\left(J_r\varphi_a + J_a\varphi_r \right)}  \ee^{iS^\varphi_{0}\left[ \varphi_r, \varphi_a\right]} \ee^{iS_{\mathrm{IF}}\left[ \varphi_r, \varphi_a\right]}.\Bigg.
\end{align}
where 
\begin{align}
    \ee^{iS_{\mathrm{IF}} \left[ \varphi_r, \varphi_a\right]} = \int_{\mathrm{BD}}^\sigma \mathcal{D}\sigma_r \int_{\mathrm{BD}}^0 \mathcal{D}\sigma_a  \ee^{iS^\sigma_{0}\left[ \sigma_r, \sigma_a\right]} \ee^{iS_{\mathrm{int}}\left[ \varphi_r, \varphi_a, \sigma_r, \sigma_a\right]}.\Bigg.
\end{align}
In the Keldysh basis, we have where the free actions 
\begin{align}
    S_0^\varphi[\varphi_r,\varphi_a] =& -\frac12 \int \dd \eta a^2(\eta)\int \dd^3 x \,
        \begin{pmatrix}
        \varphi_r & \varphi_a
        \end{pmatrix}
        \begin{pmatrix}
        G^K_\varphi & G^R_\varphi \\ G^A_\varphi & 0
        \end{pmatrix}^{-1}
        \begin{pmatrix}
        \varphi_r \\ \varphi_a
        \end{pmatrix},\\
        S_0^\sigma[\sigma_r,\sigma_a] =& -\frac12\int \dd \eta a^2(\eta)\int \dd^3 x\,
        \begin{pmatrix}
        \sigma_r & \sigma_a
        \end{pmatrix}
        \begin{pmatrix}
        G^K_\sigma & G^R_\sigma \\ G^A_\sigma & 0
        \end{pmatrix}^{-1}
        \begin{pmatrix}
        \sigma_r \\ \sigma_a
        \end{pmatrix},
\end{align}
and the interaction part
\begin{align}
    S_{\text{int}}[\varphi_r,\varphi_a,\sigma_r,\sigma_a] =&~\frac{1}{2\Lambda} \int \dd \eta a^2(\eta)\int \dd^3 x \, ( \dot\varphi_+^2\sigma_+ - \dot\varphi_-^2\sigma_-)\nonumber\\
    =&~\frac{1}{\Lambda} \int \dd \eta a^2(\eta)\int \dd^3 x\, \Big( \dot\varphi_r\dot\varphi_a\sigma_r + \frac12 \dot\varphi_r^2\sigma_a + \frac18 \dot\varphi_a^2\sigma_a \Big).
\end{align}
In practice, we follow a procedure similar to the one described in \cite{Proukakis:2024pua} by expanding in powers of $S_{\mathrm{int}}$. At second order in $1/\Lambda$ 
\begin{align}
    \ee^{iS_{\mathrm{IF}}}  = 1 + i \langle S_{\mathrm{int}} \rangle_\sigma - \frac{1}{2} \langle S^2_{\mathrm{int}}
            \rangle_\sigma + \cdots .
\end{align}
The theory being linear in $\sigma$ and having removed the tadpole contribution, we assume for the moment that $\langle S_{\mathrm{int}} \rangle_\sigma = 0$. We then identify the leading order $S_{\mathrm{IF}} \simeq \langle S^2_{\mathrm{int}} \rangle_\sigma$. 
        
    \subsection{Non-local open effective field theory}\label{subsec:NLEFT}
            
The second order effective functional is simply obtained by replacing internal $\sigma$ legs by its propagators, leading to 
\begin{align}\label{eq:IFpre}
            S_{\mathrm{IF}}[\varphi_r,\varphi_a] &= \frac{1}{2\Lambda^2} \int \dd \eta a^2(\eta)\int \dd \eta' a^2(\eta')\int \dd^3 x  \int \dd^3 y  \nonumber \\
            \bigg\{&\left[ \frac{1}{2}\varphi_r^{\prime2}(x) \varphi_r'(y) \varphi_a'(y) + \frac{1}{8} \varphi_r'(y) \varphi_a^{\prime2}(x) \varphi_a'(y)\right]G^{A}_\sigma(x,y) \nonumber \\
            +&\left[ \frac{1}{2} \varphi_r'(x) \varphi_r^{\prime2}(y) \varphi_a'(x) + \frac{1}{8} \varphi_r'(x) \varphi_a'(x) \varphi_a^{\prime2}(y)\right]G^{R}_\sigma(x,y) \nonumber \\
            +&\bigg[\varphi_r'(x)\varphi_r'(y)\varphi_a'(x)\varphi_a'(y)\bigg] G^{K}_\sigma(x,y)\bigg\}. 
\end{align}
As expected, the influence functional is non-local and further assumptions must be performed to obtain a local open EFT. 

\paragraph{Unitary subset.} We anticipate that part of these effective contributions can be captured within a unitary single-field EFT — either local or non-local — while the remaining contributions lie outside this framework and constitute genuinely non-unitary effects. For instance, the last line of \Eq{eq:IFpre} is manifestly non-unitary: it contains an even number of advanced fields and, going back to the original $+/-$ basis, can never be written in a factorised form, $S_{\mathrm{unit}}[\varphi_+] - S_{\mathrm{unit}}[\varphi_-]$. These terms are generally attributed to stochastic effects which couple operators of the system to fluctuations of the surrounding environment \cite{kamenev_2011}.
        
The first two lines of \Eq{eq:IFpre} are more subtle. Part of it should control the unitary time evolution of the system while the rest may be imputed to dissipative effects. To make these roles more explicit, we again decompose the retarded and advanced propagators in their product and difference
\begin{align}
    G^{R}_\sigma(x,y) = G^P_\sigma(x,y) + \frac{G^\Delta_\sigma(x,y)}{2}, \qquad 
    G^{A}_\sigma(x,y) = G^P_\sigma(x,y) -\frac{G^\Delta_\sigma(x,y)}{2} ,
\end{align}
that is 
\begin{align}
    G^P_\sigma(x,y) = \frac{G^{R}_\sigma(x,y) + G^{A}_\sigma(x,y)}{2}, \qquad G^\Delta_\sigma(x,y) = G^{R}_\sigma(x,y) - G^{A}_\sigma(x,y).
\end{align}
The properties of the principal-value propagator $G^P_\sigma(x,y)$ and the Pauli-Jordan propagator $G^\Delta_\sigma(x,y)$ are summarised in Tab.\,\ref{tab_prop}. Note the symmetry and anti-symmetry properties
\begin{align}\label{eq:sym}
    G^P_\sigma(y,x) = G^P_\sigma(x,y), \qquad G^\Delta_\sigma(y,x) = -G^\Delta_\sigma(x,y)
\end{align}
which directly follow from $G^{R}_\sigma(y,x) = G^{A}_\sigma(x,y)$. Contrary to the Keldysh propagator $G^K_\sigma$ that has support outside the light cone, all these propagators are supported only inside the light cone. Moreover, $G^\Delta_\sigma$ is asymmetric, whereas $G^P_\sigma$ is symmetric. This observation foreshadows the link between $G^\Delta_\sigma$ and dissipation, which generates an arrow of time, and between $G^P_\sigma$ and unitary evolution, which is time-reversible.
               
Under this rewriting, \Eq{eq:IFpre} becomes
\begin{align}
    S_{\mathrm{IF}}[\varphi_r,\varphi_a] &= \frac{1}{2\Lambda^2} \int \dd \eta a^2(\eta)\int \dd \eta' a^2(\eta')\int \dd^3 x  \int \dd^3 y  \nonumber \\
    \Bigg\{&\bigg[ \varphi_r^{\prime2}(x) \varphi_r'(y) \varphi_a'(y) +  \frac{1}{4}\varphi_r'(y) \varphi_a^{\prime2}(x) \varphi_a'(y) \bigg]G_\sigma^P(x,y) \nonumber \\
    -&\bigg[ \frac{1}{2}\varphi_r^{\prime2}(x) \varphi_r'(y) \varphi_a'(y) +\frac{1}{8} \varphi_r'(y) \varphi_a^{\prime2}(x) \varphi_a'(y) \bigg] G^\Delta_\sigma(x,y) \nonumber \\
    +&\bigg[\varphi_r'(x)\varphi_r'(y)\varphi_a'(x)\varphi_a'(y)\bigg] G^{K}_\sigma(x,y)\Bigg\},\label{eq:IFpost}
\end{align}
where we used the symmetry properties \eqref{eq:sym} to simplify the expressions. 
Written back in the $+/-$ basis, the line controlled by $G_\sigma^P(x,y)$ takes the form of a non-local unitary EFT, $S_{\mathrm{IF}}[\varphi_r,\varphi_a] \supset S_{\mathrm{unit}}[\varphi_+] - S_{\mathrm{unit}}[\varphi_-] $ with
\begin{align}
    S_{\mathrm{unit}}[\varphi_\pm] = \frac{1}{8\Lambda^2} \int \dd \eta a^2(\eta)\int \dd \eta' a^2(\eta')\int \dd^3 x  \int \dd^3 y \left[ \varphi_\pm^{\prime2}(x) \varphi_\pm^{\prime2}(y) \right]G_\sigma^P(x,y).
\end{align}
Conversely, $G^\Delta_\sigma(x,y)$ controls an orthogonal non-unitary direction we interpret as dissipation.
\begin{tcolorbox}[%
			enhanced, 
			breakable,
			skin first=enhanced,
			skin middle=enhanced,
			skin last=enhanced,
			before upper={\parindent15pt},
			]{}

            \vspace{0.05in}

\paragraph{Summary.}

\Eq{eq:IFpost} represents the second-order effects of $\sigma$ on the dynamics of $\varphi$. It can be decomposed into three distinctive effects:\\
\begin{enumerate}
    \item  The line controlled by $G_\sigma^P(x,y)$ is \textit{unitary} and corresponds to the generation of an effective (non-local) vertex in the Lagrangian of $\varphi$, sometimes called the \textit{Lamb shift}.
    \item The line controlled by $G^\Delta_\sigma(x,y)$ is \textit{non-unitary} and corresponds to the \textit{dissipative} evolution of $\varphi$ through the $\sigma$ medium.
    \item The line controlled by $G_\sigma^K(x,y)$ is \textit{non-unitary} and corresponds to the \textit{noise} generated by fluctuations of the $\sigma$ medium backreacting on the evolution of $\varphi$.
\end{enumerate}
    
\end{tcolorbox}

\paragraph{General parametrisation.} The top-down analysis of cosmological collider signals in the preceding sections provides us a guideline to build an effective parametrisation of the cosmological collider at the level of the influence functional. We now attempt to write down an open EFT that captures the leading cosmological collider signals from general massive field exchanges. We shall turn on all possible couplings between the system $\varphi$ and the environment $\sigma$, under the assumption of scale invariance and a Bunch-Davies UV completion. For simplicity, we restrict attention to the parity-preserving case, noting that the parity-violating generalisation is straightforward. The construction is based upon the late-time expansion of the propagators i.e. \eqref{eq:dSHadamardPropIRExpansion} and \eqref{eq:lateTimeKeldyshProp}. To leading order in position space, we can replace the bi-local propagators by
\begin{align}
    G^P_\sigma(x,y)&\mapsto\frac{ H^2}{2\mu} \times\sin\left(\mu\ln \frac{\eta}{\eta'}\right)\delta^3(\bm x-\bm y)(\eta \eta')^{3/2} \sgn (\eta-\eta')\,,\\
    G^K_\sigma(x,y)&\mapsto i e^{-\pi\mu}(\pi\mu)^{1/2}\frac{H^2}{(4\pi)^2}\times \cos\left(\mu \ln \frac{\eta \eta'}{|\bm x -\bm y|^2}+\frac{\pi}{4}\right)\frac{(\eta \eta')^{3/2}}{|\bm x -\bm y|^3}\,.
\end{align}
Here the terminology of local and non-local signals becomes self-explanatory: the components relevant for producing signals in the principal-value $G^P_\sigma$ and Keldysh $G^K_\sigma$ propagators are sourced by a \textit{local} Dirac $\delta$-function and a \textit{non-local} power-law function, respectively. At higher orders, the propagators systematically yield higher-derivative terms with fixed coefficients. Since an EFT framework allows for arbitrary higher-derivative interactions, we can, without loss of generality, treat these expansion coefficients as free parameters that define the EFT.
We thus obtain an infinite tower of bi-local waveforms, each dressed with independent amplitudes and phases, 
\begin{align}
	G^P_\sigma(x,y)&\mapsto \frac{ H^2}{2\mu}\left[\sum_{m,n=0}^{\infty}c_{m,n}^{(\rm L)} \sin\left(\mu\ln \frac{\eta}{\eta'}+\vartheta_{n,m}^{\rm (L)}\right)\left(\eta^2\partial_{\bm x}^2\right)^m \left(\eta^{\prime 2}\partial_{\bm y}^2\right)^n\right]\delta^3(\bm x-\bm y)(\eta \eta')^{3/2}\sgn (\eta-\eta') \,,\\
    G^K_\sigma(x,y)&\mapsto i \mu^{1/2}e^{-\pi\mu}H^2 \left[\,\sum_{m,n=0}^{\infty} c_{m,n}^{(\rm NL)} \cos\left(\mu \ln \frac{\eta \eta'}{|\bm x -\bm y|^2}+\vartheta_{n,m}^{\rm (NL)}\right)\left(\eta^2\partial_{\bm x}^2\right)^m \left(\eta^{\prime 2}\partial_{\bm y}^2\right)^n\right]\dfrac{(\eta \eta')^{3/2}}{|\bm x-\bm y|^{3}}\,.
\end{align}
Here $c_{m,n}^{(\rm L)}, c_{m,n}^{(\rm NL)}\sim (1/\mu)^{m+n}$ are real constants that behave like Wilson coefficients and parametrise higher-derivative terms.\footnote{We note in passing that this EFT expansion is essentially an expansion in powers of the ratio of the massive particle's non-relativistic kinetic energy $a^{-2}\bm{\nabla}^2/(2M)$ and the Hubble energy $H$ when $M$ is large.} In general, these bi-local functions couple to pairs of local operators in the system and make up an open EFT for cosmological collider signals,
\begin{align}\label{OEFTforCCS}
     S_{\mathrm{IF}}[\varphi_r,\varphi_a] &= \int \dd\eta \dd^3 x \, a^4(\eta)\int \dd\eta' \dd^3 y \, a^4(\eta')\\
    \nonumber\sum_{m,n,l}^{\infty}\Bigg\{&\left[ c_{m,n,l}^{(\rm L)} \sin\left(\mu\ln \frac{\eta}{\eta'}+\vartheta_{n,m,l}^{\rm (L)}\right)\mathcal{O}^P_l(x,y)\left(\eta^2\partial_{\bm x}^2\right)^m \left(\eta^{\prime 2}\partial_{\bm y}^2\right)^n\right]\delta^3(\bm x-\bm y)(\eta \eta')^{3/2}\sgn (\eta-\eta')\\
    +&\left[ i  \times e^{-\pi\mu} c_{m,n,l}^{(\rm NL)} \cos\left(\mu \ln \frac{\eta \eta'}{|\bm x -\bm y|^2}+\vartheta_{n,m,l}^{\rm (NL)}\right)\mathcal{O}^K_l(x,y)\left(\eta^2\partial_{\bm x}^2\right)^m \left(\eta^{\prime 2}\partial_{\bm y}^2\right)^n\right]\dfrac{(\eta \eta')^{3/2}}{|\bm x-\bm y|^{3}}\Bigg\}\,, \nonumber 
\end{align}
where we have also absorbed numerical factors and power-law factors of $\mu$ into the Wilson coefficients and
\begin{align}
    \mathcal{O}^P_l(x,y)&=\mathcal{O}^P_l[\varphi_r(x),\varphi_a(x);\varphi_r(y),\varphi_a(y)]\,,\\
    \mathcal{O}^K_l(x,y)&=\mathcal{O}^K_l[\varphi_r(x),\varphi_a(x);\varphi_r(y),\varphi_a(y)]\,,
\end{align}
are Hermitian bi-local operators built out of retarded and advanced massless $\varphi$ fields in the EFT. Their subscript $l$ schematically labels different derivative structures and is ordered in terms of mass dimensions. For instance, matching with our toy model \eqref{eq:IFpost} yields
\begin{align}
    \mathcal{O}^P_{\{\partial_\eta^{2},\,\partial_{\eta}^{\prime 2}\}}(x,y)&=\varphi_r^{\prime2}(x) \varphi_r'(y) \varphi_a'(y) +  \frac{1}{4}\varphi_a^{\prime2}(x) \varphi_r'(y) \varphi_a'(y) \,,\\
    \mathcal{O}^K_{\{\partial_\eta^{2},\,\partial_{\eta}^{\prime 2}\}}(x,y)&=\varphi_r'(x)\varphi_a'(x) \varphi_r'(y)\varphi_a'(y)\,.
\end{align}
Due to its sophisticated mathematical structure, this open EFT for cosmological collider signals is of little practical merit. However, it provides useful insights on the effective description of light fields propagating in an environment of heavy fields with Bunch-Davies UV completion. In particular, inspecting the structure of \eqref{OEFTforCCS} shows the following rules: 

\begin{tcolorbox}[%
			enhanced, 
			breakable,
			skin first=enhanced,
			skin middle=enhanced,
			skin last=enhanced,
			before upper={\parindent15pt},
			]{}

            \vspace{0.05in}

\paragraph{Open EFT description of (leading) local signals.} $\,$

\ 

\begin{itemize}
    \item  Local signals are captured by a tower of EFT operators that are spatially local and temporally non-local. They descend from the principal-value propagators and evolve \textit{causally} inside the light cone.
    \item  Operators sourcing the local cosmological collider signals involve the time-ordering $\sgn$ function, suggesting that they are always intertwined with contact interactions in a large-mass EFT in an in-out context.
    \item  The bi-local operators $\mathcal{O}^P_l$ are Hermitian and the Wilson coefficients $c_{m,n,l}^{(\rm L)}$ are real. 
    In addition, the contact interactions associated to the leading local cosmological collider signals can be rearranged into two copies on the two branches of the path integral, $S_{\mathrm{EFT}}[\varphi_+] - S_{\mathrm{EFT}}[\varphi_-]$. It follows that we attribute them to the \textit{unitary} effects in the low-energy open EFT. This resonates with the intuition that local cosmological collider signals come from the creation-propagation-annihilation process that occur \textit{internally} inside the system, There is no information leakage into the environment as no $\sigma$ particles eventually escape.
\end{itemize}

\paragraph{Open EFT description of non-local signals.}

\begin{itemize}
    \item  Non-local signals are captured by a tower of EFT operators that are space-time non-local and thereby evolve \textit{acausally}.
    \item  Operators sourcing the non-local cosmological collider signals enjoy completely factorised dynamics along the time-ordered and anti-time-ordered contours, 
    \item  The explicit factor of $i$ in the last line of \eqref{OEFTforCCS} shows that non-local cosmological collider signals introduce \textit{non-unitary} effects in the low-energy open EFT in the form of a noise. This matches the physical intuition that such signals come from on-shell pair-production of massive particles that leak into the environment, carrying away information.
\end{itemize}
    
\end{tcolorbox}

We stress that the above descriptions are valid only at the leading order in the Boltzmann factor $e^{-\pi\mu}$. At subleading orders, terms descending from $G^\Delta_\sigma$ and $G^K_\sigma$ also yield $\mathcal{O}(e^{-2\pi\mu})$ local signals albeit with a different physical origin. Such subleading signals are produced in processes where the system shoots an on-shell massive particle into the environment while simultaneously receiving \textit{another} on-shell particle with identical quantum numbers. Consequently, these subleading local signals are described by non-unitary operators associated with dissipation. We refer the reader to \Fig{fig:OpenPerspectiveOnCCS} for a cartoon depiction of the cosmological collider signals from an open perspective.


    \subsection{Time-convolutionless influence functional}\label{subsec:TCL}

    The single-field theory derived in \Eq{eq:IFpost} contains three non-local terms. However, when working perturbatively at order $1/\Lambda^2$, one may use the free equations of motion for $\varphi$ to rewrite operators evaluated at time $\eta'$ in terms of operators evaluated at time $\eta$ — a technique known as the time-convolutionless method in open quantum systems \cite{breuerTheoryOpenQuantum2002}. Indeed, the error made by such a manipulation scales with extra powers of $1/\Lambda$ that are discarded in the perturbative treatment. 
    
    Following the exact same approach we used in the flat space treatment in \Sec{subsec:flatEFT}, we first use the solutions to the Klein-Gordon equation to evolve the field at different times. In terms of the mode function, this reads
    \begin{align}
        \varphi_{r,a}(\bmk, \eta') &= G_{11}^\varphi(k, \eta, \eta') \varphi_{r,a}(\bmk, \eta) + G_{12}^\varphi(k, \eta, \eta')  \varphi^\prime_{r,a}(\bmk, \eta)\\
        \varphi^\prime_{r,a}(\bmk, \eta') &= G_{21}^\varphi(k, \eta, \eta') \varphi_{r,a}(\bmk, \eta) + G_{22}^\varphi(k, \eta, \eta')  \varphi^\prime_{r,a}(\bmk, \eta) \label{eq:dSevolve}
    \end{align}
    with 
    \begin{align}
        G_{11}^\varphi(k, \eta, \eta') &\equiv - i a^2(\eta) \left[u^{*\prime}_\varphi(k,\eta)u_\varphi(k,\eta') - u_\varphi^\prime(k,\eta)u^*_\varphi(k,\eta') \right] \\
        G_{12}^\varphi(k, \eta, \eta') &\equiv i a^2(\eta) \left[u^{*}_\varphi(k,\eta)u_\varphi(k,\eta') - u_\varphi(k,\eta)u^*_\varphi(k,\eta') \right] \\
        G_{21}^\varphi(k, \eta, \eta') &\equiv - i a^2(\eta) \left[u^{*\prime}_\varphi(k,\eta)u^\prime_\varphi(k,\eta') - u_\varphi^\prime(k,\eta)u^{*\prime}_\varphi(k,\eta') \right] \\
        G_{22}^\varphi(k, \eta, \eta') &\equiv i a^2(\eta) \left[u^{*}_\varphi(k,\eta)u^\prime_\varphi(k,\eta') - u_\varphi(k,\eta)u^{*\prime}_\varphi(k,\eta') \right].
    \end{align}
    Injecting the massless mode function \eqref{eq:modefctvp}, we obtain  
    \begin{align}
        G_{11}^\varphi(k, \eta, \eta') &= \frac{\eta'}{\eta} \cos\left[ k (\eta - \eta')\right]  + \frac{\sin\left[ k (\eta - \eta')\right] }{k \eta} \\ 
        G_{12}^\varphi(k, \eta, \eta') &= \frac{\eta -\eta'}{k^2 \eta^2} \cos\left[ k (\eta - \eta')\right] - \frac{1 + k^2 \eta \eta'}{k^3 \eta^2}\sin\left[ k (\eta - \eta')\right]\\ 
        G_{21}^\varphi(k, \eta, \eta') &=    \frac{k \eta'}{\eta}  \sin\left[ k (\eta - \eta')\right] \\
        G_{22}^\varphi(k, \eta, \eta') &=    \frac{\eta'}{\eta}\cos\left[ k (\eta - \eta')\right] - \frac{\eta'}{k \eta^2} \sin\left[ k (\eta - \eta')\right].
    \end{align}
    
    We then ``pinch'' the two interaction vertices to the same spacetime point, such that the influence functional is manifest as an integral of a local operator. This procedure is explicitly carried out in \App{app:EFTcoeff}. The influence functional \eqref{eq:IFpost} can then be decomposed into three terms: 
    \begin{align}
    S_{\mathrm{IF}}[\varphi_r,\varphi_a] =&~ S_{\mathrm{IF}}^P[\varphi_r,\varphi_a]+S_{\mathrm{IF}}^\Delta[\varphi_r,\varphi_a]+S_{\mathrm{IF}}^K[\varphi_r,\varphi_a],
    \end{align}
    where, using the shorthand $\varphi_{i} \equiv \varphi(\eta,\bm k_i)$, the contribution from $G^P_\sigma$ is 
    \begin{align}
    \label{eq_dSIFH}
     S_{\mathrm{IF}}^P&[\varphi_r,\varphi_a] = \int_{-\infty}^0 \dd \eta \int\mathcal D\bm k\, \\
    \bigg\{&a^2(\eta) c_1(k_1,k_2,s;\eta)\left[ \varphi_{r,1}\varphi_{a,2}
    \Big( \varphi'_{r,3}\varphi'_{r,4}+\frac14\varphi'_{a,3}\varphi'_{a,4}\Big)
    +\Big(\varphi_{r,1}\varphi_{r,2}+\frac14 \varphi_{a,1}\varphi_{a,2}\Big)\varphi'_{r,3}\varphi'_{a,4} \right]\nonumber\\
    +&c_2(k_1,k_2,s;\eta)\left[ \varphi'_{r,1}\varphi'_{a,2}
    \Big( \varphi'_{r,3}\varphi'_{r,4}+\frac14\varphi'_{a,3}\varphi'_{a,4}\Big)
    +\Big(\varphi'_{r,1}\varphi'_{r,2}+\frac14 \varphi'_{a,1}\varphi'_{a,2}\Big)\varphi'_{r,3}\varphi'_{a,4} \right] \nonumber \\
    +&a(\eta)c_3(k_1,k_2,s;\eta)\bigg[ \left( \varphi_{r,1}\varphi'_{a,2} + \varphi_{a,1}\varphi'_{r,2} \right)
    \Big( \varphi'_{r,3}\varphi'_{r,4}+\frac14\varphi'_{a,3}\varphi'_{a,4}\Big) \nonumber \\
    &\qquad \qquad \quad   +2\Big(\varphi_{r,1}\varphi'_{r,2}
    +\frac14 \varphi_{a,1}\varphi'_{a,2}\Big)\varphi'_{r,3}\varphi'_{a,4} \bigg]\bigg\}, \nonumber 
    \end{align}
    the contribution from $G^\Delta_\sigma$ is:
    \begin{align}
    \label{eq_dSIFD}
    S_{\mathrm{IF}}^\Delta[\varphi_r,\varphi_a] &= - \int_{-\infty}^0 \dd \eta \int\mathcal D\bm k\, \\
    \bigg\{&a^2(\eta)c_1(k_1,k_2,s;\eta)\left[ \varphi_{r,1}\varphi_{a,2}
    \Big( \varphi'_{r,3}\varphi'_{r,4}+\frac14\varphi'_{a,3}\varphi'_{a,4}\Big)
    -\Big(\varphi_{r,1}\varphi_{r,2}+\frac14 \varphi_{a,1}\varphi_{a,2}\Big)\varphi'_{r,3}\varphi'_{a,4} \right]\nonumber\\
    +&c_2(k_1,k_2,s;\eta)\left[ \varphi'_{r,1}\varphi'_{a,2}
    \Big( \varphi'_{r,3}\varphi'_{r,4}+\frac14\varphi'_{a,3}\varphi'_{a,4}\Big)
    -\Big(\varphi'_{r,1}\varphi'_{r,2}+\frac14 \varphi'_{a,1}\varphi'_{a,2}\Big)\varphi'_{r,3}\varphi'_{a,4} \right] \nonumber \\
    +&a(\eta)c_3(k_1,k_2,s;\eta)\bigg[ \left( \varphi_{r,1}\varphi'_{a,2} + \varphi_{a,1}\varphi'_{r,2}\right)
    \Big( \varphi'_{r,3}\varphi'_{r,4}+\frac14\varphi'_{a,3}\varphi'_{a,4}\Big)
    \nonumber \\
    &\qquad \qquad \quad -2\Big(\varphi_{r,1}\varphi'_{r,2}+\frac14 \varphi_{a,1}\varphi'_{a,2}\Big)\varphi'_{r,3}\varphi'_{a,4} \bigg]\bigg\}, \notag 
    \end{align}
    and the contribution from $G^K_\sigma$ is:
    \begin{align}
    \label{eq_dSIFK}
    S_{\mathrm{IF}}^K[\varphi_r,\varphi_a]  &= ~ i\int_{-\infty}^0 \dd \eta \int\mathcal D\bm k\, \varphi'_{r,3}\varphi'_{a,4} \bigg[a^2(\eta) \tilde{c}_1(k_1,k_2,s;\eta)\varphi_{r,1}\varphi_{a,2} \\
    +&\tilde{c}_2(k_1,k_2,s;\eta)\varphi'_{r,1}\varphi'_{a,2} +a(\eta)\tilde{c}_3(k_1,k_2,s;\eta)\left(\varphi_{r,1}\varphi'_{a,2} 
    +\varphi_{a,1}\varphi'_{r,2} \right)\bigg]. \notag
    \end{align}
    Note the $i$ prefactor in the last expression, chosen such that $c_i(k_1,k_2,s; \eta)$ are manifestly real, following the non-equilibrium constraint $S_{\mathrm{IF}}[\varphi_r,\varphi_a] = S^*_{\mathrm{IF}}[\varphi_r,-\varphi_a]$ \cite{Liu:2018kfw}. 
    Here we have only considered the $s$ channel and $\int\mathcal D\bm k$ is defined as in \eqref{eq_intKmeasure},    
    while the EFT coefficients are given by:
    \begin{align}
    &c_1(k_i,k_j,s;\eta) \equiv \frac{1}{2\Lambda^2} \int_{-\infty}^{\eta} \dd \eta' a^2(\eta') G^{P}_\sigma(s;\eta,\eta') G^\varphi_{21}(k_i, \eta,\eta') G^\varphi_{21}(k_j, \eta,\eta') ,\\
    &c_2(k_i,k_j,s;\eta) \equiv \frac{a^2(\eta)}{2\Lambda^2} \int_{-\infty}^{\eta} \dd \eta' a^2(\eta') G^{P}_\sigma(s;\eta,\eta') G^\varphi_{22}(k_i, \eta,\eta') G^\varphi_{22}(k_j, \eta,\eta') ,\\
    &c_3(k_i,k_j,s;\eta) \equiv \frac{a(\eta)}{2\Lambda^2} \int_{-\infty}^{\eta} \dd \eta' a^2(\eta') G^{P}_\sigma(s;\eta,\eta') G^\varphi_{21}(k_i, \eta,\eta') G^\varphi_{22}(k_j, \eta,\eta') ,
    \end{align}    
    and 
    \begin{align}
    &\tilde{c}_1(k_i,k_j,s;\eta) \equiv \frac{1}{i\Lambda^2} \int_{-\infty}^{\eta} \dd \eta' a^2(\eta') G^{K}_\sigma(s;\eta,\eta') G^\varphi_{21}(k_i, \eta,\eta') G^\varphi_{21}(k_j, \eta,\eta') ,\\
    &\tilde{c}_2(k_i,k_j,s;\eta) \equiv \frac{a^2(\eta)}{i\Lambda^2} \int_{-\infty}^{\eta} \dd \eta' a^2(\eta') G^{K}_\sigma(s;\eta,\eta') G^\varphi_{22}(k_i, \eta,\eta') G^\varphi_{22}(k_j, \eta,\eta') ,\\
    &\tilde{c}_3(k_i,k_j,s;\eta) \equiv \frac{a(\eta)}{i\Lambda^2} \int_{-\infty}^{\eta} \dd \eta' a^2(\eta') G^{K}_\sigma(s;\eta,\eta') G^\varphi_{21}(k_i, \eta,\eta') G^\varphi_{22}(k_j, \eta,\eta') .
    \end{align}
    The explicit scale factors kept in \Eqs{eq_dSIFH}, \eqref{eq_dSIFD} and \eqref{eq_dSIFK} depend on the number of time derivatives on the operator. 

    While the generic expression of these coefficients is involved and does not bring much light to the physics, their soft limit, when $k_{34} > k_{12} \gg s$ is rather compact. In this squeezed limit, we use the soft expansion of $G^P_\sigma$ found in Eq.~\eqref{eq:dSHadamardPropIRExpansion}, and we obtain
    \begin{align}
    \lim_{s\to 0}c_1(k_1,k_2,s;\eta) =& \left[ -\frac{i k_1k_2\eta^2}{32\Lambda^2\mu} e^{i k_{12}\eta}E_{i\mu-3/2}(i k_{12}\eta) +(\mu\to-\mu) \right] \nonumber\\
    & \qquad +(k_1\to-k_1) +(k_2\to-k_2) + (k_{1,2} \to -k_{1,2})\,, \label{eq:c1dS} \\
    \lim_{s\to 0}c_2(k_1,k_2,s;\eta) =& \left[ -\frac{i (1-i k_1\eta)(1-i k_2\eta)}{32\Lambda^2H^2\mu k_1k_2\eta^2} e^{i k_{12}\eta}E_{i\mu-3/2}(i k_{12}\eta) +(\mu\to-\mu) \right] \nonumber\\
    & \qquad +(k_1\to-k_1) +(k_2\to-k_2) + (k_{1,2} \to -k_{1,2})\,, \label{eq:c2dS}\\
    \lim_{s\to 0}c_3(k_1,k_2,s;\eta) =& \left[ -\frac{i k_1(1-i k_2\eta)}{32\Lambda^2H\mu k_2} e^{i k_{12}\eta}E_{i\mu-3/2}(i k_{12}\eta) +(\mu\to-\mu) \right] \nonumber\\
    & \qquad +(k_1\to-k_1) +(k_2\to-k_2) + (k_{1,2} \to -k_{1,2})\,. \label{eq:c3dS}
    \end{align}
    These coefficients control the unitary part of the EFT. They are analytic in the exchanged momenta $s$. The exponential integral $E_{i \mu - 3/2}$ includes the Boltzmann suppression factor $e^{-\pi\mu}$ one expects to recover in the local signal. On the contrary, the stochastic part of the EFT follows from
        \begin{align}
    \lim_{s\to 0}\tilde c_1(k_1,k_2,s;\eta) =& \bigg\{\frac{k_1k_2\eta^2}{16\pi\Lambda^2}  \left[ \frac{\pi\coth \pi\mu}{\mu} - \left(-\frac{s\eta}{2}\right)^{-2i\mu}\Gamma^2(i\mu)  \right] e^{i k_{12}\eta}E_{i\mu-3/2}(i k_{12}\eta) \nonumber\\
    & \qquad+(\mu\to-\mu) \bigg\}  +(k_1\to-k_1) +(k_2\to-k_2) + (k_{1,2} \to -k_{1,2})\,, \label{eq:c1tdS}\\
    \lim_{s\to 0}\tilde c_2(k_1,k_2,s;\eta) =& \bigg\{\frac{(1-i k_1\eta)(1-i k_2\eta)}{16\pi\Lambda^2H^2k_1k_2\eta^2}  \left[ \frac{\pi\coth \pi\mu}{\mu} -\left(-\frac{s\eta}{2}\right)^{-2i\mu}\Gamma^2(i\mu) \right] e^{i k_{12}\eta}E_{i\mu-3/2}(i k_{12}\eta) \nonumber\\
    & \qquad +(\mu\to-\mu) \bigg\}  +(k_1\to-k_1) +(k_2\to-k_2) + (k_{1,2} \to -k_{1,2})\,, \label{eq:c2tdS}\\
    \lim_{s\to 0}\tilde c_3(k_1,k_2,s;\eta) =& \bigg\{\frac{k_1(1-i k_2\eta)}{16\pi\Lambda^2Hk_2} \left[ \frac{\pi\coth \pi\mu}{\mu} -\left(-\frac{s\eta}{2}\right)^{-2i\mu}\Gamma^2(i\mu) \right] e^{i k_{12}\eta}E_{i\mu-3/2}(i k_{12}\eta) \nonumber\\
    & \qquad +(\mu\to-\mu) \bigg\} +(k_1\to-k_1) +(k_2\to-k_2) + (k_{1,2} \to -k_{1,2})\,. \label{eq:c3tdS}
    \end{align}
    One finds a similar structure, on top of which the factor $(-s\eta/2)^{-2i\mu}$ emerges. The associated branch cut yields the non-analyticity in the exchanged momenta that one expects from the non-local signal. The rest of the coefficients proportional to $\coth(\pi \mu)/\mu$ that are analytic in $s$ leads to a large spurious local signal cancelled by the contribution from $G^\Delta_\sigma$.
    
\paragraph{Wilson coefficients in large mass.}
It is natural to consider the asymptotic behaviour of these Wilson coefficients in the large mass limit $m\to\infty$, where the curvature of the spacetime background is negligible, and one expects the (open) EFT and the associated effective operators to recover the flat space result.

Let us start from $c_i$ in the soft limit given in Eqs.~\eqref{eq:c1dS}, \eqref{eq:c2dS} and \eqref{eq:c3dS}. Notice that the $c_i$ depend on the mass through $\mu \to m/H \to\infty$, and the only complicated dependence is encoded in the exponential integral, say $E_{i\mu-3/2}(i k_{12}\eta)$. To handle this, we rewrite the exponential integral into a series
\begin{align}
E_{i\mu-3/2}(i k_{12}\eta) = (i k_{12}\eta)^{-5/2+i\mu}\Gamma\left(\frac52- i\mu\right) + \sum_{n=0}^\infty \frac{-(-i k_{12}\eta)^n}{(\frac52- i\mu +n )n!}.
\end{align}
As we will see, the first part is associated with signals that will be ultimately suppressed exponentially by the heavy mass. On the other hand, the second part is associated with the analytic background, and we can easily take the large mass limit:
\begin{align}
\lim_{\mu\to\infty} \sum_{n=0}^\infty \frac{-(-i k_{12}\eta)^n}{(\frac52- i\mu +n )n!} = \sum_{n=0}^\infty \frac{-(-i k_{12}\eta)^n}{(- i\mu)n!} = \frac{-i}{\mu} e^{-ik_{12}\eta}.
\end{align}
Now if we replace the exponential function by its analytic part in the large mass limit, we find $c_1$ and $c_3$ both vanish due to the permutation $k_1\leftrightarrow -k_1$, while $c_2$ becomes finite and at the order of $1/m^2$:
\begin{align}
\lim_{\mu\to\infty} \lim_{s\to 0}c_2 \supset& \left[ -\frac{i (1-i k_1\eta)(1-i k_2\eta)}{32\Lambda^2H^2\mu k_1k_2\eta^2} \times \frac{-i}{\mu}+(\mu\to-\mu) \right] \nonumber\\
    & \qquad +(k_1\to-k_1) +(k_2\to-k_2) + (k_{1,2} \to -k_{1,2})\nonumber\\
=&~\frac{1}{4\Lambda^2m^2}.
\end{align}
Similarly, all of $\tilde c_i$ vanish due to $\mu\to-\mu$. Therefore, if we go back to the influence functional, we find the only term that survives will be:
\begin{align}
\lim_{m\to\infty}S_{\mathrm{IF}}[\varphi_r,\varphi_a] \supset & \int_{-\infty}^0 \dd \eta \int\mathcal D\bm k\,
    c_2 \left[ \varphi'_{r,1}\varphi'_{a,2}
    \Big( \varphi'_{r,3}\varphi'_{r,4}+\frac14\varphi'_{a,3}\varphi'_{a,4}\Big)
    +\Big(\varphi'_{r,1}\varphi'_{r,2}+\frac14 \varphi'_{a,1}\varphi'_{a,2}\Big)\varphi'_{r,3}\varphi'_{a,4} \right] \nonumber \\
    =&~\frac{1}{2\Lambda^2m^2}\int_{-\infty}^0 \dd \eta \int\mathcal D\bm k\,
     \varphi'_{r,1}\varphi'_{a,2}
    \Big( \varphi'_{r,3}\varphi'_{r,4}+\frac14\varphi'_{a,3}\varphi'_{a,4}\Big)\nonumber\\
    =&~ \frac{1}{8\Lambda^2m^2} \int d^4x\, \left[ \varphi^{\prime 4}_+(x) - \varphi^{\prime 4}_-(x) \right]\,.
    \end{align}
which implies that this part can be traced back to a unitary theory,
\begin{align}
S_{\mathrm{EFT}}[\varphi] = \frac{1}{8\Lambda^2m^2} \int d^4x\, \varphi^{\prime 4}(x)\,
\end{align}
which meets our intuition. As expected, the heavy mass expansion softens the non-analyticity of the EFT coefficients and the effective theory becomes local, both in space and time. If one wants to extend to higher orders in $1/M$, one also needs to include higher-order corrections in the soft expansion. 

    
 \subsection{Recovering the cosmological collider signal}\label{subsec:recover}

We conclude this analysis by demonstrating how the leading CC signals arise from the open EFT derived above.

\subsubsection{Local signal}

Let us first focus on the local signal generated from the principal-value propagator $G^P_\sigma$. The full computation is rather complicated, so for simplicity, we assume the configuration to be squeezed, $k_{34} > k_{12} \gg s$.
We notice that the local signal typically has the standard form $(k_{12}/k_{34})^{\pm i\omega} \times (s^\#k_{12}^\#/k_{34}^\#)$ as $k_{34}\to\infty$, with $\omega$ being the oscillation frequency. Furthermore, although it is analytic within physical configuration, the local signal is associated with a branch cut on the $k_{34}$-complex plane starting from $k_{34}\to-\infty$ due to the imaginary exponent $\pm i\omega$. Therefore, we can extract the local signal by identifying only the singular part of the correlator as $k_{34}\to-\infty$.

From $S_{\mathrm{IF}}^P$ derived in \Eq{eq_dSIFH}, we compute the corresponding correlator following the standard Feynman rules. In particular, the contribution of $G^P_\sigma$ associated with $c_1$ found in \Eq{eq:c1dS} is
\begin{align}
\mathcal I^P \supset i  \int_{-\infty}^0 \dd \eta\,a^2(\eta)c_1(k_1,k_2,s;\eta) \bigg[& G^K_1G^R_2\Big({G^K_3}'{G^K_4}'+\frac14 {G^R_3}'{G^R_4}' \Big)\nonumber \\
&+ \Big( G^K_1G^K_2 + \frac14 G^R_1G^R_2 \Big) {G^K_3}'{G^R_4}'\bigg]\,,\label{eq_IPc1}
\end{align}
where we have used shorthand $G^{R/K}_i \equiv G_\varphi^{R/K}(k_i;0,\eta)$ for brevity. One should also include permutations $k_1\leftrightarrow k_2$, $k_3\leftrightarrow k_4$ and $(k_1,k_2)\leftrightarrow(k_3,k_4)$, but let us first focus on the contribution explicitly shown in Eq.~\eqref{eq_IPc1}.
The key observation is that the bracketed term in the integrand of Eq.~\eqref{eq_IPc1} — namely, the combination of $G_i^{R/K}$ — consists of polynomial and trigonometric functions. Consequently, it can be recast in the form $\sum \eta^{\#} e^{iK\eta}/k_i^{\#}$, where $K$ corresponds either to the total momentum $\pm k_T$ or to the folded configurations $\pm (k_T - 2k_i)$, but never to $\pm (k_T - 2k_{ij})$. The resulting time integral therefore takes the form
\begin{align}
\int_{-\infty}^0 \dd \eta (-\eta)^{p-1} e^{i(K+K_{12})\eta}E_\nu(i K_{12}\eta) = \frac{\Gamma(p)(-iK_{12})^{-p}}{p+\nu-1} {}_2\mathrm F_1\left[
\begin{matrix} p, p+\nu-1 \\ p+\nu
\end{matrix}\middle| 1+\frac{K-i\epsilon}{K_{12}}
\right]\,,\label{eq_typicalint}
\end{align}
with $K_{12}$ being either $\pm k_{12}$ or $\pm \bar k_{12}$, and $p\in \mathbb Z$, $\nu\in\mathbb C$. Also notice that we have added the $i\epsilon$-prescription to make the integral convergent.\footnote{Since the integrand behaves as $e^{i K\eta}$ as $\eta\to -\infty$, we choose the $i\epsilon$-prescription to be $K \to K- i \epsilon $ to make it exponentially damped.}
Now we aim to isolate the local signal, namely nonanalyticity in $k_{34}\to -\infty$. Since the only nonanalyticity of the integral \eqref{eq_typicalint} comes from the branch cut of the hypergeometric function, when its argument $z$ lies in $[1,+\infty)$, the only possibility to generate a local signal is when
$K_{12} = \pm k_{12}$ and $K=\mp k_T$,
so that the argument for the hypergeometric becomes
\begin{align}
z\equiv 1+\frac{\mp k_T - i \epsilon |k_T|}{\pm k_{12}} = - \frac{k_{34}}{k_{12}} \mp \frac{i\epsilon}{k_{12}}\,,
\end{align}
which hits the branch point $z\to+\infty$ as $k_{34}\to-\infty$.
On the other hand, for any physical configuration $0<k_{12}<k_{34}$, we always have $z<0$, away from the branch cut of the hypergeometric function, so that the prescription can be neglected. In particular, we have the following expansion around $k_{34}\to+\infty$:
\begin{align}\label{eq:2F1gamma}
\lim_{k_{34}\to+\infty } {}_2\mathrm F_1\left[
\begin{matrix} p, p+\nu-1 \\ p+\nu
\end{matrix}\middle| -\frac{k_{34}}{k_{12}}
\right] = \frac{\Gamma(p+\nu)\Gamma(1-\nu)}{\Gamma(p)} \left(\frac{k_{12}}{k_{34}}\right)^{1-p-\nu} + \mathcal O(k_{34}^{-p})\,.
\end{align}
The first term in the right-hand side of \Eq{eq:2F1gamma} is exactly of the form of the local signal $(k_{12}/k_{34})^{\pm i\omega}$, while the remaining terms $\mathcal O(k_{34}^{-p})$ are purely analytic as $p$ is an integer. Therefore, to extract the local signal, we only keep the first term above in the time integral of Eq.~\eqref{eq_typicalint}, leading to
\begin{align}
\int_{-\infty}^0 \dd \eta (-\eta)^{p-1} e^{\mp i k_{34}\eta}E_\nu(\pm i k_{12}\eta) \supset  \Gamma(p+\nu-1)\Gamma(1-\nu)(\mp ik_{12})^{-p}\left(\frac{k_{34}}{k_{12}}\right)^{1-p-\nu}\,.
\label{eq_singpart}
\end{align}

Importantly, this reasoning holds even prior to carrying out the integral explicitly. From the perspective of Landau analysis, this singular term for the integral \eqref{eq_typicalint} must originate from either pinched poles in the integrand, or singularities of the integrand itself at the endpoints $\eta=-\infty,\,0$. One can easily check from
\begin{align}
E_\nu(\pm i k_{12}\eta) = ( \pm i k_{12}\eta)^{\nu-1}\Gamma(1-\nu) + \sum_{n=0}^\infty \frac{(-1)^n (\pm i k_{12}\eta)^n}{n!(\nu-n)}\,,
\end{align}
that the only possibility comes from the exponential integral $E_\nu(\pm i k_{12}\eta)$ around $\eta=0$ (as long as the kinematic configuration is away from folded/total energy singularities). 
It follows that, in practice, one may replace $E_\nu(\pm i k_{12}\eta)$ with its singular part $(\pm i k_{12}\eta)^{\nu-1}\Gamma(1-\nu)$ in the coefficient $c_1$ appearing in \Eq{eq:c1dS}, combine this with the term proportional to $e^{\mp i k_T \eta}$ in the product of propagators $G_i^{R/K}$, and then perform the integral to recover the local signal. As a consistency check, this procedure correctly reproduces the non-analytic contribution in the integral \eqref{eq_typicalint}:
\begin{align}
\int_{-\infty}^0 \dd\eta(-\eta)^{p-1}e^{\mp i k_{34} \eta} (\pm i k_{12}\eta)^{\nu-1}\Gamma(1-\nu) = \Gamma(p+\nu-1)\Gamma(1-\nu) ( \mp i k_{12})^{-p}\left(\frac{k_{12}}{k_{34}}\right)^{1-p-\nu}\,,
\end{align}
which matches exactly the result found in Eq.~\eqref{eq_singpart}.

Following this procedure, one can extract the local signal hidden in Eq.~\eqref{eq_IPc1} by retaining only the relevant part of the Wilson coefficient $c_1$. Explicitly,
\begin{align}
\mathcal I^P \supset &~ i  \int_{-\infty}^0 \dd\eta\,a^2(\eta) \times \Big( -\frac{ik_1k_2\eta^2}{32\Lambda^2\mu} \Big) \sum_{\pm} \Big[ e^{\pm i k_{12}\eta}E_{i\mu-3/2}(\pm ik_{12}\eta) \nonumber\\
& \qquad \times \frac{\mp H^8(1\pm ik_1\eta)(1\pm ik_2\eta)}{32k_1^3k_2^3k_3k_4}\eta^2 e^{\mp ik_T\eta} \Big]  + (\mu\to-\mu) \nonumber\\
\supset &~ \frac{H^6}{32\Lambda^2\mu k_1^2k_2^2k_3k_4} \sum_\pm \mp\int_{-\infty}^0 \dd\eta\,(1\pm ik_1\eta)(1\pm ik_2\eta)\eta^2 (\pm ik_{12}\eta)^{i\mu-5/2}e^{\mp i k_{34}\eta} \Gamma\Big(\frac52-i\mu\Big)\nonumber\\
& \qquad  + (\mu\to-\mu)\,, \bigg.
\end{align}
which leads to 
\begin{align}
\mathcal I^P \supset &~\frac{iH^6\Gamma(\frac52-i\mu)\Gamma(\frac52+i\mu)}{512\Lambda^2\mu k_1k_2k_3k_4 k_{12}^{5/2}k_{34}^{5/2}}\left(\frac{k_{12}}{k_{34}}\right)^{i\mu} \left[1+\frac{2k_{12}k_{34}}{(3+2i\mu)k_1k_2} + \frac{4k_{34}^2}{(1+2i\mu)(3+2i\mu)k_1k_2}\right] \,.
\end{align}
To obtain the full contribution from $c_1$, one should also include all the permutations. However,  notice that under the permutation $(k_1,k_2)\leftrightarrow(k_3,k_4)$, the argument for the hypergeometric function in the integral \eqref{eq_typicalint} becomes $1+K/K_{34}$, which cannot uniformly hit the branch points ($1$ or $\infty$) when $k_{34}\to\infty$, and thus will not contribute to the local signal. Therefore, we only need to include the permutations $k_1\leftrightarrow k_2$ and $k_3\leftrightarrow k_4$, giving a factor of $4$:
\begin{align}
\mathcal I^P_{L,1} &= \frac{iH^6\Gamma(\frac52-i\mu)\Gamma(\frac52+i\mu)}{128\Lambda^2\mu k_1k_2k_3k_4 k_{12}^{5/2}k_{34}^{5/2}}\left(\frac{k_{12}}{k_{34}}\right)^{i\mu} \left[1+\frac{2k_{12}k_{34}}{(3+2i\mu)k_1k_2} + \frac{4k_{34}^2}{(1+2i\mu)(3+2i\mu)k_1k_2}\right]\nonumber\\
& \qquad  + (\mu\to-\mu)\,. \bigg.
\end{align}
Similarly, we can calculate the local signal contributed from $c_2$ and $c_3$ given in \Eqs{eq:c2dS} and \eqref{eq:c3dS} respectively, leading to 
\begin{align}
\mathcal I^P_{L,2} &= \frac{iH^6\Gamma(\frac52-i\mu)\Gamma(\frac52+i\mu)}{128\Lambda^2\mu k_1k_2k_3k_4 k_{12}^{5/2}k_{34}^{5/2}}\left(\frac{k_{12}}{k_{34}}\right)^{i\mu} \left[1-\frac{2k_{12}k_{34}}{(3+2i\mu)k_1k_2} + \frac{4k_{34}^2}{(1+2i\mu)(3+2i\mu)k_1k_2}\right]\nonumber\\
& \qquad  + (\mu\to-\mu)\,, \bigg.
\end{align}
and 
\begin{align}
&\mathcal I^P_{L,3} = \frac{iH^6\Gamma(\frac52-i\mu)\Gamma(\frac52+i\mu)}{64\Lambda^2\mu k_1k_2k_3k_4 k_{12}^{5/2}k_{34}^{5/2}}\left(\frac{k_{12}}{k_{34}}\right)^{i\mu} \left[1- \frac{4k_{34}^2}{(1+2i\mu)(3+2i\mu)k_1k_2}\right] + (\mu\to-\mu)\,.
\end{align}
Summing up all three contributions, we recover the local signal 
\begin{align}
\mathcal I^P_{L} = \frac{iH^6\Gamma(\frac52-i\mu)\Gamma(\frac52+i\mu)}{32\Lambda^2\mu k_1k_2k_3k_4 k_{12}^{5/2}k_{34}^{5/2}}\left(\frac{k_{12}}{k_{34}}\right)^{i\mu}  + (\mu\to-\mu)\,.
\end{align}
At last, notice that we assumed $k_{34}>k_{12}$. For local signal in the region where $k_{12}>k_{34}$, one just flips the momenta by $k_{12}\leftrightarrow k_{34}$. Therefore, we finally obtain
\begin{align}
\mathcal I^P_{L} = \left[\frac{iH^6\Gamma(\frac52-i\mu)\Gamma(\frac52+i\mu)}{32\Lambda^2\mu k_1k_2k_3k_4 k_{12}^{5/2}k_{34}^{5/2}}\left(\frac{k_{12}}{k_{34}}\right)^{i\mu}  + (\mu\to-\mu)\right]\theta(k_{34}-k_{12}) + (k_{12}\leftrightarrow k_{34})\,,
\end{align}
which perfectly matches the top-down calculation of the signal part from $\mathcal I^P$ in \Eq{eq_topdownIPsignal} in the squeezed limit $s\to 0$.


\subsubsection{Non-local signal}

Next, we turn to the non-local signal, which arises solely from the contribution of $G^K_\sigma$ due to its non-locality in the exchanged momentum $s$. We again focus on the squeezed limit where $s\ll k_{12},k_{34}$ for simplicity, and the logic closely parallels the analysis of the local signal. The main difference is that, we can simply replace the Wilson coefficients \eqref{eq:c1tdS}-\eqref{eq:c3tdS} by their non-analytic part in $s$ to select the non-local signal, that are:
\begin{align}
    \lim_{s\to 0}\tilde c_1(k_1,k_2,s;\eta) \supset & \left[-\frac{k_1k_2\eta^2}{16\pi\Lambda^2} \left(-\frac{s\eta}{2}\right)^{-2i\mu}\Gamma^2(i\mu)  e^{i k_{12}\eta}E_{i\mu-3/2}(i k_{12}\eta) +(\mu\to-\mu) \right] \nonumber\\
    & \qquad +(k_1\to-k_1) +(k_2\to-k_2) + (k_{1,2} \to -k_{1,2})\,,\\
    \lim_{s\to 0}\tilde c_2(k_1,k_2,s;\eta) \supset & \left[-\frac{(1-i k_1\eta)(1-i k_2\eta)}{16\pi\Lambda^2H^2k_1k_2\eta^2} \left(-\frac{s\eta}{2}\right)^{-2i\mu}\Gamma^2(i\mu)  e^{i k_{12}\eta}E_{i\mu-3/2}(i k_{12}\eta) +(\mu\to-\mu) \right] \nonumber\\
    & \qquad +(k_1\to-k_1) +(k_2\to-k_2) + (k_{1,2} \to -k_{1,2})\,,\\
    \lim_{s\to 0}\tilde c_3(k_1,k_2,s;\eta) \supset & \left[-\frac{k_1(1-i k_2\eta)}{16\pi\Lambda^2Hk_2} \left(-\frac{s\eta}{2}\right)^{-2i\mu}\Gamma^2(i\mu)  e^{i k_{12}\eta}E_{i\mu-3/2}(i k_{12}\eta) +(\mu\to-\mu) \right] \nonumber\\
    & \qquad +(k_1\to-k_1) +(k_2\to-k_2) + (k_{1,2} \to -k_{1,2})\,.
    \end{align}
After the time integral and all permutations, we obtain
\begin{align}
\mathcal I^K \supset &~ \bigg[ \frac{i H^6}{32\pi \Lambda^2 k_1k_2k_3k_4 k_{12}^5}\frac{\Gamma(5+2i\mu)\Gamma^2(-i\mu)}{5+2i\mu}\left(\frac{s}{2k_{12}}\right)^{i\mu} \nonumber\\
&\times \left( e^{\pi\mu}{}_2\mathrm F_1\left[\begin{matrix}
\frac52+i\mu, 5+2i\mu \\  \frac72+i\mu
\end{matrix}\middle| -\frac{k_{34}-i\epsilon}{k_{12}}\right] - e^{-\pi\mu}{}_2\mathrm F_1\left[\begin{matrix}
\frac52+i\mu, 5+2i\mu \\  \frac72+i\mu
\end{matrix}\middle| -\frac{k_{34}+i\epsilon}{k_{12}}\right] \right. \nonumber\\
&~~~~~+\left. 
e^{-\pi\mu}{}_2\mathrm F_1\left[\begin{matrix}
\frac52+i\mu, 5+2i\mu \\  \frac72+i\mu
\end{matrix}\middle| \frac{k_{34}-i\epsilon}{k_{12}}\right] - e^{\pi\mu}{}_2\mathrm F_1\left[\begin{matrix}
\frac52+i\mu, 5+2i\mu \\  \frac72+i\mu
\end{matrix}\middle| \frac{k_{34}+i\epsilon}{k_{12}}\right]
\right) \nonumber\\
&+ (k_{1,2}\leftrightarrow k_{3,4}) \bigg] + (\mu\to-\mu)\,.
\label{eq_IK_expr1}
\end{align}
This result can be further simplified using the identity \cite{ItNISTDigital},
\begin{align}
{}_2\mathrm F_1\left[\begin{matrix} a,b\\c \end{matrix}\middle| z\right] =  \frac{(-z)^{-a}\Gamma(b-a)\Gamma(c)}{\Gamma(b)\Gamma(c-a)}{}_2\mathrm F_1\left[\begin{matrix} a,a-c+1\\a-b+1 \end{matrix}\middle| \frac1z \right] +  \frac{(-z)^{-b}\Gamma(a-b)\Gamma(c)}{\Gamma(a)\Gamma(c-b)}{}_2\mathrm F_1\left[\begin{matrix} b,b-c+1\\b-a+1 \end{matrix}\middle| \frac1z \right]\,.
\label{eq_2F1_id}
\end{align}
We can apply this equality to all hypergeometric functions with arguments $(k_{12} \pm i\epsilon) /k_{34}$ and $-(k_{12} \pm i\epsilon) /k_{34}$, for instance\footnote{Notice that we have used ${}_2\mathrm F_1[0,b,c;z]=1$, and have kept track of all $i\epsilon$-prescriptions which are essential to obtain the correct result.}
\begin{align}
    {}_2\mathrm F_1\left[\begin{matrix}
\frac52+i\mu, 5+2i\mu \\  \frac72+i\mu
\end{matrix}\middle| \frac{k_{12}+i\epsilon}{k_{34}}\right] = &~\frac{\Gamma(\frac52+i\mu)\Gamma(\frac72+i\mu)}{\Gamma(5+2i\mu)}\left(-\frac{k_{12}+i\epsilon}{k_{34}}\right)^{-5/2-i\mu} \nonumber\\
&- \left(-\frac{k_{12}+i\epsilon}{k_{34}}\right)^{-5-2i\mu} {}_2\mathrm F_1\left[\begin{matrix}
\frac52+i\mu, 5+2i\mu \\  \frac72+i\mu
\end{matrix}\middle| \frac{k_{34}}{k_{12}+i\epsilon}\right]\,.
\end{align}
Crucially, all the hypergeometric functions cancel in Eq.~\eqref{eq_IK_expr1}, simplifying the final result to
\begin{align}
\mathcal I^K_{\mathrm{NL}} = \frac{H^6(1+i\sinh\pi\mu)\Gamma^2(\frac52+i\mu)\Gamma^2(-i\mu)}{32\pi\Lambda^2k_1k_2k_3k_4k_{12}^{5/2}k_{34}^{5/2}}\left(\frac{s^2}{4k_{12}k_{34}}\right)^{i\mu} + (\mu\to-\mu)\,,
\end{align}
which again matches the non-local signal part of top-down calculation \eqref{eq_topdownIK} in the squeezed limit $s\to 0$. This completes the recovery of the leading CC signals at order $\Lambda^{-2} e^{-\pi\mu}$ from the open EFT derived in \Sec{subsec:TCL}.


\subsection{Connection with entropy measures}\label{subsec:entropy}

    Whenever a system interacts with an environment, the openness of the dynamics entails the exchange of information between each side of the bipartition. This exchange of information can be monitored through the computation of entropy measures such as the purity or the entanglement entropy. These quantities are informative if one desires to assess the validity of the unitary approximation when deriving a single-field EFT — in other words, how good it is to approximate the system as being closed. 

    We follow the approach of \cite{Colas:2024ysu} to provide an expression for the leading order correction to the linear entropy of a mode, a quantity which characterises the amount of information shared between the two fields. Formally, the reduced state of a system $\varphi$ is obtained by partially tracing the full density matrix $\widehat{\rho}$ over the Hilbert space of the environment $\sigma$, 
    \begin{equation} \label{red_rho_def}
    \widehat{\rho}_{\varphi}(\eta_0) \equiv \mathrm{Tr}_{\sigma} \left[\widehat{\rho}(\eta_0)\right]. 
    \end{equation}
    The reduced density matrix $\widehat{\rho}_{\varphi}(\eta_0)$ is the object we commonly use when we compute correlators of the $\varphi$ field at time $\eta_0$, 
    \begin{align}
        \langle \prod_{i = 1}^n \widehat{\varphi}(\bmk_i, \eta_0) \rangle =\mathrm{Tr}_{\varphi} \left[\prod_{i = 1}^n \widehat{\varphi}(\bmk_i, \eta_0) \widehat{\rho}_{\varphi}(\eta_0) \right]  = \int \dd \varphi   \prod_{i = 1}^n \varphi(\bmk_i, \eta_0) \langle \varphi| \widehat{\rho}_{\varphi}(\eta_0) | \varphi \rangle,
    \end{align}
    even though it often only appears implicitly in the in-in formalism \cite{Chen:2017ryl}. When the state is pure, $\widehat{\rho}_{\varphi}(\eta_0) = | \Psi_\varphi(\eta_0) \rangle\langle \Psi_\varphi(\eta_0) |$ and we recover the standard wavefunction formalism \cite{Salcedo:2022aal}
    \begin{align}
        \langle \prod_{i = 1}^n \widehat{\varphi}(\bmk_i, \eta_0) \rangle = \int \dd \varphi   \prod_{i = 1}^n \varphi(\bmk_i, \eta_0) |\Psi_\varphi(\eta_0) |^2.
    \end{align}
    On the contrary, when $\varphi$ interacts with $\sigma$, the two field entangle and the reduced state $\widehat{\rho}_{\varphi}(\eta_0)$ becomes mixed. The amount by which $\widehat{\rho}_{\varphi}(\eta_0)$ fails to be pure is controlled by entropy measures such as the linear entropy, defined through
    \begin{align}\label{eq:Renyi2}
	S_2\left[\widehat{\rho}_{\varphi}(\eta_0)\right]  \equiv -  \ln\left\{\mathrm{Tr}_{\varphi} \left[\widehat{\rho}_{\varphi}^2(\eta_0)\right] \right\} = \delta(\bm{0})\int_{\bms\in\mathbb{R}^{3+}}\dd^3\bms \; S_2(s,\eta_0)\,.
    \end{align}
    In the right-hand side, we have expressed the linear entropy as an integral over Fourier space, and identified the integrand as a spectral density $S_2(s, \eta_0)$, characterising the amount of information shared at each scale. 
    In \cite{Colas:2024ysu}, we worked out the formal perturbative expansion of this quantity and expressed it in terms of the free propagators of the theory. For the interaction considered in our toy model \eqref{eq:S_dSL}, at second order in $1/\Lambda$, $S_2(s,\eta_0)$ is given by \cite{Colas:2024ysu}
    \begin{align}\label{eq:Slinexp}
        S_2(s,\eta_0) = \frac{4}{(2\pi)^6} \frac{1} {\Lambda^2} \int^{\eta_0}_{-\infty} \dd \eta a^2(\eta) \int^{\eta_0}_{-\infty} \dd \eta' a^2(\eta') \mathcal{K}_{\varphi}(s,\eta,\eta') \mathcal{K}_{\sigma}(s,\eta,\eta'),
    \end{align}
    where $\mathcal{K}_{\varphi/\sigma}(s,\eta,\eta')$ are the system and environment \textit{memory kernels}
    \begin{align}
        \mathcal{K}_{\varphi}(s,\eta,\eta')  &\equiv  \langle \widehat{ \varphi}^{\prime2}(\bms, \eta)  \widehat{ \varphi}^{\prime2}(-\bms, \eta') \rangle \label{eq:sysmem}  \\
        \mathcal{K}_{\sigma}(s,\eta,\eta')  &\equiv \langle \widehat{\sigma}(\bms, \eta) \widehat{\sigma}(-\bms, \eta') \rangle.
    \end{align}
    These expectation values are computed within the free theory, \Eq{eq:Slinexp} already being second order in $1/\Lambda$. The environment contribution is easy to compute, leading to 
    \begin{align}
        \mathcal{K}_{\sigma}(s,\eta,\eta')  =  D^\sigma_{-+}(s,\eta,\eta') = - i G^K_\sigma(s,\eta,\eta') - \frac{i}{2} G^\Delta_\sigma(s,\eta,\eta'),
    \end{align}
    which confirms the expectation that the principal-value propagator $G^P_\sigma$, which controls the unitary part of the single-field EFT, does not appear in the expression of the linear entropy — the unitary evolution preserving entropy measures. 

\begin{figure}[tbp]
    \centering
    \includegraphics[width=0.7\textwidth]{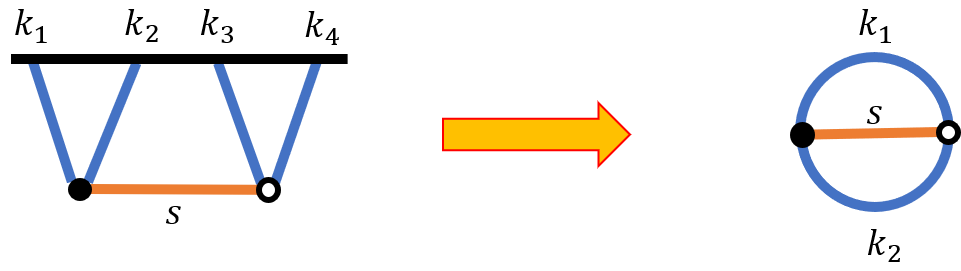}
    \caption{In-in diagrammatics of the $-+$ diagram $\mathcal{I}_{-+}$ (\textit{Left}) and its incidence on the computation of linear purity $S_2$ (\textit{Right}). One obtain the latter from the former by closing the momenta loops.}
        \label{fig:purdiag}
    \end{figure}

    On the other hand, the system contribution is more involved. $\mathcal{K}_{\varphi}(s,\eta,\eta')$ is obtained from the mean-removed four-point function
    \begin{align}
        &\mathcal{K}_{\varphi}(s,\eta,\eta') = \int \frac{\dd^3 \bmk_1}{(2\pi)^3} \int \frac{\dd^3 \bmk_2}{(2\pi)^3} \int \frac{\dd^3 \bmk_3}{(2\pi)^3} \int \frac{\dd^3 \bmk_4}{(2\pi)^3} \delta(\bmk_1 + \bmk_2 - \bm{s}) \delta(\bmk_3 + \bmk_4 + \bm{s}) \nonumber \\
        &\bigg[ \langle \widetilde{p}_\varphi(\bmk_1, \eta)  \widetilde{p}_\varphi(\bmk_2, \eta)  \widetilde{p}_\varphi(\bmk_3, \eta')  \widetilde{p}_\varphi(\bmk_4, \eta') \rangle - \langle \widetilde{p}_\varphi(\bmk_1, \eta)  \widetilde{p}_\varphi(\bmk_2, \eta)  \rangle \langle \widetilde{p}_\varphi(\bmk_3, \eta')  \widetilde{p}_\varphi(\bmk_4, \eta') \rangle\bigg]
    \end{align}
    where $\widetilde{p}_\varphi$ is the conjugate momentum of $\varphi$ in the interaction picture. Performing the Wick contractions and using the mode functions 
    \begin{align}
        p_\varphi (k, \eta) = \partial_\eta K^\varphi_-(k,\eta)\qquad \mathrm{and} \qquad p^*_\varphi (k, \eta) = \partial_\eta K^\varphi_+(k,\eta), 
    \end{align}
    we obtain 
    \begin{align}
        &\mathcal{K}_{\varphi}(s,\eta,\eta') = \frac{2}{(2\pi)^6}\int \frac{\dd^3 k_1}{(2\pi)^3} \int \frac{\dd^3 k_2}{(2\pi)^3} \delta(\bmk_1 + \bmk_2 - \bm{s}) \nonumber \\
        & \quad \left[\partial_\eta K^\varphi_-(k_1,\eta) \partial_\eta K^\varphi_-(k_2,\eta) \partial_{\eta'} K^\varphi_+(k_1,\eta') \partial_{\eta'} K^\varphi_+(k_2,\eta') \right].
    \end{align}
    Combining the above results, we finally obtain the linear entropy expression from
    \begin{align}
        &S_2(s,\eta_0)  = \frac{8}{(2\pi)^{12}} \frac{1} {\Lambda^2} \int \frac{\dd^3 k_1}{(2\pi)^3} \int \frac{\dd^3 k_2}{(2\pi)^3} \bigg[\int^{\eta_0}_{-\infty} \dd \eta a^2(\eta)  \partial_\eta K^\varphi_-(k_1,\eta) \partial_\eta K^\varphi_-(k_2,\eta) \nonumber \\
        &\quad \times \int^{\eta_0}_{-\infty} \dd \eta' a^2(\eta')  \partial_{\eta'} K^\varphi_+(k_1,\eta') \partial_{\eta'} K^\varphi_+(k_2,\eta')  
        \times D^\sigma_{-+}(s,\eta,\eta') \bigg] \delta(\bmk_1 + \bmk_2 - \bm{s}) .
    \end{align}
    Interestingly, this expression is very reminiscent of 
    the $-+$ diagram of the in-in diagrammatics \cite{Chen:2017ryl}, see \Eq{eq_dS4ptinin}. It clearly appears that one can relate the two expression by closing the external legs of 
    the $-+$ diagram and performing the loop over momenta, that is
    \begin{align}
        &S_2(s,\eta_0)  = \frac{2}{(2\pi)^{12}} \int \frac{\dd^3 k_1}{(2\pi)^3} \int \frac{\dd^3 k_2}{(2\pi)^3} \mathcal{I}_{-+}(k_1,k_2; k_1,k_2) \delta(\bmk_1 + \bmk_2 - \bm{s}),
    \end{align}
    which we represented diagrammatically in \Fig{fig:purdiag}. 
    
    A few remarks are in order. First, we recover the well-known observation that the leading variation of linear entropy (or equivalently, linear entropy) is associated with the $-+$ diagram, commonly referred to as the Cut-In-the-Middle (CIM) diagram \cite{Senatore:2009cf}, in agreement with \cite{Colas:2024ysu, Colas:2024xjy, Burgess:2024eng, Burgess:2024heo, Colas:2022hlq, Colas:2022kfu, DuasoPueyo:2024rsa, Sano:2025ird}. Furthermore, \cite{Burgess:2024heo} highlights that this contribution ultimately governs changes in the occupation number of the system’s state. This aligns with the intuition that variations in entropy measures are connected to on-shell particle production.
    Second, unless the momentum loop integral amplifies the signal by altering the Boltzmann factor, the $-+$ diagram remains proportional to $\exp(-2 \pi \mu)$, see \Eq{eq:suppr}. Consequently, this contribution is subleading compared to the $++$ diagram, which encapsulates both the dominant local and non-local CC signals. As a result, variations of the linear entropy represent a subleading effect, making them challenging to detect in realistic experiments. This Boltzmann factor is also consistent with the recoherence mechanism identified in \cite{Colas:2022kfu}.


\section{Conclusions}\label{sec:conclu}

Building on the flat-space intuition that a large mass hierarchy $m \gg H$ simplifies the late-time, long-wavelength dynamics of massless modes, we have joined earlier efforts to clarify the connection between cosmological collider signals and single-field effective field theories \cite{Cespedes:2012hu, Mukohyama:2020lsu, Aoki:2021ffc, Salcedo:2022aal, Green:2024cmx, DuasoPueyo:2025lmq, Cespedes:2025ple}, with the aim of uncovering distinctive features of cosmological dynamics along the way. \\

\paragraph{Summary.}  We have investigated the origin of the cosmological collider signal within the framework of open EFTs, aiming to make the underlying physics more transparent.
The flat-space example of \Sec{sec:prelude} illustrates the procedure we later apply in de Sitter space. It captures the key features: (i) the trispectrum splits into distinct, physically interpretable components when the massive propagators are decomposed into principal value $G^P_\sigma$, Pauli–Jordan $G^\Delta_\sigma$, and Keldysh $G^K_\sigma$ contributions; (ii) the single-field EFT obtained after integrating out the massive field contains both unitary and non-unitary operators, the latter governing the dissipative and stochastic extensions of the theory; and (iii) the EFT coefficients are \textit{a priori} non-analytic in the kinematics, although this non-analyticity can be removed by imposing a scale hierarchy, such as in the heavy-mass expansion.
          
In de Sitter space, the Keldysh basis naturally aligns with the distinct kinematic components of the signal — the local and non-local contributions. In \Sec{sec:Keldysh}, we show that the Keldysh propagator $G^K_\sigma$ generates the non-local cosmological signal in both the parity-even and parity-odd sectors. The local signal, however, requires more care, as it receives contributions from all types of propagators. Nevertheless, we find that the leading local signal arises from the principal value propagator $G^P_\sigma$ in the parity-even sector, and from the Keldysh propagator $G^K_\sigma$ in the parity-odd sector, pointing to distinct generation mechanisms.
           
The lack of direct observation of the $\sigma$ field — probed only through correlators of the massless scalar $\varphi$ — together with the large mass hierarchy, motivates the derivation of a single-field EFT for $\varphi$, in which $\sigma$ has been integrated out. Deriving this EFT from a top-down perspective in \Sec{sec:EFT}, we characterised the patterns that give rise to the cosmological collider signal. The resulting effective theory is intrinsically open, featuring dissipative and stochastic operators in addition to unitary ones. Interestingly, these operators correlate with the distinct kinematic regimes — local and non-local — discussed above. The non-local signal originates from noise operators in the single-field EFT for $\varphi$, sensitive to the occupation number of the $\sigma$ field. In contrast, the local signal in parity-conserving theories arises from unitary operators in the single-field EFT, and is insensitive to the on-shell statistics of $\sigma$.
           
Finally, using time-convolutionless methods, we recast the operators in a time-local form, absorbing the memory of past interactions into the EFT coefficients. By doing so, we can investigate the emergence of non-analytic features in these coefficients and their eventual removal upon imposing the scale hierarchy. This procedure establishes consistency requirements for open EFTs that generate vacuum cosmological collider signals. It also clarifies the connection between cosmological collider signals, entropy production, and the de(re)coherence mechanisms explored in \cite{Colas:2022hlq, Colas:2022kfu, Colas:2024xjy, Burgess:2024eng, Colas:2024ysu}, linking observational features such as the non-local signal to intrinsic properties of the quantum state, including its degree of mixedness. \\

\paragraph{Discussion.} Throughout this article, we have made several intriguing observations that may motivate further investigation.
A few comments can be made regarding the flat-space results.
Decomposing the trispectrum signal into contributions from each propagator, $\mathcal{I} = \mathcal{I}^P + \mathcal{I}^\Delta + \mathcal{I}^K$, we uncover an interesting kinematic structure. 
First, it is worth noting that the $\mathcal{I}^P$ contribution, given in \Eq{eq_flatIH}, contains a $k_T$ pole, whereas the $\mathcal{I}^\Delta$ and $\mathcal{I}^K$ contributions, given in \Eqs{eq_flatID} and \eqref{eq_flatIK} respectively, do not exhibit this singularity. 
It is well known that the $k_T$ pole in cosmological correlators encodes the analogue of the scattering amplitude result (see \cite{Goodhew:2020hob, Salcedo:2022aal, Cespedes:2025dnq} and references therein). Since the $\mathcal{I}^P$ contribution corresponds to the unitary part of the signal in the single-field perspective, this observation is consistent with expectations. 
More striking, however, is the fact that $\mathcal{I}^\Delta$, given in \Eq{eq_flatID}, vanishes in the $k_T \rightarrow 0$ limit. 
These two observations are likely related. While scattering amplitudes arise from virtual exchange of the $\sigma$ field through the $s$-channel — a feature shared by the unitary sector of the effective theory generating $\mathcal{I}^P$ — the dissipative and noise contributions, identifiable with $\mathcal{I}^\Delta$ and $\mathcal{I}^K$ respectively, are associated with on-shell particle production \cite{Lau:2024mqm}. In \cite{Burgess:2024heo}, the authors investigated how these unitary results are recovered within the in-in formalism. It is tempting to interpret the $k_T \rightarrow 0$ limit as a way to restore time-translation symmetry. 
By enforcing this limit, one effectively imposes a vacuum state in the asymptotic future, thereby suppressing dissipative contributions — a plausible explanation for the vanishing of $\mathcal{I}^\Delta$. One might then wonder why, although $\mathcal{I}^\Delta$ vanishes as $k_T \rightarrow 0$, $\mathcal{I}^K$ does not. This may be attributed to the fact that even in the vacuum, quantum fluctuations generate a residual noisy environment for the system. A possible test of these hypotheses would be to replace the vacuum initial state with an excited state and examine whether these qualitative features persist.

It is also notable that the structure of the EFT coefficients $c_1(k_i,k_j,E_s), \, c_2(k_i,k_j,E_s)$, and $c_3(k_i,k_j,E_s)$ given in \eqref{eq_flatc1}-\eqref{eq_flatc3} closely resembles the computation of mixed local bispectra in the Keldysh basis — see Appendix D of \cite{Salcedo:2024smn}. In particular, they exhibit the same singularity structure: a partial energy pole $E_s + k_{ij} = 0$ followed by three folded singularities, $E_s + k_i - k_j$, $E_s - k_i + k_j$, and $E_s - k_{ij}$, which encode the kinematic transfer of one momentum into two others \cite{Green:2020whw}. If these EFT coefficients could be obtained by computing $\langle \varphi_{\bmk_i} \varphi_{\bmk_j} \sigma_{\bms} \rangle'$ from the appropriate effective cubic vertices, 
the $s$-channel tree-level trispectrum could be expressed as the product of a local trispectrum $\langle \varphi_{\bmk_1} \varphi_{\bmk_2} \varphi_{\bmk_3} \varphi_{\bmk_4} \rangle'$, obtained from the effective quartic operators $\varphi^4$ appearing in the influence functional given in \eqref{eq_flatIFH}, \eqref{eq_flatIFD}, and \eqref{eq_flatIFK}, weighted by EFT coefficients that are themselves mixed tree-level bispectra arising from operators of the form $\varphi^2 \sigma$. 
Given that a similar emergence of subgraphs in the computation of in-in correlators has been reported in \cite{Cespedes:2025dnq}, it would be interesting to understand whether this decomposition — \textit{a priori} different from the usual cutting rules \cite{Goodhew:2020hob, Goodhew:2021oqg, Ema:2024hkj} — can be systematically formulated and leveraged to uplift flat-space correlators to de Sitter.

On the cosmological side, the following observations may invite further investigation.
It is quite fortunate that the Keldysh basis makes manifest both the unitary/non-unitary splitting of the EFT and the kinematic structure of the local and non-local signals. One may then ask whether an underlying physical principle can explain this observation.
We also observed that, while the original $+/-$ basis remains the most practical choice for early-time sub-Hubble computations, the Keldysh basis is particularly convenient for analyzing the emerging simplicity of late-time super-Hubble dynamics. Further work is needed to fully understand whether this basis can shed light on recent discussions regarding the emergence of anomalous dimensions for operators in de Sitter \cite{Cohen:2020php, Chowdhury:2023arc, Cohen:2024anu, Baumann:2025tkm, Sleight:2025dmt}. At a technical level, this will require addressing conceptual questions, such as how to perform a Wick rotation in the Keldysh basis, as well as developing a clearer understanding of the $i \epsilon$ prescription. The latter has proven to be subtle in this project, and while convergence may be a necessary condition, it is not sufficient to guarantee the physical consistency of the results.
Finally, it is fair to say that the open EFT for cosmological collider signals that we derived serves less as a convenient framework for practical computations and more as an informative guide toward the general description of light fields in an environment of heavy fields with a Bunch-Davies UV completion. Whether the non-analytic features appearing in the EFT coefficients can be systematically classified and controlled to broaden the class of physical effects that cosmological EFTs can describe, while remaining theoretically controllable, remains an open question.

At last, a few remarks apply more broadly to non-equilibrium QFT and the Schwinger–Keldysh formalism, regardless of their specific domain of application.
We have shown that non-locality in the operators can be traded for non-analyticity in the EFT coefficients via time-convolutionless methods. It would be interesting to examine whether cutting rules derived from the locality of the partial UV completion can impose further structure.
In \Sec{subsec:entropy}, we illustrated how entropy measures such as the linear entropy relate to observable signals. Beyond this cosmological context, it would be interesting to examine whether the linear entropy can be computed within the single-field open EFT and expressed directly in terms of its coefficients. \\

\paragraph{Outlook.} We conclude with a few remarks and possible directions for future work. 
One of the most important outcomes of this study is the realisation that reproducing the trispectrum from a massive exchange requires more than simply writing down an effective action. One must instead consider an influence functional that includes both dissipative and stochastic effects in order to capture the correct result. More generally, this implies that model-agnostic explorations of low-energy EFTs in cosmology must incorporate all operators allowed by the Schwinger–Keldysh contour, including those associated with dissipation and noise. Understanding the appropriate power-counting in the presence of such operators remains an open and largely unexplored direction.
        
In this article, we focused on the simplest implementation that makes the underlying physics most transparent. Looking ahead, our approach can be readily extended to settings in which specific features of the signal are enhanced. For instance, introducing a chemical potential \cite{Sou:2021juh, Tong:2023krn, Bodas:2025wuk} would plausibly increase the occupation number of the massive field, thereby \textit{in fine} amplifying the dissipative and stochastic effects that scale with it.  
One may also explore the impact of non-Bunch Davies initial conditions, where the initial state starts with a finite occupation number.
At last, having a lighter $\sigma$ field, lying in the complementary series rather than in the principal series, would also substantially enhance the non-unitary sector of the theory, which may ultimately dominate the production of the signal \cite{Cespedes:2025ple}. In such regimes, one may be forced to abandon strict locality, although the super-Hubble gradient expansion could still provide a pathway to identifying an appropriate Markovian regime \cite{Burgess:2024eng}.

Ultimately, one may ask what an \emph{open cosmological collider} looks like: what new features arise when the massive exchange in a partial UV completion is itself governed by an open quantum field theory? The interplay between the heavy mass scale and the damping/sourcing induced by the coupling of this two-field system to its surrounding bath may reveal a rich structure, with potential relevance for models of gauge–axion inflation \cite{Anber:2009ua, Barnaby:2011qe, Peloso:2022ovc, Creminelli:2023aly} and warm inflation \cite{Tong:2018tqf}. Furthermore, one can consider integrating out all massive fields in a open environment, and ask what is the most general single-field open EFT that encompasses signals even in an open cosmological collider. Such general frameworks of open cosmological collider physics could eventually provide a model-agnostic parametrisation of a broad spectrum of new physics beyond the single-field unitary lamppost.


\subsection*{Acknowledgements:} We thank Sebastian Cespedes, Laura Engelbrecht, Carlos Duaso Pueyo, Yohei Ema, Sadra Jazayeri, Enrico Pajer, Sébastien Renaux-Petel, David Stefanyszyn, Dong-Gang Wang and Zhong-Zhi Xianyu for insightful discussions. This work has been supported by STFC consolidated grant ST/X001113/1, ST/T000694/1, ST/X000664/1 and EP/V048422/1. ZQ is supported by NSFC under Grant No.\ 12275146, the National Key R\&D Program of China (2021YFC2203100), and the Dushi Program of Tsinghua University.
 


\appendix

\section{Flat space trispectrum from local open effective field theory}\label{app:recover}

In this appendix, we start from the non-local influence functional \eqref{eq_flatIFHD} to derive a local open EFT, and then recover the three contributions \eqref{eq_flatIH}-\eqref{eq_flatIK} of the trispectrum \eqref{eq_flatIresult}.

\subsection{Local open effective field theory}\label{app:localEFT}

To proceed, we first turn to the three-dimensional Fourier space to manifest the spatial translation symmetry since we are interested in the equal-time correlator. In this Fourier space, the influence functional \eqref{eq_flatIFHD} becomes
\begin{align}
            S_{\mathrm{IF}}[\varphi_r,\varphi_a] \supset & ~\frac{1}{2\Lambda^2} \int_{-\infty}^0 \dd t_1 \int_{-\infty}^0 \dd t_2\int \mathcal D\bm k  \nonumber \\
            &\bigg\{ \dot\varphi_r(t_1,\bm k_1) \dot\varphi_a(t_1,\bm k_2) 
              \left[ \dot\varphi_r(t_2,\bm k_3)\dot\varphi_r(t_2,\bm k_4) + \frac{1}{4}\dot\varphi_a(t_2,\bm k_3)\dot\varphi_a(t_2,\bm k_4)\right] G^{P}_\sigma(s;t_1,t_2) \nonumber \\
            &+\frac12 \dot\varphi_r(t_1,\bm k_1) \dot\varphi_a(t_1,\bm k_2) 
              \left[ \dot\varphi_r(t_2,\bm k_3)\dot\varphi_r(t_2,\bm k_4) + \frac{1}{4}\dot\varphi_a(t_2,\bm k_3)\dot\varphi_a(t_2,\bm k_4)\right] G^{\Delta}_\sigma(s;t_1,t_2)\nonumber \\
            &+\dot\varphi_r(t_1,\bm k_1) \dot\varphi_a(t_1,\bm k_2) \dot\varphi_r(t_2,\bm k_3)\dot\varphi_a(t_2,\bm k_4)G^{K}_\sigma(s;t_1,t_2)
            \bigg\},
        \end{align}
where we only focus on the $s$ channel and have defined 
\begin{equation}
\label{eq_intKmeasure}
\int \mathcal D\bm k \equiv (2\pi)^3\delta^3(\bm k_1+\bm k_2+\bm k_3+\bm k_4) \times \int \prod_{i=1}^4 \frac{\dd^3\bm k_i}{(2\pi)^3} \times \int \dd^3\bm s\, \delta^3(\bm k_1+\bm k_2 - \bm s).
\end{equation}

Let us consider the contribution from $G^K_\sigma$ first:
\begin{align}
 S_{\mathrm{IF}}^K[\varphi_r,\varphi_a] = \frac{1}{2\Lambda^2} \int_{-\infty}^0 \dd t_1 \int_{-\infty}^0 \dd t_2\int \mathcal D\bm k\,
            \dot\varphi_r(t_1,\bm k_1) \dot\varphi_a(t_1,\bm k_2)
            \dot\varphi_r(t_2,\bm k_3) \dot\varphi_a(t_2,\bm k_4) G^{K}_\sigma(s;t_1,t_2).
\end{align}
As mentioned in the main text, the next step is to evolve fields at different times using free equations of motion \eqref{eq_flatevolve}. Naively, one would expect to express all fields at $t_1$ in terms of fields at $t_2$ (or vice versa) using \Eq{eq_flatevolve}, and then integrate over $t_1$ ($t_2$) to obtain an EFT local in time. However, there is a caveat. If we do so, we obtain
\begin{align}
S_{\mathrm{IF}}^K[\varphi_r,\varphi_a] =&~ \frac{1}{2\Lambda^2}\int_{-\infty}^0\dd t_2\int_{-\infty}^0\dd t_1
\int\mathcal D\bm k\, \frac{i \cos E_s(t_1-t_2)}{2E_s} \times \dot\varphi_r(t_2,\bm k_3)\dot\varphi_a(t_2,\bm k_4)\nonumber\\
&\times\Big[k_1 \sin[k_1(t_2 - t_1)]\varphi_r (t_2,\bm k_1)+\cos[k_1(t_2-t_1)]\dot\varphi_r(t_2,\bm k_1)\Big]\nonumber\\
&\times\Big[k_2 \sin[k_2(t_2 - t_1)]\varphi_a (t_2,\bm k_2)+\cos[k_2(t_2-t_1)]\dot\varphi_a(t_2,\bm k_2)\Big]\nonumber\\
=&~i\int_{-\infty}^0\dd t\int\mathcal D\bm k\, \Big[
\alpha_1(k_1,k_2,E_s;t)\varphi_{r,1}\varphi_{a,2}
+ \alpha_2(k_1,k_2,E_s;t)\dot\varphi_{r,1}\dot\varphi_{a,2}
\nonumber\\
&+ \alpha_3(k_1,k_2,E_s;t)(\varphi_{r,1}\dot\varphi_{a,2} + \varphi_{a,1}\dot\varphi_{r,2})\Big] \dot\varphi_{r,3}\dot\varphi_{a,4},
\label{eq_SK2}
\end{align}
Here we have renamed $t_2$ as $t$ in the second step and used the shorthand $\varphi_{r/a,i}\equiv\varphi_{r/a}(t,\bm k_i)$, and also used the symmetry $(k_1,k_2)\leftrightarrow (k_2,k_1)$. The EFT coefficients are given by:
\begin{align}
&\alpha_1(k_i,k_j,E_s;t) \equiv \frac{k_ik_j}{16\Lambda^2E_s}\left[ \frac{\sin(E_s+k_{12})t}{E_s+k_{12}} - \frac{\sin(E_s+\bar k_{12})t}{E_s+\bar k_{12}} - \frac{\sin(E_s-\bar k_{12})t}{E_s-\bar k_{12}} + \frac{\sin(E_s-k_{12})t}{E_s-k_{12}} \right],\\
&\alpha_2(k_i,k_j,E_s;t) \equiv \frac{-1}{16\Lambda^2E_s}\left[ \frac{\sin(E_s+k_{12})t}{E_s+k_{12}} + \frac{\sin(E_s+\bar k_{12})t}{E_s+\bar k_{12}} + \frac{\sin(E_s-\bar k_{12})t}{E_s-\bar k_{12}} + \frac{\sin(E_s-k_{12})t}{E_s-k_{12}} \right],\\
&\alpha_3(k_i,k_j,E_s;t) \equiv \frac{k_i}{16\Lambda^2E_s}\left[ \frac{\cos(E_s+k_{12})t}{E_s+k_{12}} + \frac{\cos(E_s+\bar k_{12})t}{E_s+\bar k_{12}} -\frac{\cos(E_s-\bar k_{12})t}{E_s-\bar k_{12}} - \frac{\cos(E_s-k_{12})t}{E_s-k_{12}} \right].
\end{align}
At first glance, these EFT coefficients are time dependent, which is an artifact since the theory itself is invariant under time translation. Furthermore, they also depend on the final time at which the correlators are computed. Indeed, if we are aiming to compute equal-time correlators at $t_0$, the coefficients become $\alpha_i(k_i,k_j,E_s;t-t_0)$, and their dependence only on the time difference $t-t_0$ is a hint of translation symmetry.

Fortunately, there is one way to make these EFT coefficients independent of the final time at which we make observations. The trick is to insert the identity $\mathbb{I} \equiv \theta(t_2-t_1)+\theta(t_1-t_2)$ in the integrand to avoid reaching the final time in the first layer of integral where the EFT coefficients are computed. In particular, for the first term for which $t_1 < t_2$, we still use \Eq{eq_flatevolve} to express $\dot\varphi_{r,a}(t_1,\bm k)$ in terms of $\varphi_{r,a}(t_2,\bm k)$ and $\dot\varphi_{r,a}(t_2,\bm k)$ and integrate over $t_1$ to obtain 
\begin{align}
S_{\mathrm{IF}}^K[\varphi_r,\varphi_a] \supset &~ \frac{1}{2\Lambda^2}\int_{-\infty}^0\dd t_2\int_{-\infty}^{t_2}\dd t_1
\int\mathcal D\bm k\, \frac{i \cos E_s(t_1-t_2)}{2E_s} \times \dot\varphi_r(t_2,\bm k_3)\dot\varphi_a(t_2,\bm k_4)\nonumber\\
&\times\Big[k_1 \sin[k_1(t_2 - t_1)]\varphi_r (t_2,\bm k_1)+\cos[k_1(t_2-t_1)]\dot\varphi_r(t_2,\bm k_1)\Big]\nonumber\\
&\times\Big[k_2 \sin[k_2(t_2 - t_1)]\varphi_a (t_2,\bm k_2)+\cos[k_2(t_2-t_1)]\dot\varphi_a(t_2,\bm k_2)\Big]\nonumber\\
=&~i\int_{-\infty}^0\dd t\int\mathcal D\bm k\, \alpha_3(k_1,k_2,E_s;0)(\varphi_{r,1}\dot\varphi_{a,2} + \varphi_{a,1}\dot\varphi_{r,2}) \dot\varphi_{r,3}\dot\varphi_{a,4}.
\end{align}
For the second term, we switch to express fields at $t_1$ by fields at $t_2$ and integrate over $t_2$:
\begin{align}
S_{\mathrm{IF}}^K[\varphi_r,\varphi_a] \supset &~ \frac{1}{2\Lambda^2}\int_{-\infty}^0\dd t_1\int_{-\infty}^{t_1}\dd t_2
\int\mathcal D\bm k\, \frac{i \cos E_s(t_1-t_2)}{2E_s} \times \dot\varphi_r(t_1,\bm k_1)\dot\varphi_a(t_1,\bm k_2)\nonumber\\
&\times\Big[k_3 \sin[k_3(t_1 - t_2)]\varphi_r (t_1,\bm k_3)+\cos[k_3(t_1-t_2)]\dot\varphi_r(t_1,\bm k_3)\Big]\nonumber\\
&\times\Big[k_4 \sin[k_4(t_1 - t_2)]\varphi_a (t_1,\bm k_4)+\cos[k_4(t_1-t_2)]\dot\varphi_a(t_1,\bm k_4)\Big]\nonumber\\
=&~i\int_{-\infty}^0\dd t\int\mathcal D\bm k\, \alpha_3(k_3,k_4,E_s;0)\dot\varphi_{r,1}\dot\varphi_{a,2}(\varphi_{r,3}\dot\varphi_{a,4} + \varphi_{a,3}\dot\varphi_{r,4}).
\end{align}
Combining the two contributions and using the symmetry under $(\bm k_1,\bm k_2,\bm k_3,\bm k_4,\bm s)\leftrightarrow(\bm k_3,\bm k_4,\bm k_1,\bm k_2,-\bm s)$ in $\int\mathcal D\bm k$, we finally obtain:
\begin{align}
S_{\mathrm{IF}}^K[\varphi_r,\varphi_a] = i\int_{-\infty}^0\dd t\int\mathcal D\bm k\, c_3(k_1,k_2,E_s)(\varphi_{r,1}\dot\varphi_{a,2} + \varphi_{a,1}\dot\varphi_{r,2}) \dot\varphi_{r,3}\dot\varphi_{a,4},
\end{align}
with the main text EFT coefficient defined as
\begin{align}
c_3(k_i,k_j,E_s) \equiv \frac{k_i}{8\Lambda^2E_s} \left(\frac{1}{E_s+ k_{ij}} + \frac{1}{E_s + \bar k_{ij}} - \frac{1}{E_s-\bar k_{ij}} - \frac{1}{E_s-k_{ij}}\right),
\end{align}
which now is independent of time and thus implies time translation symmetry.

Following the same procedure, we can rewrite the contributions from $G^P_\sigma$ and $G^\Delta_\sigma$ as integrals of local operators, which are
\begin{align}
S_{\mathrm{IF}}^P[\varphi_r,\varphi_a] &= \int_{-\infty}^0 \dd t\int\mathcal D\bm k\,\\
\bigg\{&c_1(k_1,k_2,E_s)\left[ \varphi_{r,1}\varphi_{a,2}
\Big( \dot\varphi_{r,3}\dot\varphi_{r,4}+\frac14\dot\varphi_{a,3}\dot\varphi_{a,4}\Big)
+\Big(\varphi_{r,1}\varphi_{r,2}+\frac14 \varphi_{a,1}\varphi_{a,2}\Big)\dot\varphi_{r,3}\dot\varphi_{a,4} \right]\nonumber\\
+&c_2(k_1,k_2,E_s)\left[ \dot\varphi_{r,1}\dot\varphi_{a,2}
\Big( \dot\varphi_{r,3}\dot\varphi_{r,4}+\frac14\dot\varphi_{a,3}\dot\varphi_{a,4}\Big)
+\Big(\dot\varphi_{r,1}\dot\varphi_{r,2}+\frac14 \dot\varphi_{a,1}\dot\varphi_{a,2}\Big)\dot\varphi_{r,3}\dot\varphi_{a,4} \right]\bigg\},\notag
\end{align}
and
\begin{align}
S_{\mathrm{IF}}^\Delta[\varphi_r,\varphi_a] &= -\int_{-\infty}^0 \dd t\int\mathcal D\bm k\,\\
\bigg\{&c_1(k_1,k_2,E_s)\left[ \varphi_{r,1}\varphi_{a,2}
\Big( \dot\varphi_{r,3}\dot\varphi_{r,4}+\frac14\dot\varphi_{a,3}\dot\varphi_{a,4}\Big)
-\Big(\varphi_{r,1}\varphi_{r,2}+\frac14 \varphi_{a,1}\varphi_{a,2}\Big)\dot\varphi_{r,3}\dot\varphi_{a,4} \right]\nonumber\\
+&c_2(k_1,k_2,E_s)\left[ \dot\varphi_{r,1}\dot\varphi_{a,2}
\Big( \dot\varphi_{r,3}\dot\varphi_{r,4}+\frac14\dot\varphi_{a,3}\dot\varphi_{a,4}\Big)
-\Big(\dot\varphi_{r,1}\dot\varphi_{r,2}+\frac14 \dot\varphi_{a,1}\dot\varphi_{a,2}\Big)\dot\varphi_{r,3}\dot\varphi_{a,4} \right]\bigg\},
\end{align}
respectively, with the other two coefficients giving by
\begin{align}
&c_1(k_i,k_j,E_s) \equiv \frac{-k_ik_j}{16\Lambda^2E_s}\left(\frac{1}{E_s+ k_{ij}} - \frac{1}{E_s + \bar k_{ij}} - \frac{1}{E_s-\bar k_{ij}} + \frac{1}{E_s-k_{ij}}\right),\\
&c_2(k_i,k_j,E_s) \equiv \frac{1}{16\Lambda^2E_s}\left(\frac{1}{E_s+ k_{ij}} + \frac{1}{E_s + \bar k_{ij}} + \frac{1}{E_s-\bar k_{ij}} + \frac{1}{E_s-k_{ij}}\right).
\end{align}
Notice that there are infinite possibilities to write the theory in terms of local-in-time operators with time-dependent coefficients. One can consider linear combinations of $\alpha_i$ and $c_i$, or equivalently, to insert the identity $\mathbb I = \lambda[\theta(t_2-t_1)+\theta(t_1-t_2)] + (1-\lambda)\mathbb I$ into the integrand. All of these theories are equivalent to each other when computing observables (four-point correlation functions) at time $t_0$. However, there is only one choice of $\lambda$ that can make all coefficients independent of time (and thus independent of $t_0$), that is, $\lambda=1$.


\subsection{Recovering the trispectrum}\label{app:trispectrum}

With the influence functional derived in \App{app:localEFT}, it is easy to derive the trispectrum $\mathcal I$  found in \Eq{eq_flatI} following the Feynman rules \cite{Salcedo:2024smn}. In short, we assert only $\varphi_r$ at the future boundary, while bulk fields could be either $\varphi_r$ or $\varphi_a$. Each vertex represents an interaction (at time $t$ which we will integrate) that comes with an extra $i$, while each line represents a contraction between $\varphi_r$ and $\varphi_{r,a}$, see \Eq{eq_GRAK}. Explicitly,
\begin{subequations}
\begin{align}
&\langle \varphi_r(t_1,\bm k) \varphi_a(t_2,-\bm k) \rangle' = -i G_\varphi^R(k;t_1,t_2),\\
&\langle \varphi_r(t_1,\bm k) \varphi_r(t_2,-\bm k) \rangle' = -i G_\varphi^K(k;t_1,t_2),
\end{align}
\end{subequations}
where the retarded and Keldysh propagators of $\varphi$ are given by:
\begin{subequations}
\begin{align}
&G_\varphi^R(k;t_1,t_2) = \frac{\sin k(t_1-t_2)}{k}\theta(t_1-t_2),\\
&G_\varphi^K(k;t_1,t_2) = \frac{i\cos k(t_1-t_2)}{2k}.
\end{align}
\end{subequations}

Let us consider the effect of the Keldysh propagator as an example. Following the Feynman rule, we have
\begin{align}
\mathcal I^K =&~ i \times ic_3(k_1,k_2,E_s) \times (-i)^4\times \int_{-\infty}^0 \dd t\,
\Big[ G_\varphi^K(k_1;0,t)\dot G_\varphi^R(k_2;0,t) + G_\varphi^{R}(k_1;0,t)\dot G_\varphi^K(k_2;0,t) \Big] \nonumber \\
&\qquad \qquad \qquad \qquad \qquad \qquad \qquad \times \dot G_\varphi^R(k_3;0,t) \dot G_\varphi^K(k_4;0,t) + \text{7 perms} \bigg. \,.
\end{align}
Injecting the expression of the free propagators for $\varphi$, we obtain 
\begin{align}
\mathcal I^K =&~ -c_3 (k_1,k_2,E_s) \times \int_{-\infty}^0 \dd t\, \Big[ \frac{i\cos k_1 t}{2k_1} \times (-\cos k_2t)
+ \frac{-\sin k_1t}{k_1}\times \frac{-i \sin k_2 t}{2}\Big]\notag\\
&\qquad \qquad \qquad \qquad \qquad   \times (-\cos k_3t)\times \Big( \frac{-i\sin k_4 t}{2}\Big) + \text{7 perms}\\
=& -\frac{k_1}{8\Lambda^2E_s} \Big(\frac{1}{E_s+ k_{12}} + \frac{1}{E_s + k_1 -k_2} - \frac{1}{E_s-k_1+k_2} - \frac{1}{E_s-k_{12}}\Big)\notag\\
&\qquad  \times \frac{1}{16k_1}\Big(\frac{1}{k_{12}-k_{34}} + \frac{1}{k_{123}-k_4} - \frac{1}{k_{124}-k_3} -\frac{1}{k_{1234}}\Big) + \text{7 perms} \, ,
\end{align}
which simplifies to 
\begin{align}
\mathcal I^K =&~\frac{k_{12}k_{34}}{8\Lambda^2E_s(E_s^2-k_{12}^2)(E_s^2-k_{34}^2)}.
\end{align}
Here the factor $(-i)^4$ on the first line comes from the four propagators, and the permutations include combinations of $(k_1,k_2)\leftrightarrow (k_2,k_1)$, $(k_3,k_4)\leftrightarrow (k_4,k_3)$, and $(k_1,k_2,k_3,k_4)\leftrightarrow (k_3,k_4,k_1,k_2)$. As we can see, the above result reproduces exactly \Eq{eq_flatIK}.

The calculations of $\mathcal I^P$ and $\mathcal I^\Delta$ are basically the same but more tedious. Let us define the total energy $k_T\equiv k_{1234}$, and with the shorthands $G_i^{R,K} \equiv G^{R,K}_\varphi(k_i;0,t)$ and $c_{1,2}\equiv c_{1,2}(k_1,k_2,E_s)$. We have
\begin{align}
\mathcal I^P =& ~ i \times (-i)^4\times \int_{-\infty}^0 \dd t\,
c_1
\bigg[ G^K_1G^R_2\Big(\dot G^K_3\dot G^K_4 + \frac14 \dot G^R_3\dot G^R_4\Big)
+\Big(G^K_1G^K_2 + \frac14 G^R_1 G^R_2\Big)\dot G^K_3\dot G^R_4
\bigg] \notag\\
& \qquad + c_2 \bigg[
\dot G^K_1\dot G^R_2\Big(\dot G^K_3\dot G^K_4+\frac14\dot G^R_3\dot G^R_4\Big)
+\Big(\dot G^K_1\dot G^K_2+\frac14 \dot G^R_1 \dot G^R_2\Big)\dot G^K_3\dot G^R_4
\bigg] + \text{7 perms}\\
=&~i \bigg[ \frac{i c_1}{32k_1k_2}\Big( \frac{2}{k_T} + \frac{1}{k_T-2k_1} - \frac{1}{k_T-2k_2} - \frac{1}{k_T-2k_3} +\frac{1}{k_T-2k_4} \Big)\notag\\
&\qquad -\frac{ic_2}{32}\Big( \frac{2}{k_T} - \frac{1}{k_T-2k_1} + \frac{1}{k_T-2k_2} - \frac{1}{k_T-2k_3} +\frac{1}{k_T-2k_4} \Big)\bigg] + \text{7 perms} \, ,
\end{align}
leading to 
\begin{align}
\mathcal I^P =&-\frac{1}{16\Lambda^2k_T}\Big(\frac{1}{k_{12}^2-E_s^2}+\frac{1}{k_{34}^2-E_s^2}\Big),
\end{align}
and
\begin{align}
\mathcal I^\Delta =& -i \times (-i)^4\times \int_{-\infty}^0 \dd t\,
c_1
\bigg[ G^K_1G^R_2\Big(\dot G^K_3\dot G^K_4 + \frac14 \dot G^R_3\dot G^R_4\Big)
-\Big(G^K_1G^K_2 + \frac14 G^R_1 G^R_2\Big)\dot G^K_3\dot G^R_4
\bigg] \notag\\
& \qquad + c_2 \bigg[
\dot G^K_1\dot G^R_2\Big(\dot G^K_3\dot G^K_4+\frac14\dot G^R_3\dot G^R_4\Big)
-\Big(\dot G^K_1\dot G^K_2+\frac14 \dot G^R_1 \dot G^R_2\Big)\dot G^K_3\dot G^R_4
\bigg] + \text{7 perms}\\
=&-i \bigg[ \frac{i c_1}{32k_1k_2}\Big( \frac{2}{k_{12}-k_{34}} + \frac{1}{k_T-2k_1} - \frac{1}{k_T-2k_2} + \frac{1}{k_T-2k_3} - \frac{1}{k_T-2k_4} \Big)\notag\\
&\qquad-\frac{ic_2}{32}\Big( \frac{2}{k_{12}-k_{34}} - \frac{1}{k_T-2k_1} + \frac{1}{k_T-2k_2} + \frac{1}{k_T-2k_3} -\frac{1}{k_T-2k_4} \Big)\bigg] + \text{7 perms} \,,
\end{align}
which reduces to 
\begin{align}
\mathcal I^\Delta =&-\frac{k_T}{16\Lambda^2(k_{12}^2-E_s^2)(k_{34}^2-E_s^2)}.
\end{align}
These expressions are the same as \Eqs{eq_flatIH} and \eqref{eq_flatID}, respectively.

\section{Soft analyticity of the causal propagators}\label{zeroMomentumAnalyticityAppendix}

In this appendix, we show that the scalar causal propagators $G^{R/A/P/\Delta}$ are always analytic in the zero-momentum limit $k\to 0$, as long as the free-theory Lagrangian is local. We begin by discussing the retarded propagator, and the rest follow automatically since they are simple variants of the retarded propagator. The retarded propagator $G^R$ in a local quantum field theory enjoys a nice property in momentum space, i.e. it is analytic in the zero momentum limit $k\to 0$,
\begin{align}
	G^R(k;\eta,\eta')=G^R(0;\eta,\eta')+k \left[\partial_{k}G^R(k;\eta,\eta')\right]_{k=0}+\frac{k^2}{2!} \left[\partial_{k}^2G^R(k;\eta,\eta')\right]_{k=0}+\mathcal{O}(k^3)\,.
\end{align}
For instance, in Minkowski spacetime, we have for a Lorentz-invariant scalar field with mass $M$,
\begin{align}
	\nonumber G^R(k;t,t')&=\theta(t-t')\frac{\sin E_k(t-t')}{E_k}\\
	&=\theta(t-t')\left[\frac{\sin M (t-t')}{M}+k^2\frac{M(t-t')\cos M (t-t')-\sin M (t-t')}{2M^3}+\mathcal{O}(k^4)\right]\,.\label{eq:MinkRetardedPropIRExpansion}
\end{align}
In de Sitter space, a de-Sitter invariant massive field of mass $\mu$ have
\begin{align}
	\nonumber G^R(k;\eta,\eta')&=-\theta(\eta-\eta')\frac{i \pi  H^2(\eta \eta')^{3/2}}{2\sinh \pi\mu}  \Bigg(J_{i \mu }(-k  \eta ) J_{-i \mu }(-k \eta')-J_{-i \mu }(-k \eta ) J_{i \mu }(-k \eta')\Bigg)\\
	\nonumber&=\theta(\eta-\eta')H^2 (\eta \eta')^{3/2}\Bigg\{\frac{\sin \mu\ln \frac{\eta}{\eta'}}{\mu}\\
	\nonumber&\qquad\qquad+k^2\frac{1}{4\mu(1+\mu^2)}\left[\mu \left(\eta^2-\eta^{\prime 2}\right)\cos\left( \mu\ln\frac{\eta}{\eta'}\right)-\left(\eta^2+\eta^{\prime 2}\right)\sin\left( \mu\ln\frac{\eta}{\eta'}\right)\right]\\
	&\qquad\qquad+\mathcal{O}(k^4)\Bigg\}\,.\label{eq:dSRetardedPropIRExpansion}
\end{align}
In passing, we notice the fascinating structural similarity between \eqref{eq:MinkRetardedPropIRExpansion} and \eqref{eq:dSRetardedPropIRExpansion}, with $\eta/\eta'$ now playing the role of the time difference $t-t'$. Note also that in a free theory, both expressions are independent of the state choice, because the free-theory field commutator is just a c-number which acts trivially on the state, i.e. $\langle \Psi|[\phi(x),\phi(x')]|\Psi\rangle=[\phi(x),\phi(x')]=G^\Delta(x,x')$ for any $|\Psi\rangle$ in the Fock space. In particular, \eqref{eq:dSRetardedPropIRExpansion} remains the same for non-Bunch-Davies vacua.
    
The analyticity of the retarded propagator can be traced back to the locality of the theory Lagrangian. The defining equation for the retarded propagator in a local quantum field theory in Friedmann–Lemaître–Robertson–Walker (FLRW) spacetimes take the general form
\begin{align}
	\mathcal{\hat E}(k;\eta)G^R(k;\eta,\eta')=\sqrt{-g(\eta')}\delta(\eta-\eta')\,,\label{generalEOMFRW}
\end{align}
where the EoM operator must be an analytic function of $k$ around the origin,
\begin{align}
	\mathcal{\hat E}(k;\eta)=\mathcal{\hat E}(0;\eta)+k \left[\partial_{k}\mathcal{\hat E}(k;\eta)\right]_{k=0}+\frac{k^2}{2!} \left[\partial_{k}^2\mathcal{\hat E}(k;\eta)\right]_{k=0}+\mathcal{O}(k^3)\,,
\end{align}
since only a non-negative integer number of derivatives can appear in the Lagrangian. However, this condition alone does not guarantee the analyticity of $G^R$ around $k=0$, since the Feynman propagator $D^{++}$ satisfy the same equation \eqref{generalEOMFRW} but it is not analytic around $k=0$. The crucial difference is that the retarded propagator satisfies a standard unit boundary condition:
\begin{subequations}
	\begin{align}
		\left[G^R(k;\eta,\eta')\right]_{\eta=\eta'}&=0\,,\\
		\left[\partial_\eta G^R(k;\eta,\eta')\right]_{\eta=\eta'}&=-i\mathcal{W}_\eta(v(k;\eta'),v^*(k;\eta'))=\frac{[\phi(k;\eta'),\pi(k;\eta')]}{i\sqrt{-g(\eta')}g^{00}(\eta')}=\frac{1}{\sqrt{-g(\eta')}g^{00}(\eta')}\,,\label{generalBCFRW}
    \end{align}
\end{subequations}
where in the second line we have used the canonical quantisation condition to replace the Wronskian by a quantity that only depends on the geometry of spacetime at $\eta'$. This is possible only if the theory is described by a local Lagrangian/Hamiltonian (or Linbladian) density in spacetime. Since both the equation \eqref{generalEOMFRW} and the boundary conditions \eqref{generalBCFRW} are analytic at $k=0$, we expect their unique solution $G^R$ to be analytic at $k=0$. In contrast, the Feynman propagator is symmetric and satisfies the Bunch-Davies boundary condition, 
\begin{subequations}
	\begin{align}
		D^{++}(k;\eta,\eta')&=D^{++}(k;\eta',\eta)\,,\\
		\lim_{\eta'\to-\infty (1-i\epsilon)} D^{++}(k;\eta,\eta')&=0\,.
	\end{align}
\end{subequations}
Here the second condition is an asymptotic condition that does not commute with taking $k\to 0$ (since $k\eta'$ appear together on the exponential), and therefore leads to the non-analyticity of $D^{++}$ near $k=0$.
    
The advanced propagator $G^A$, being the reversal of the retarded propagator, is also analytic at $k=0$. Thus any combinations of $G^R$ and $G^A$ are also analytic at $k=0$, including the principal-value propagator $G^P$ and the Pauli-Jordan propagator $G^\Delta$.

    \section{de Sitter effective field theory coefficients}\label{app:EFTcoeff}

    In this appendix, we use the free equations of motion for $\varphi$ to make the influence functional manifest as an integral of a local operator. In Fourier space, the influence functional \eqref{eq:IFpost} becomes: 
    \begin{align}
            S_{\mathrm{IF}}[\varphi_r,\varphi_a] \supset & ~\frac{1}{2\Lambda^2} \int_{-\infty}^0 \dd \eta_1 a^2(\eta_1) \int_{-\infty}^0 \dd \eta_2 a^2(\eta_2) \int \mathcal D\bm k  \nonumber \\
            &\bigg\{ \varphi'_r(\eta_1,\bm k_1) \varphi'_a(\eta_1,\bm k_2) 
              \left[ \varphi'_r(\eta_2,\bm k_3)\varphi'_r(\eta_2,\bm k_4) + \frac{1}{4}\varphi'_a(\eta_2,\bm k_3)\varphi'_a(\eta_2,\bm k_4)\right] G^{P}_\sigma(s;\eta_1,\eta_2) \nonumber \\
            &+ \varphi'_r(\eta_1,\bm k_1) \varphi'_a(\eta_1,\bm k_2) 
              \left[ \frac12\varphi'_r(\eta_2,\bm k_3)\varphi'_r(\eta_2,\bm k_4) + \frac{1}{8}\varphi'_a(\eta_2,\bm k_3)\varphi'_a(\eta_2,\bm k_4)\right] G^{\Delta}_\sigma(s;\eta_1,\eta_2)\nonumber \\
            &+\varphi'_r(\eta_1,\bm k_1) \varphi'_a(\eta_1,\bm k_2) \varphi'_r(\eta_2,\bm k_3)\varphi'_a(\eta_2,\bm k_4)G^{K}_\sigma(s;\eta_1,\eta_2)
            \bigg\},
        \end{align}
where we only focus on the $s$ channel and have defined 
\begin{equation}
\int \mathcal D\bm k \equiv (2\pi)^3\delta^3(\bm k_1+\bm k_2+\bm k_3+\bm k_4) \times \int \prod_{i=1}^4 \frac{\dd^3\bm k_i}{(2\pi)^3} \times \int \dd^3\bm s\, \delta^3(\bm k_1+\bm k_2 - \bm s).
\end{equation}

Let us consider the contribution from $G^P_\sigma$ first:
\begin{align}
 S_{\mathrm{IF}}^P[\varphi_r,\varphi_a] = & ~\frac{1}{2\Lambda^2} \int_{-\infty}^0 \dd \eta_1 a^2(\eta_1) \int_{-\infty}^0 \dd \eta_2 a^2(\eta_2) \int \mathcal D\bm k\,
            \varphi'_r(\eta_1,\bm k_1) \varphi'_a(\eta_1,\bm k_2)   \nonumber \\
&\times              \left[ \varphi'_r(\eta_2,\bm k_3)\varphi'_r(\eta_2,\bm k_4) + \frac{1}{4}\varphi'_a(\eta_2,\bm k_3)\varphi'_a(\eta_2,\bm k_4)\right] G^{P}_\sigma(s;\eta_1,\eta_2).
\end{align}
We can insert identity $\mathbb{I} \equiv \theta(\eta_1-\eta_2)+\theta(\eta_2-\eta_1)$ in the integrand. For first term for which $\eta_1 > \eta_2$, we use \Eq{eq:dSevolve} to express $\varphi'_{r,a}(\eta_2,\bm k)$ in terms of $\varphi_{r,a}(\eta_1,\bm k)$ and $\varphi'_{r,a}(\eta_1,\bm k)$ and integrate over $\eta_2$ to obtain
\begin{align}
S_{\mathrm{IF}}^P[\varphi_r,\varphi_a] \supset & ~\frac{1}{2\Lambda^2} \int_{-\infty}^0 \dd \eta_1 a^2(\eta_1)\int_{-\infty}^{\eta_1} \dd \eta_2 a^2(\eta_2)\int \mathcal D\bm k\,G^{P}_\sigma(s;\eta_1,\eta_2)
\times \varphi'_r(\eta_1,\bm k_1)\varphi'_a(\eta_1,\bm k_2)\nonumber\\
\times\bigg\{\Big[ &G^\varphi_{21}(k_3, \eta_1,\eta_2)\varphi_r (\eta_1,\bm k_3)+G^\varphi_{22}(k_3, \eta_1,\eta_2)\varphi'_r(\eta_1,\bm k_3)\Big]\nonumber\\
\times\Big[&G^\varphi_{21}(k_4, \eta_1,\eta_2)\varphi_r (\eta_1,\bm k_4)+G^\varphi_{22}(k_4, \eta_1,\eta_2)\varphi'_r(\eta_1,\bm k_4)\Big]\nonumber\\
+\frac14 \Big[ &G^\varphi_{21}(k_3, \eta_1,\eta_2)\varphi_a (\eta_1,\bm k_3)+G^\varphi_{22}(k_3, \eta_1,\eta_2)\varphi'_a(\eta_1,\bm k_3)\Big]\nonumber\\
\times\Big[& G^\varphi_{21}(k_4, \eta_1,\eta_2)\varphi_a (\eta_1,\bm k_4)+G^\varphi_{22}(k_4, \eta_1,\eta_2)\varphi'_a(\eta_1,\bm k_4)\Big]\bigg\},
\end{align}
that is 
\begin{align}
    S_{\mathrm{IF}}^P[\varphi_r,\varphi_a]
\supset&\int_{-\infty}^0 \dd \eta_1 \int\mathcal D\bm k\, \varphi'_r(\eta_1,\bm k_1)\varphi'_a(\eta_1,\bm k_2)\nonumber\\
\times \bigg\{&a^2(\eta_1)c_1(k_3,k_4,s;\eta_1)
\left[ \varphi_r(\eta_1,\bm k_3)\varphi_r(\eta_1,\bm k_4) +\frac14 \varphi_a(\eta_1,\bm k_3)\varphi_a(\eta_1,\bm k_4)\right] \nonumber\\
+ &c_2(k_3,k_4,s;\eta_1)\left[ \varphi'_r(\eta_1,\bm k_3)\varphi'_r(\eta_1,\bm k_4) + \frac14 \varphi'_a(\eta_1,\bm k_3)\varphi'_a(\eta_1,\bm k_4)\right]\bigg\} \nonumber \\
+ &a(\eta_1)c_3(k_3,k_4,s;\eta_1)\left[ \varphi_r(\eta_1,\bm k_3)\varphi'_r(\eta_1,\bm k_4) + \frac14 \varphi_a(\eta_1,\bm k_3)\varphi'_a(\eta_1,\bm k_4)\right]\bigg\} \nonumber \\
+ &a(\eta_1)c_4(k_3,k_4,s;\eta_1)\left[ \varphi'_r(\eta_1,\bm k_3)\varphi_r(\eta_1,\bm k_4) + \frac14 \varphi'_a(\eta_1,\bm k_3)\varphi_a(\eta_1,\bm k_4)\right]\bigg\}.
\end{align}
where the coefficients $c_{1,2,3,4}$ are given by
\begin{align}
&c_1(k_i,k_j,s;\eta) \equiv \frac{1}{2\Lambda^2} \int_{-\infty}^{\eta} \dd \eta' a^2(\eta') G^{P}_\sigma(s;\eta,\eta') G^\varphi_{21}(k_i, \eta,\eta') G^\varphi_{21}(k_j, \eta,\eta') ,\\
&c_2(k_i,k_j,s;\eta) \equiv \frac{a^2(\eta)}{2\Lambda^2} \int_{-\infty}^{\eta} \dd \eta' a^2(\eta') G^{P}_\sigma(s;\eta,\eta') G^\varphi_{22}(k_i, \eta,\eta') G^\varphi_{22}(k_j, \eta,\eta') ,\\
&c_3(k_i,k_j,s;\eta) \equiv \frac{a(\eta)}{2\Lambda^2} \int_{-\infty}^{\eta} \dd \eta' a^2(\eta') G^{P}_\sigma(s;\eta,\eta') G^\varphi_{21}(k_i, \eta,\eta') G^\varphi_{22}(k_j, \eta,\eta') ,\\
&c_4(k_i,k_j,s;\eta) \equiv \frac{a(\eta)}{2\Lambda^2} \int_{-\infty}^{\eta} \dd \eta' a^2(\eta') G^{P}_\sigma(s;\eta,\eta') G^\varphi_{22}(k_i, \eta,\eta') G^\varphi_{21}(k_j, \eta,\eta') .
\end{align}
From the definition one can see that $c_4(k_i,k_j,s;\eta) = c_3(k_j,k_i,s;\eta)$. Similarly, for the second term with $\theta(\eta_2-\eta_1)$, we express $\varphi'_{r,a}(\eta_1,\bm k)$ in terms of $\varphi_{r,a}(\eta_2,\bm k)$ and $\dot\varphi'_{r,a}(\eta_2,\bm k)$, then integrate over $\eta_1$ to obtain
\begin{align}
 S_{\mathrm{IF}}^P[\varphi_r,\varphi_a] \supset & ~\frac{1}{2\Lambda^2} \int_{-\infty}^0 \dd \eta_2 a^2(\eta_2) \int_{-\infty}^{\eta_2}\dd \eta_1 a^2(\eta_1)  \int \mathcal D\bm k\, \nonumber \\
 &\times \Big[G^\varphi_{21}(k_1, \eta_2,\eta_1)\varphi_r (\eta_2,\bm k_1)+G^\varphi_{22}(k_1, \eta_2,\eta_1)\varphi'_r(\eta_2,\bm k_1)\Big] \nonumber \\
 &\times \Big[ G^\varphi_{21}(k_2, \eta_2,\eta_1)\varphi_a (\eta_2,\bm k_2)+G^\varphi_{22}(k_2, \eta_2,\eta_1)\varphi'_a(\eta_2,\bm k_2)\Big]  \nonumber \\
&\times   \left[ \varphi'_r(\eta_2,\bm k_3)\varphi'_r(\eta_2,\bm k_4) + \frac{1}{4}\varphi'_a(\eta_2,\bm k_3)\varphi'_a(\eta_2,\bm k_4)\right] G^{P}_\sigma(s;\eta_1,\eta_2). 
\end{align}
Expressed in terms of the EFT coefficients defined above, this is 
\begin{align}
 &S_{\mathrm{IF}}^P[\varphi_r,\varphi_a] \supset  \int_{-\infty}^0 \dd \eta_2 \int\mathcal D\bm k\, 
 \left[ \varphi'_r(\eta_2,\bm k_3)\varphi'_r(\eta_2,\bm k_4) + \frac{1}{4}\varphi'_a(\eta_2,\bm k_3)\varphi'_a(\eta_2,\bm k_4)\right] \nonumber \\
 \times \bigg[& a^2(\eta_2) c_1(k_1,k_2,s;\eta_2) \varphi_r (\eta_2,\bm k_1) \varphi_a (\eta_2,\bm k_2) + c_2(k_1,k_2,s;\eta_2) \varphi'_r (\eta_2,\bm k_1) \varphi'_a (\eta_2,\bm k_2) \nonumber \\
 +& a(\eta_2) c_3(k_1,k_2,s;\eta_2) \varphi_r (\eta_2,\bm k_1) \varphi'_a (\eta_2,\bm k_2) + a(\eta_2) c_4(k_1,k_2,s;\eta_2) \varphi'_r (\eta_2,\bm k_1) \varphi_a (\eta_2,\bm k_2) \bigg] ,
\end{align}
where we used the symmetry property of the principal-value propagator. Combining the two parts and using the symmetry $(\bm k_1,\bm k_2,\bm k_3,\bm k_4,\bm s) \leftrightarrow(\bm k_3,\bm k_4,\bm k_1,\bm k_2,-\bm s)$ in $\int \mathcal D\bm k$, we finally have: 
\begin{align}
S_{\mathrm{IF}}^P&[\varphi_r,\varphi_a] = \int_{-\infty}^0 \dd \eta \int\mathcal D\bm k\, \\
\bigg\{&a^2(\eta) c_1(k_1,k_2,s;\eta)\left[ \varphi_{r,1}\varphi_{a,2}
\Big( \varphi'_{r,3}\varphi'_{r,4}+\frac14\varphi'_{a,3}\varphi'_{a,4}\Big)
+\Big(\varphi_{r,1}\varphi_{r,2}+\frac14 \varphi_{a,1}\varphi_{a,2}\Big)\varphi'_{r,3}\varphi'_{a,4} \right]\nonumber\\
+&c_2(k_1,k_2,s;\eta)\left[ \varphi'_{r,1}\varphi'_{a,2}
\Big( \varphi'_{r,3}\varphi'_{r,4}+\frac14\varphi'_{a,3}\varphi'_{a,4}\Big)
+\Big(\varphi'_{r,1}\varphi'_{r,2}+\frac14 \varphi'_{a,1}\varphi'_{a,2}\Big)\varphi'_{r,3}\varphi'_{a,4} \right] \nonumber \\
+&a(\eta) c_3(k_1,k_2,s;\eta)\left[ \varphi_{r,1}\varphi'_{a,2}
\Big( \varphi'_{r,3}\varphi'_{r,4}+\frac14\varphi'_{a,3}\varphi'_{a,4}\Big)
+\Big(\varphi_{r,1}\varphi'_{r,2}+\frac14 \varphi_{a,1}\varphi'_{a,2}\Big)\varphi'_{r,3}\varphi'_{a,4} \right]\nonumber \\
+& a(\eta) c_4(k_1,k_2,s;\eta)\left[ \varphi'_{r,1}\varphi_{a,2}
\Big( \varphi'_{r,3}\varphi'_{r,4}+\frac14\varphi'_{a,3}\varphi'_{a,4}\Big)
+\Big(\varphi'_{r,1}\varphi_{r,2}+\frac14 \varphi'_{a,1}\varphi_{a,2}\Big)\varphi'_{r,3}\varphi'_{a,4} \right]\bigg\}, \notag
\end{align}
where we have defined $\varphi_{i} \equiv \varphi(\eta,\bm k_i)$ for convenience. Using that $c_4(k_i,k_j,s;\eta) = c_3(k_j,k_i,s;\eta)$, we obtain the main text result.

Following the same procedure, we can rewrite the contribution from $G^\Delta_\sigma$ and $G^K_\sigma$ as integrals of local operators, which are:
\begin{align}
   S_{\mathrm{IF}}^\Delta[\varphi_r,\varphi_a] &= - \int_{-\infty}^0 \dd \eta \int\mathcal D\bm k\, \\
\bigg\{&a^2(\eta)c_1(k_1,k_2,s;\eta)\left[ \varphi_{r,1}\varphi_{a,2}
\Big( \varphi'_{r,3}\varphi'_{r,4}+\frac14\varphi'_{a,3}\varphi'_{a,4}\Big)
-\Big(\varphi_{r,1}\varphi_{r,2}+\frac14 \varphi_{a,1}\varphi_{a,2}\Big)\varphi'_{r,3}\varphi'_{a,4} \right]\nonumber\\
+&c_2(k_1,k_2,s;\eta)\left[ \varphi'_{r,1}\varphi'_{a,2}
\Big( \varphi'_{r,3}\varphi'_{r,4}+\frac14\varphi'_{a,3}\varphi'_{a,4}\Big)
-\Big(\varphi'_{r,1}\varphi'_{r,2}+\frac14 \varphi'_{a,1}\varphi'_{a,2}\Big)\varphi'_{r,3}\varphi'_{a,4} \right] \nonumber \\
+&a(\eta)c_3(k_1,k_2,s;\eta)\left[ \varphi_{r,1}\varphi'_{a,2}
\Big( \varphi'_{r,3}\varphi'_{r,4}+\frac14\varphi'_{a,3}\varphi'_{a,4}\Big)
-\Big(\varphi_{r,1}\varphi'_{r,2}+\frac14 \varphi_{a,1}\varphi'_{a,2}\Big)\varphi'_{r,3}\varphi'_{a,4} \right]\nonumber \\
+&a(\eta) c_4(k_1,k_2,s;\eta)\left[ \varphi'_{r,1}\varphi_{a,2}
\Big( \varphi'_{r,3}\varphi'_{r,4}+\frac14\varphi'_{a,3}\varphi'_{a,4}\Big)
-\Big(\varphi'_{r,1}\varphi_{r,2}+\frac14 \varphi'_{a,1}\varphi_{a,2}\Big)\varphi'_{r,3}\varphi'_{a,4} \right]\bigg\}, \notag
\end{align}
where we used the fact that $G^\Delta_\sigma(s; \eta_1,\eta_2) = 2G^P_\sigma(s; \eta_1,\eta_2) \sgn(\eta_1-\eta_2)$ (see \Eqs{eq_dSGH} and \eqref{eq_dSGD}) and
\begin{align}
S_{\mathrm{IF}}^K[\varphi_r,\varphi_a] &= ~ i\int_{-\infty}^0 \dd \eta \int\mathcal D\bm k\, \\
\bigg[&a^2(\eta)\tilde{c}_1(k_1,k_2,s;\eta)\varphi_{r,1}\varphi_{a,2} \varphi'_{r,3}\varphi'_{a,4}
+\tilde{c}_2(k_1,k_2,s;\eta)\varphi'_{r,1}\varphi'_{a,2} \varphi'_{r,3}\varphi'_{a,4} \nonumber \\
+&a(\eta)\tilde{c}_3(k_1,k_2,s;\eta)\varphi_{r,1}\varphi'_{a,2} \varphi'_{r,3}\varphi'_{a,4}
+ a(\eta)\tilde{c}_4(k_1,k_2,s;\eta)\varphi'_{r,1}\varphi_{a,2} \varphi'_{r,3}\varphi'_{a,4}\bigg], \notag
\end{align}
respectively, where we have defined new coefficients:
\begin{align}
&\tilde{c}_1(k_i,k_j,s;\eta) \equiv \frac{a^2(\eta)}{i\Lambda^2} \int_{-\infty}^{\eta} \dd \eta' a^2(\eta') G^{K}_\sigma(s;\eta,\eta') G^\varphi_{21}(k_i, \eta,\eta') G^\varphi_{21}(k_j, \eta,\eta') ,\\
&\tilde{c}_2(k_i,k_j,s;\eta) \equiv \frac{1}{i\Lambda^2} \int_{-\infty}^{\eta} \dd \eta' a^2(\eta') G^{K}_\sigma(s;\eta,\eta') G^\varphi_{22}(k_i, \eta,\eta') G^\varphi_{22}(k_j, \eta,\eta') ,\\
&\tilde{c}_3(k_i,k_j,s;\eta) \equiv \frac{a(\eta)}{i\Lambda^2} \int_{-\infty}^{\eta} \dd \eta' a^2(\eta') G^{K}_\sigma(s;\eta,\eta') G^\varphi_{21}(k_i, \eta,\eta') G^\varphi_{22}(k_j, \eta,\eta') ,\\
&\tilde{c}_4(k_i,k_j,s;\eta) \equiv \frac{a(\eta)}{i\Lambda^2} \int_{-\infty}^{\eta} \dd \eta' a^2(\eta') G^{K}_\sigma(s;\eta,\eta') G^\varphi_{22}(k_i, \eta,\eta') G^\varphi_{21}(k_j, \eta,\eta') .
\end{align}
Noticing that $\tilde{c}_4(k_i,k_j,s;\eta) = \tilde{c}_3(k_j,k_i,s;\eta)$, we obtain the main text results.

    
\bibliographystyle{JHEP}
\bibliography{biblio}

\end{document}